\documentclass[final,authoryear,5p]{elsarticle} 
\usepackage[nodisplayskipstretch]{setspace}
\usepackage{amssymb}
\usepackage{flushend}
\usepackage[
breaklinks=true,%
colorlinks=true,%
pdfauthor={M. J. Darnley \& M. Henze},%
pdftitle={On a century of extragalactic novae and the rise of the rapid recurrent novae}%
]{hyperref}
\usepackage{mathrsfs}
\usepackage{microtype}
\makeatletter
\def\@to{to}
\makeatother

\journal{Advances in Space Research}

\begin{document}

\begin{frontmatter}

\title{On a century of extragalactic novae and the rise of the rapid recurrent novae}

\author{Matthew J. Darnley} 
\address{Astrophysics Research Institute, Liverpool John Moores University, Liverpool, L3 5RF, UK}
\ead{M.J.Darnley@ljmu.ac.uk}

\author{Martin Henze\corref{cor2}} 
\address{Department of Astronomy, San Diego State University, San Diego, CA 92182, USA}

\begin{abstract}
Novae are the observable outcome of a transient thermonuclear runaway on the surface of an accreting white dwarf in a close binary system. Their high peak luminosity renders them visible in galaxies out beyond the distance of the Virgo Cluster. Over the past century, surveys of extragalactic novae, particularly within the nearby Andromeda Galaxy, have yielded substantial insights regarding the properties of their populations and sub-populations. The recent decade has seen the first detailed panchromatic studies of individual extragalactic novae and the discovery of two probably related sub-groups: the `faint--fast' and the `rapid recurrent' novae. In this review we summarise the past 100 years of extragalactic efforts, introduce the rapid recurrent sub-group, and look in detail at the remarkable faint--fast, and rapid recurrent, nova M31N\,2008-12a. We end with a brief look forward, not to the next 100 years, but the next few decades, and the study of novae in the upcoming era of wide-field and multi-messenger time-domain surveys.

\end{abstract}


\end{frontmatter}

\parindent=0.5 cm

\section{Introduction}\label{intro}

A long time ago in a galaxy far, far away....\ \citep{StarWars}:\ in 1909 a pair of nova eruptions in the 2.5\,Mly distant Andromeda Galaxy (M\,31) were discovered and followed photographically by George \citet{1917PASP...29..210R}.

Fast forward a hundred years to the present day and we have discovered over 1,100 nova eruptions in M\,31, have studied novae in almost two dozen galaxies, and have gained tremendous insights into nova physics and evolution from dedicated multi-wavelength surveys of extragalactic nova populations. At the threshold of a new golden age of multi-messenger time-domain surveys, we present a detailed review of many of the lessons learnt on the road so far.

The last extensive review of extragalactic nova populations was presented by \citet{2014ASPC..490...77S}, but this predated major multi-wavelength surveys, the rise of amateur observers, and the discovery of the `rapid recurrent novae' (RRNe). \citet{2017ASPC..509..515D} presented a brief review of the prototype RRN M31N\,2008-12a, but much has been discovered since then. In this review we will summarise the last century of extragalactic nova work, focussing in more detail on the last decade. We will introduce the rapidly expanding field of extragalactic novae, particular the newly discovered RRN subgroup with recurrence periods $P_\mathrm{rec}\leq10$\,yrs, along with the annually erupting M31N\,2008-12a. We will end with a look forward to the next few decades.

\section{Prerequisites}\label{pre}

Here we briefly summarise those aspects of nova astrophysics that will not be covered in detail in this review, yet are a necessary foundation and context for understanding the following sections.

\subsection{Nova physics:\ interacting binaries}

A classical nova (CN) eruption is the result of a thermonuclear runaway (TNR) on the surface of an accreting white dwarf \citep[WD; see][for the early history and recent reviews]{1949AnAp...12..281S,1951AnAp...14..294S,1957IAUS....3...77G,1959ApJ...130..916C,1972ApJ...176..169S,1976IAUS...73..155S,Sta08,2016PASP..128e1001S,1978A&A....62..339P,Jos16}. The TNR occurs following the accumulation of hydrogen-rich material from a donor star onto the WD within a close binary system.

Novae are a class of cataclysmic variable \citep[CV; see][]{1949ApJ...109...81S,1954ApJ...120..377J,1964ApJ...139..457K,1995cvs..book.....W}, where the donor is typically a late-type main sequence star and mass transfer usually proceeds through an accretion disk surrounding the WD. There are (observationally) small sub-classes where magnetic accretion or accretion columns play a role, and systems with further evolved donors; sub-giants or red giants \citep{2012ApJ...746...61D}.

\subsection{Multi-wavelength emission}

The TNR drives the ejection of material from the WD's surface at relatively high velocities. The expanding pseudo-photosphere (PP) of the initially optically-thick ejecta results in a rapid increase in luminosity \citep[see][for anthologies of recent reviews]{2008clno.book.....B,2014ASPC..490.....W}. What one observes depends upon the structure and geometry of those ejecta, and the emission and absorption processes within. In general, the observed radio, infrared (IR), optical, and ultraviolet (UV) emissions reflect the characteristics of the ejected shell. 

Following the TNR, nuclear burning continues on the WD surface in quasi-hydrostatic equilibrium, until the accreted fuel source is exhausted \citep{1978A&A....62..339P}. As the ejecta become optically thin, the PP recedes back to the WD, with peak emission migrating to shorter wavelengths. If the ejecta become fully transparent before the nuclear burning ceases, the underlying `super-soft X-ray source' (SSS) may be revealed \citep[see, for e.g.,][]{2006ApJ...651L.141H,2008ASPC..401..139K,2015JHEAp...7..117O}. Importantly, the visibility windows for the SSS emission are typically much longer (years to decades) than for the optical light  \citep[weeks to months; see, for e.g.,][]{2014A&A...563A...2H}.

\subsection{Nova evolution and the supernova connection}

It is widely accepted that all novae inherently recur. Following each eruption the WD and donor remain (relatively) unscathed and accretion may soon reestablish -- allowing the cycle to begin anew. An observationally small sub-set, dubbed the recurrent novae (RNe), have been observed to undergo multiple eruptions. Observed values of $P_\mathrm{rec}$ range from 1\,yr \citep{2014A&A...563L...9D} up to 98\,yrs \citep{2009AJ....138.1230P}. It seems most likely that both ends of this scale are simply due to current observational limits. 

Novae have long been heralded as one of the possible single-degenerate pathways toward type Ia supernovae \citep[SNe\,Ia; see, for e.g.,][]{1973ApJ...186.1007W,1999ApJ...522..487H,1999ApJ...519..314H,2000ARA&A..38..191H}, as, to be absolutely fair, have almost all scenarios that allow a WD to increase in mass. But a number of questions regarding the viability of the nova pathway have been posed:

Do the WDs in novae increase in mass? This is particularly important as only CO WDs grow to produce SNe\,Ia; their ONe cousins result in an accretion-induced collapse to a neutron star \citep{1996ApJ...459..701G}, once the \citet{1931ApJ....74...81C} mass is surpassed. Pioneering multi-cycle nova eruption models have now shown that the WDs do indeed increase in mass, with little or no tuning of the initial parameters \citep{2014MNRAS.437.1962H,2015MNRAS.446.1924H,2016ApJ...819..168H}. A number of other authors have arrived at similar results \citep[see, for e.g.,][]{2008NewAR..52..386H,2012BASI...40..419S,Sum19,doi:10.1063/1.4866984,2017ApJ...844..143K}. 

Are the WDs in the RN systems -- those already close to the Chandrasekhar mass -- CO or ONe? To date, there remains no published evidence for super-Solar abundances of Ne in the ejecta of RNe \citep[see, e.g.,][]{2013A&A...556C...2M}. 

Finally, are there simply enough novae, accreting at a high enough rate, to impart a measurable impact as a SN\,Ia pathway? We don't know (see Section~\ref{sec:rrne}). But, if novae do provide a significant channel then they hold an advantage over other progenitors, they are by far the most luminous, allowing their populations to be studied out to $\sim20$\,Mpc \citep[see, for e.g.,][]{2015ApJ...811...34C}, and beyond.

\subsection{The advantages and drawbacks of Galactic novae}

Novae in the Galaxy, and even in the Magellanic Clouds, have been studied individually in increasingly exquisite detail across the electromagnetic spectrum \citep[see, for e.g.,][]{2010ApJ...724..480H,2016ApJ...820..104H,2016ApJ...818..145B,2018MNRAS.474.2679A}. Even early-eruption $\gamma$-ray emission is now routinely observed from Galactic novae \citep[see, for e.g.,][]{2010Sci...329..817A,2014Sci...345..554A}, although the underlying mechanism is yet to be fully understood \citep[see][]{2014Natur.514..339C}. With current capabilities, any $\gamma$-rays can only be detected from nearby Galactic novae and hard X-ray detections from the eruptions (not to mention during quiescence) are almost exclusively restricted to Milky Way systems \citep[there is some evidence for early post-eruption hard X-ray emission from the 2016 eruption of the RN LMC 1968;][]{2016ATel.8587....1D,2019arXiv190903281K}.  

Our {\it privileged} location within the spiral structure of the Milky Way (and the irregular nature of the LMC and SMC) severely limits the ability to undertake unbiased studies of the population(s) of Galactic or similarly nearby novae. While the second data release (DR2) from the {\it Gaia} mission \citep{2016A&A...595A...1G,2018A&A...616A...1G} may have removed some ambiguity from Galactic distance estimates (see the discussions in \citealt{2018MNRAS.481.3033S} and \citealt{2019A&A...622A.186S}), we must wait for at least the fourth release (DR4) until the potential systematics \citep[in part due to the orbital motion of the unresolved nova binaries; see, for e.g.,][]{2018A&A...616A...2L} can be investigated.

There remain uncertainties on the gas and dust columns toward each Galactic nova that severely impact their (individual and) population studies, with potentially only a small fraction of Galactic novae  observable \citep[and even less observed;][]{2014AAS...22430604S,2017ApJ...834..196S}. The trials and tribulations of inferring the Galactic rate from a small spatially constrained sample has led to estimates that range from 11\,yr$^{-1}$ \citep{1990AJ.....99.1079C} to 260\,yr$^{-1}$ \citep{1972SvA....16...41S}. The most plausible estimate of the Galactic nova rate is perhaps the most recent (but relatively unconstrained) of $50^{+31}_{-23}\,\mathrm{yr}^{-1}$ \citep{2017ApJ...834..196S}.

\section{Extragalactic novae}

To minimise the effects of distance and extinction uncertainties, we turn to the study of extragalactic nova populations. And while still far from ideal, the close to edge-on M\,31 \citep[$77^\circ$ inclination;][]{1958ApJ...128..465D} is the preferred laboratory for such studies. At a distance of $752\pm17$\,kpc \citep{2001ApJ...553...47F} and experiencing a foreground reddening of $E\left(B-V\right)\approx0.1$ \citep{1992ApJS...79...77S}, eruptions of the entire peak-luminosity range of M\,31 novae are readily accessible to professional and amateur astronomers alike. Techniques, such as narrowband H$\alpha$ imaging \citep{1987ApJ...318..520C}, or difference image analysis \citep{2010MNRAS.409..247K}, allow the recovery of transients down to the central $\sim\!10^{\prime\prime}$ ($\sim\!40$\,pc) of the bright M\,31 bulge.

\subsection{A century of M31 novae:\ surveys and nova rates}\label{M31cent}

As stated by Edwin \citet{1929ApJ....69..103H}, ``In 1885 interest in (M\,31) was stimulated by the appearance of a nova very close to the nucleus''. That `nova', S\,Andromedae \citep{1885AN....112..355H,1885AReg...23..242W}, turned out to be a SN explosion \citep[SN\,1885A; see discussion by][]{1985ApJ...295..287D}. While a handful of M\,31 nova candidates were retroactively found in 1909 data \citep{1917PASP...29..210R,1929ApJ....69..103H}, the first confirmed eruptions were a pair  discovered (by Hubble) in 1932 and observed spectroscopically by Milton \citet{1932PASP...44..381H} from the Mount Wilson Observatory\footnote{From the description given in \citet{1932PASP...44..381H}, it is possible that M31N\,1925-09a may have been spectroscopically confirmed via a slit-less spectrum taken by that author.}. In the following century, the number of nova candidates in M\,31 has grown beyond 1,100\footnote{According to the on-line extragalactic nova database of \citet{2010AN....331..187P}:\ \url{http://www.mpe.mpg.de/~m31novae/index.php}}. The number of spectroscopically confirmed M\,31 novae now exceeds 200 \citep{2011ApJ...734...12S,Ransome2019}.

The most famous M\,31 nova survey was the first, but not due to the novae. Along with the first extragalactic nova sample, \citet {1929ApJ....69..103H} published the first catalogue of Cepheid variables in M\,31. The latter of course led directly to a distance determination toward M\,31 \citep{1929ApJ....69..103H} and ultimately the understanding of the scale of the Universe and the essence of galaxies --- settling the `Great Debate' of 1920 between Harlow Shapley and Heber Curtis \citep[see][for a transcript of that debate]{1921BuNRC...2..171S}. From the 85 nova candidates included in his catalogue, \citeauthor{1929ApJ....69..103H}\ estimated a global M\,31 rate of $\sim30\,\mathrm{yr}^{-1}$.

Subsequent surveys of novae in M\,31 are summarised in Table~\ref{nova_rate} together with the evolution of estimates of the galaxy-wide eruption rate. Around half the M\,31 nova candidates have been discovered by these surveys; the remainder by all-sky (particularly SN) surveys or by individuals. Special mention must be made of the exceptional efforts of  Kamil Hornoch, and the amateur astronomer team of Koichi Nishiyama and Fujio Kabashima, who between them have discovered in excess of 200 M\,31 novae. 

\begin{table}
\caption{A summary of the principal M\,31 nova surveys.\label{nova_rate}}
\begin{center}
\begin{tabular}{lll}
\hline
\hline
{Survey} & {Novae} &  {Rate} \\
{} & {} & {[yr$^{-1}$]} \\
\hline
\citet{1929ApJ....69..103H} & 85 & $\sim30$ \\
\citet{1956AJ.....61...15A} & 30 & $24\pm4$ \\
\citet{1989AJ.....97.1622C}$^{\dag}$ & 142 & $29\pm4$ \\
\citet{1987ApJ...318..520C,1990ApJ...356..472C} & 40 & \ldots \\
\citet{1991ApSS.180..273S,1992ApSS.190..119S} & 33 & \ldots \\
\citet{1992ApJS...81..683T} & 9 & \ldots \\
\citet{1999AAS...195.3608R} & 44 & \ldots \\
\citet{2001ApJ...563..749S} & 72 & $37^{+12}_{-8}$ \\
\citet{2004MNRAS.353..571D,2006MNRAS.369..257D}$^{\ddag}$ & 20 & $65^{+16}_{-15}$ \\
\citet{2005AJ....130...84F}$^{\ddag}$ & 19 & \ldots\\
\citet{2011ApJ...734...12S} & 44 & \ldots \\
\citet{2011ApJ...735...94K} & 6 & \ldots \\
\citet{2012ApJ...752..133C} & 29 & \ldots \\
\citet{2012AA...537A..43L} & 91 & \ldots \\
\citet{Ransome2019} & 180 & \ldots \\
\hline
\end{tabular}
\end{center}
\begin{minipage}{\columnwidth}
\setstretch{0.75}
{\footnotesize $^{\dag}${\citet{1989AJ.....97.1622C} presents a combined analysis of the three Asiago surveys \citep{1964AnAp...27..498R,1973A&AS....9..347R,1989AJ.....97...83R}.}} \\
{\footnotesize $^{\ddag}${\citet{2004MNRAS.353..571D,2006MNRAS.369..257D} and \citet{2005AJ....130...84F} reported independent analyses of the the POINT-AGAPE microlensing survey data \citep[see][]{2001ApJ...553L.137A}.}}
\end{minipage}
\end{table}

The most recent observational determination of the M\,31 nova rate was produced by \citet{2004MNRAS.353..571D,2006MNRAS.369..257D}, who used a high-cadence, multi-colour survey to estimate a rate of $65^{+16}_{-15}$\,yr$^{-1}$ --- almost twice that of previous studies. Being the first to implement an `automated' nova survey, that exclusively used algorithms to detect and classify novae, \citeauthor{2006MNRAS.369..257D}\ found that the M\,31 nova distribution closely followed a combination of the bulge and disk light of the host. They also reported that while the novae were therefore clustered around the central bulge, the disk contribution to the overall population was also significant:\ with rates of $38^{+15}_{-12}$ (bulge)\footnote{Note that the reported bulge rate is similar to the M\,31-wide `bulge dominated' rate of \citet{2001ApJ...563..749S}.} and  $27^{+19}_{-15}$ (disk), see Section~\ref{two_pops} for further discussion.

This high global rate is consistent with the large numbers of novae now routinely discovered each year in M\,31, particularly as larger area detectors and all-sky surveys have improved spatial and temporal coverage of the galaxy.

There are a number of reasons why the earlier surveys resulted in relatively low determined rates. \citet{1956AJ.....61...15A} and \citet{1964AnAp...27..498R,1973A&AS....9..347R} both reported a substantial decrease in the nova population toward the centre of the bulge -- however, this was a selection effect due to surface brightness limitations at the time. Using narrowband H$\alpha$ observations, which are not as affected by the central surface brightness, \citet{1987ApJ...318..520C} was the first to propose that the nova distribution followed the M\,31 galactic light all the way into the centre. Many of the earlier surveys concentrated on the bulge (and therefore simply missed the disk novae) or may have over estimated completeness \citep[see][for a relevant discussion]{2006MNRAS.369..257D}. To address this, \citet{2001ApJ...563..749S} extended their survey to cover the M\,31 disk, but reported that the nova distribution is (still) consistent with an association to the bulge.

\subsection{Selection effects and corrections}

Despite its advances, when we look in more detail at the \citet{2004MNRAS.353..571D,2006MNRAS.369..257D} work, we note that their survey did not detect any novae with speed classes \citep[the time taken for a nova to decay by two magnitudes from peak brightness;][]{1964gano.book.....P} $t_2\lesssim10$\,days, nor any with $t_2\gtrsim215$\,days\footnote{This was, in part, due to the choices made when designing the detection algorithms, as \citet{2004ApJ...601..845A} and \citet{2005AJ....130...84F}, who both also used the POINT-AGAPE data set, discovered a handful of faster novae that were not in the \citeauthor{2004MNRAS.353..571D}\ catalogue.}. The subsequent completeness analysis did not include any novae with speed classes beyond the observed range. The computed rates are therefore only applicable to the quoted $t_2$ range and as such \textit{are lower limits when considering the entire eruptive population}. 

With that limitation in mind, \citet{2016MNRAS.455..668S} utilised the \citet[which contains numerous fast, $t_2\leq20$\,d, novae]{1956AJ.....61...15A} catalogue with the \citeauthor{2004MNRAS.353..571D}\ catalogue to attempt to correct for the latter's completeness bias. \citeauthor{2016MNRAS.455..668S}\ assume that the M\,31 novae follow the galactic light and found that $\sim30\%$ of M\,31 novae must be fast ($t_2<10$\,days), yielding a corrected rate $\approx106$\,yr$^{-1}$. \citet{2016MNRAS.458.2916C} coupled a population synthesis approach with the nova eruption model from \citet[also see Section~\ref{two_pops}]{2006MNRAS.369..257D} to derive an M\,31 rate of 97\,yr$^{-1}$, again indicating a `missing' population of the fastest novae. To date a large population of very fast M\,31 novae has not been uncovered (also see Section~\ref{sec_mmrd}), but we do note that there has not been a dedicated campaign to detect such eruptions.

\subsection{Studies of individual extragalactic novae}

The last decade has seen a rapid development in the scope of observations toward individual extragalactic novae. Historically, observations of M\,31 novae typically consisted of sparsely populated light curves and the occasional spectrum. Now, Local Group novae are routinely spectroscopically confirmed, often have detailed multi-colour optical light curves, plus the inclusion of UV and X-ray observations, even late-time infrared observations have been attempted utilising {\it Spitzer} \citep{2011ApJ...727...50S}. 

This era of extensive panchromatic studies of extragalactic novae began in earnest in 2007 when four separate eruptions were examined in detail. M31N\,2007-11a was one of the first M\,31 novae to be studied extensively in the optical and X-ray \citep[the latter via {\it Chandra} and {\it XMM-Newton};][]{2009A&A...498L..13H}. The slowly rising yet luminous M31N\,2007-11d was among the first to be studied in detail optically and with multiple epochs of spectroscopy \citep{2009ApJ...690.1148S}. A study of the RN {\it candidate} M31N\,2007-12b quickly followed \citep{2009ApJ...705.1056B}, which combined photometric and spectroscopic evolution with a Neil Gehrels {\it Swift} Observatory detection of the SSS, and the first recovery of a nova progenitor system (which contains a red giant donor) beyond the Milky Way. Further analysis of 2007-12b by \citet{2011A&A...531A..22P} reported additional SSS observations, likely measured the WD rotation period (and potentially the orbital period), and proposed that the system may be an intermediate polar (see \citealt{1977ivsw.conf..238K} and \citealt{1983ASSL..101..155W}). Finally, M31N\,2007-06b became the first CN in an M\,31 Globular Cluster (GC) discovered optically \citep{2007ApJ...671L.121S} and subsequently detected in X-rays \citep{2009A&A...500..769H}.

But it is M31N\,2008-12a, first discovered optically the following year, that has become by far the best studied extragalactic nova to date --- we devote Section~\ref{12a} entirely to that remarkable system.

\subsection{Multiple populations within a single host\label{two_pops}}

The proposal that multiple nova populations may coexist in the same galaxy was initially postulated by \citet{1990LNP...369...34D} and was expanded by \citet{1992A&A...266..232D}. A two-population model was formulated due to evidence for bright--fast novae showing association with the Milky Way `thin disk', whereas the faint--slow novae arose from a more spatially extended `thick disk' or bulge population. \citet{1998ApJ...506..818D} further proposed that the bright--fast novae all belonged to the He/N taxonomic spectral class  \citep[see][]{1992AJ....104..725W,2012AJ....144...98W} and were all located at scale heights within 100\,pc of the Galactic plane, contained high mass WDs ($M_\mathrm{WD}$) and were related to Population\,I (relatively young). In contrast, the faint--slow novae were typically Fe\,{\sc ii} novae that extended up to $\sim1$\,kpc beyond the plane, contain low $M_\mathrm{WD}$ and are Population\,II (relatively old). This result has been questioned by \citet{2018MNRAS.476.4162O} who did not find evidence for slow or fast, or Fe\,{\sc ii} or He/N, novae having different Galactic scale height distributions, but instead found that all novae are largely concentrated within the Galactic disk -- a result that they predominantly put down to advances in catalogue completeness, particularly spectroscopically. The spatial distribution of Galactic novae has, however, yet to be studied in a post-Gaia era, so the apparent contradiction between these two studies may soon be understood.

In M\,31, the work by \citet{1987ApJ...318..520C}, \citet{1989AJ.....97.1622C}, and \citet{2001ApJ...563..749S} reported a strong association between the bulge light and the nova distribution -- with little evidence for a {\it substantial} disk contribution\footnote{The disk contribution required to match observations has evolved upward with time.}. Of course, selection effects may have played some part. If faster novae do tend to reside in the disk \citep[as proposed by][]{1992A&A...266..232D}, then survey cadence could impact the ability to detect disk novae. 

\citet{1987ApJ...318..520C}, \citet{2001ApJ...563..749S}, and \citet{2006MNRAS.369..257D} each presented a single parameter model for the nova distribution within M\,31:

\begin{equation}
\Psi_i=\frac{\theta\mathscr{L}_i^d+\mathscr{L}_i^b}{\theta\sum_i\mathscr{L}_i^d+\sum_i\mathscr{L}_i^b},\label{nova_mod}
\end{equation}

\noindent where $\Psi_i$ is the probability of a nova erupting at a given location, $i$, that has a contribution $\mathscr{L}^d_i+\mathscr{L}^b_i$ from the disk and bulge light, respectively. The wavelength dependant parameter $\theta$ is the ratio of the disk and bulge eruption rates per unit light. This approach allows exploration of the population distribution(s) without explicitly assigning a `bulge' or `disk' origin to individual novae. Due to a limited number of novae and a bulge dominated survey, \citeauthor{1987ApJ...318..520C}\ were restricted to placing an upper limit of $\theta<0.1$ --- i.e.\ a bulge dominated population ($\theta=0$ represents a bulge {\it only} population). \citeauthor{2001ApJ...563..749S}\ derived $\theta=0.41^{+0.40}_{-0.25}$ by considering the $B$-band galactic light. 

The \citet{2006MNRAS.369..257D} analysis led to a determination of $\theta=0.18^{+0.24}_{-0.10}$ (when considering the $r'$-band M\,31 light\footnote{$1\sigma$ confidence limits, the distribution is non-Gaussian.}), i.e.\ the bulge nova rate per unit light is $\sim5$ times that of the disk, and ruled out that the novae follow the $r'$-band light (i.e.\ $\theta=1$) of M\,31 at beyond the 95\% level --- thereby lending strong support to separate `bulge' and `disk' populations.

\citet{2011ApJ...734...12S} presented a spectroscopic and photometric catalogue of 46 M\,31 novae, bringing (at the time) the number of spectroscopically confirmed systems up to 91. This work confirmed that the M\,31 proportion of Fe\,{\sc ii} (82\%) and He/N (18\%) novae was consistent with that measured in the Milky Way \citep{1998ApJ...506..818D,2007AAS...211.5115S}. By combining their data set with that of \citet{1989AJ.....97.1622C}, \citeauthor{2011ApJ...734...12S}\ demonstrated that the M\,31 `fast novae' ($t_2\le25$\,d) were more spatially extended than their slower counterparts ($t_2>25$\,d), as might be expected if a younger disk population contained novae with on average higher $M_\mathrm{WD}$ \citep[as proposed by][for the Milky Way]{1992A&A...266..232D}. However, \citeauthor{2011ApJ...734...12S}\ were unable to find compelling evidence for a difference in the spatial distribution of the M\,31 Fe\,{\sc ii} and He/N novae.

Combining the nova catalogue of \citet{2011ApJ...734...12S} and multi-waveband {\it Hubble Space Telescope (HST)} imaging of the north-eastern half of M\,31 \citep[the PHAT survey;][]{2012ApJS..200...18D}, \citet{2014ApJS..213...10W} undertook the first extragalactic survey for nova progenitor systems. From an input catalogue of 38 novae, \citeauthor{2014ApJS..213...10W}\ recovered the progenitors of 11 systems -- those harbouring giant donors and/or bright accretion disks (both potential indicators of a high mass accretion rate, $\dot{M}$). The subsequent statistical analysis found the proportion of M\,31 novae with luminous progenitors is $30^{+13}_{-10}\%$ \citep[$>10\%$ at the 99\% confidence level;][]{2016ApJ...817..143W}. This analysis also indicated that these luminous progenitors were more likely to be associated with the disk population, and the authors could not formally exclude the possibility that all of these systems were disk novae \citep{2016ApJ...817..143W}.

\subsection{The X-ray properties of Andromeda Galaxy novae}\label{xrayprop}

A new and crucial angle was added to the nova population research when \citet{2005A&A...442..879P} used their existing large {\it XMM-Newton} surveys of M\,31 and M\,33 to specifically identify nova X-ray counterparts --- increasing the M\,31 sample size by more than a factor of four\footnote{They also utilised archival M\,31 data from {\it ROSAT} \citep[see surveys by][]{1996LNP...472...75G,2004ApJ...610..261G} and {\it Chandra}.}. While the nova rate in M\,33 is too low to allow a population approach, the M\,31 numbers were significant. \citet{2005A&A...442..879P} concluded that nova eruptions are the main source of transient SSSs in M\,31. In a follow-up study, \citet{2007A&A...465..375P} analysed more recent archival {\it Chandra} and {\it XMM-Newton} data to find additional novae --- among them objects with unexpectedly short SSS states of only a few months alongside novae that remained X-ray bright almost a decade post-eruption. The superior performance of this new generation of large X-ray telescopes, {\it XMM-Newton} and {\it Chandra}, was promising strong synergies with the high nova rate of M\,31.

Building upon those pioneering surveys, \citet{2010A&A...523A..89H,2011A&A...533A..52H,2014A&A...563A...2H} undertook a series of X-ray surveys between 2006 and 2012. These surveys were designed specifically for nova discovery:\ they used cadences of 10 days to study short SSS phases, focussed only on the bulge of M\,31 where most novae are found, and used a coordinated observing strategy of {\it XMM-Newton} and {\it Chandra} pointings to cover the galaxy during a continuous 3-4 months\footnote{The small Sun angle of M\,31 during part of the year strongly affects visibility especially for {\it XMM-Newton}.}. The unparalleled spatial resolution of {\it Chandra} allowed the first X-ray detections of novae close to the M\,31 core. The superior effective area of {\it XMM-Newton} provided the depth to detect faint sources and perform low-resolution spectroscopy for the brighter ones.

By the final paper of the series, \citet{2014A&A...563A...2H} had increased the sample size of M\,31 novae with X-ray detections to 79 and derived a large set of SSS parameters alongside optical properties from support or community observations. \textit{For the first time, it was possible to study the X-ray vs optical parameters of novae using population statistics.} \citet{2010A&A...523A..89H} had found a correlation between the optical decline time and the duration of the SSS phase, confirming a similar result found for Galactic novae \citep{2011ApJS..197...31S}. \citet{2010A&A...523A..89H} also reported tentative evidence for differing X-ray properties between M\,31 bulge and disk novae.

\begin{figure}[ht!]
\begin{center}
\includegraphics[width=0.9\columnwidth]{{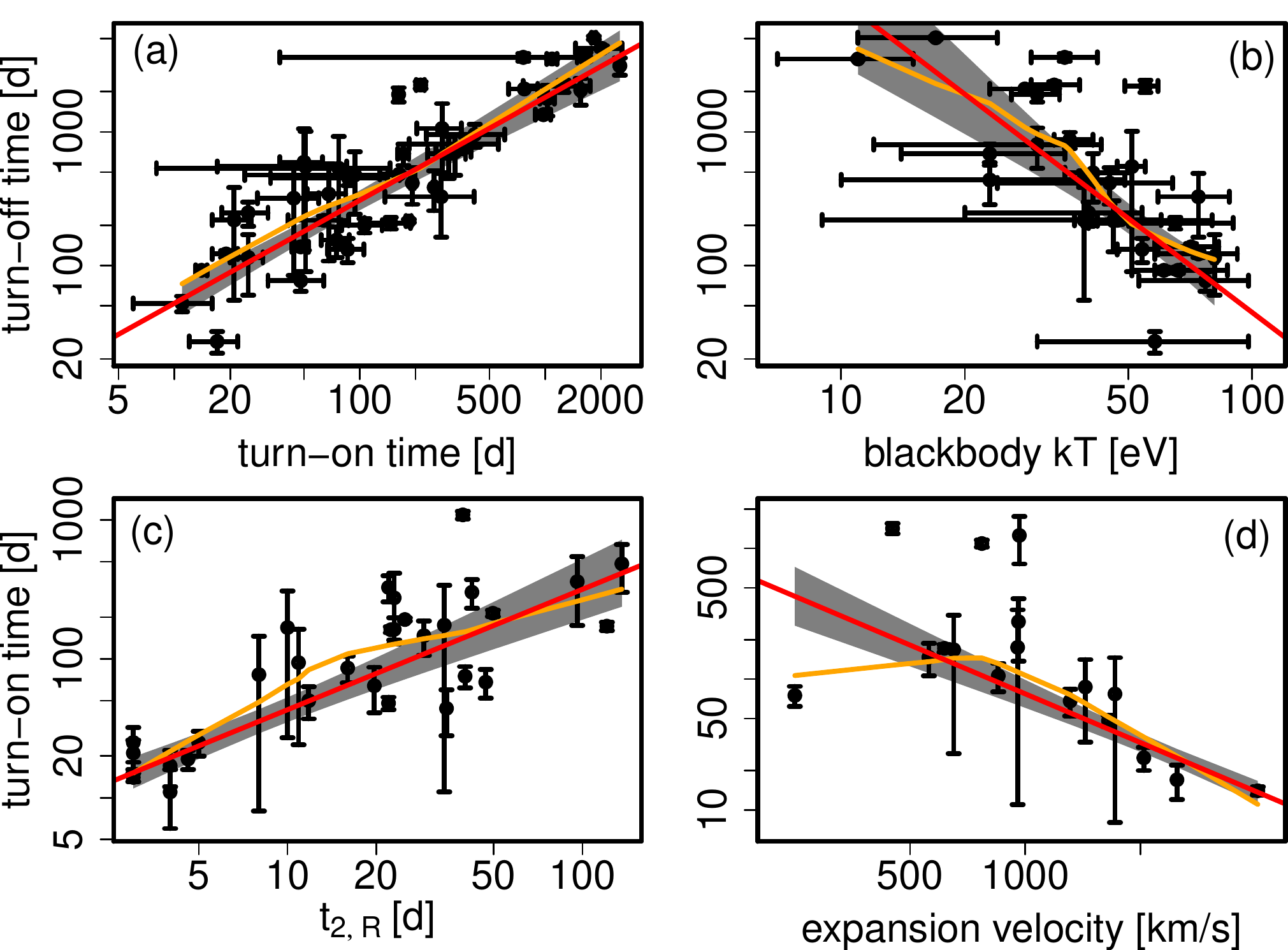}}
\end{center}
\caption{M\,31 nova X-ray vs optical correlations based on \citet{2014A&A...563A...2H}. Black:\ data; orange:\ smooth fit for visualisation; red:\ robust power-law fit with corresponding 95\% confidence regions.\label{fig:xcor}}
\end{figure}

Using the complete sample, \citet{2014A&A...563A...2H} discovered strong correlations between five fundamental observable nova parameters:  the `turn-on' ($t_\mathrm{on}$) and `turn-off' ($t_\mathrm{off}$) times of the SSS, the SSS black-body (BB) effective temperature\footnote{SSS spectra are not BBs \citep[see, for e.g.,][among many others]{2013A&A...559A..50N}, yet BB fits can serve as a consistent parametrisation.} ($k_\mathrm{B}T$), the optical decline time ($t_2$), and the ejecta expansion velocity ($v_\mathrm{exp}$) as derived from optical spectra. Many of these relations are now routinely used in the planning of extragalactic X-ray observations of novae. In essence:\ \textit{novae that decline fast in the optical have short and high-temperature SSS states.} In Figure~\ref{fig:xcor} we show correlations based on \citet{2014A&A...563A...2H} and here we reproduce the corresponding best fits:

\begin{equation}
 t_\mathrm{on} = 10^{\left(0.8\pm0.1\right)} \cdot t_{2,R}^{\left(0.9\pm0.1\right)}\,\left[\mathrm{days}\right],
\end{equation}

\begin{equation}
 t_\mathrm{on} = 10^{\left(5.6\pm0.5\right)} \cdot v_\mathrm{exp}^{\left(-1.2\pm0.1\right)}\,\left[\mathrm{days}\right],
\end{equation}

\begin{equation}
t_\mathrm{off} = 10^{\left(0.9\pm0.1\right)}\cdot t_\mathrm{on}^{\left(0.8\pm0.1\right)}\,\left[\mathrm{days}\right],
\end{equation}

\begin{equation}
 t_\mathrm{off} = 10^{\left(6.3\pm0.5\right)} \cdot \left(k_\mathrm{B}T\right)^{\left(-2.3\pm0.3\right)}\,\left[\mathrm{days}\right],
\end{equation}

Beyond being a powerful tool for understanding nova population physics, X-ray observations are crucial for discovering a rare subset of novae:\ those found in GCs. While the intrinsic brightness of (extragalactic) GCs renders optical nova detections difficult, there are no bright SSSs in GCs other than novae (but many harder X-ray sources). With two confirmed plus one likely GC novae \citep{2013A&A...549A.120H}, M\,31 hosts three of the six known GC novae.

The first M\,31 GC nova, M31N\,2007-06b, was discovered in the optical by \citet{2007ApJ...671L.121S} and soon after in X-ray observations by \citet[during their large nova survey]{2009A&A...500..769H}. In the same survey season these authors discovered another GC SSS. Note that SSSs in GCs are also a very rare occurrence but that no optical counterpart was found for this object (yet a nova could not be excluded). The latest M\,31 GC nova, M31N\,2010-10f, was first found in a serendipitous X-ray observation and later confirmed optically \citep{2013A&A...549A.120H}. It is noteworthy that all three of the M\,31 GC novae (candidates) exhibited a short SSS phase. An exploration of this observational property with theoretical models was carried out by \citet{2013ApJ...779...19K}.

\subsection{Beyond Andromeda}

Nova eruptions have been detected in many of the Local Group galaxies, with nova rates determined for the largest constituents (see Table~\ref{LG_rates} for a summary of the most recently published rates). However, the populations of M\,31 and the Milky Way, which constitute the vast majority of the Local Group stellar mass and therefore the vast majority of the nova eruptions, are by far the best studied. 

\begin{table}
\caption{Nova rates for Local Group galaxies.\label{LG_rates}}
\begin{center}
\begin{tabular}{lll}
\hline
\hline
{Galaxy} & {Rate} &  {Reference} \\
{} & {[yr$^{-1}$]}\\
\hline
Milky Way & $50^{+31}_{-23}$ & \citet{2017ApJ...834..196S}\\
LMC & $2.4\pm0.8$ & \citet{2016ApJS..222....9M} \\
SMC & $0.9\pm0.4$ & \citet{2016ApJS..222....9M} \\
M\,31$^{\dag}$ & $65^{+16}_{-15}$ & \citet{2006MNRAS.369..257D}\\
M\,32$^{\ddag}$ & $2^{+2}_{-1}$ & \citet{2005AJ....129.1873N}\\
M\,33 & $2.5\pm1.0$ & \citet{2004ApJ...612..867W} \\
M\,110$^{\ddag}$ & $2^{+2}_{-1}$ & \citet{2005AJ....129.1873N}\\
NGC\,147$^{\ddag}$ & $<2$ & \citet{2005AJ....129.1873N} \\
NGC\,185$^{\ddag}$ & $<1.8$ & \citet{2005AJ....129.1873N} \\
\hline
\end{tabular}
\end{center}
\begin{minipage}{\columnwidth}
\setstretch{0.75}{\footnotesize $^{\dag}${See discussion about a possible elevated rate in Section~\ref{M31cent}.}}\\
{\footnotesize $^{\ddag}${The  \citeauthor{2005AJ....129.1873N}\ rates are estimates based upon a single nova in each of M\,32 and M\,110, and no detections in NGC\,147 or NGC\,185.}}
\end{minipage}
\end{table}

\citet{2012ApJ...752..156S} published the first photometric and spectroscopic analysis of the nova population of M\,33. This catalogue contained 36 novae (the majority drawn from the literature) of which 8 yielded spectra (6 newly reported), and directly compared the M\,33 population to that of M\,31 \citep[largely following][]{2011ApJ...734...12S}. Unlike M\,31, \citeauthor{2012ApJ...752..156S}\ found that most M\,33 novae (5/8) belonged to the He/N spectral class, and that only two novae were clearly Fe\,{\sc ii} (cf.\ 82\% for M\,31). Those authors concluded that the spectroscopic mix of M\,33 novae differed from that of M\,31 at the 99\% confidence level. 

In the LMC, \citet{2013AJ....145..117S} again confirmed the connection between spectral type and decline time. As with M\,33, only around half of the LMC novae were classified as Fe\,{\sc ii}, and the LMC nova population is more rapidly evolving than that of the Milky Way and M\,31. \citeauthor{2013AJ....145..117S}\ proposed that the LMC nova population is younger than that of the M\,31 bulge, and therefore contains {\it on average} higher mass WDs that evolve more rapidly. \citeauthor{2013AJ....145..117S}\ also comments on the large proportion of known RNe within the LMC population ($\sim10\%$ of systems, or $\sim16\%$ of eruptions).

Recently, individual novae have been studied in detail in IC\,1613 \citep{2017MNRAS.472.1300W} and NGC\,6822 \citep{2019MNRAS.486.4334H}, both hosts are dwarf irregular galaxies in the Local Group and, like the Magellanic Clouds, provide further examples of novae in low metallicity environments \citep[see the discussion within][]{2013AstRv...8a..71O}. 

\subsection{Beyond the Local Group and the `LSNR'}

The study of extragalactic novae is not constrained to the Local Group. In Table~\ref{nonLG_rates} we provide a summary of some of the more distant extragalactic nova work -- out to, and including, the Virgo Cluster. A recent highlight within that realm is the results of a {\it HST} survey toward M\,87 
by \citet{2016ApJS..227....1S}. In a similar vein to the POINT-AGAPE survey of M\,31 (see Section~\ref{M31cent}), a micro-lensing survey was repurposed to study nova eruptions. In a result entirely independent of an earlier yet similar one presented in \citet{2002AIPC..637..457S}, \citet{2016ApJS..227....1S} derived a nova rate for that giant elliptical galaxy of $\sim400$\,yr$^{-1}$. As noted by \citet{2017RNAAS...1a..11S}, that rate is over double those derived from ground-based observations \citep[see][]{2000ApJ...530..193S,2015ApJ...811...34C}. \citeauthor{2017RNAAS...1a..11S}\ undertook an independent analysis of the {\it HST} data reporting that their results ``are in general agreement'' with  \citet{2016ApJS..227....1S}. But \citeauthor{2017RNAAS...1a..11S}\ urge caution, particularly when deriving nova rates using unconfirmed (spectroscopically) nova eruptions and especially when extrapolating a rate beyond the constraints of the survey data.

\begin{table}
\caption{Nova rates for galaxies beyond the Local Group.\label{nonLG_rates}}
\begin{center}
\begin{tabular}{lll}
\hline
\hline
{Galaxy} & {Rate} &  {Reference} \\
{} & {[yr$^{-1}$]}\\
\hline
M\,49 & $189_{-22}^{+26}$ & \citet{2015ApJ...811...34C} \\
M\,51 & $18\pm7$ & \citet{2000ApJ...530..193S} \\
M\,81 & $ 33^{+13}_{-8}$ & \citet{2004AJ....127..816N} \\
M\,84 & $95_{-14}^{+15}$ & \citet{2015ApJ...811...34C} \\
M\,87 & $363_{-45}^{+33}$ & \citet{2016ApJS..227....1S}\\
M\,94 & $5.0^{+1.8}_{-1.4}$ & \citet{2010ApJ...720.1155G} \\
M\,100 & $\sim25$ & \citet{1996ApJ...468L..95F} \\
M\,101 & $11.7^{+1.9}_{-1.5}$ & \citet{2008ApJ...686.1261C} \\
NGC\,1316 & $135\pm45$&\citet{2002Sci...296.1275D} \\
NGC\,2403 & $2.0^{+0.5}_{-0.3}$ & \citet{2012ApJ...760...13F}\\
NGC\,5128 & $8.0\pm2.8$ & \citet{1990AJ.....99.1079C}\\
\hline
\end{tabular}
\end{center}
\end{table}

\begin{figure}
\begin{center}
\includegraphics[width=0.9\columnwidth]{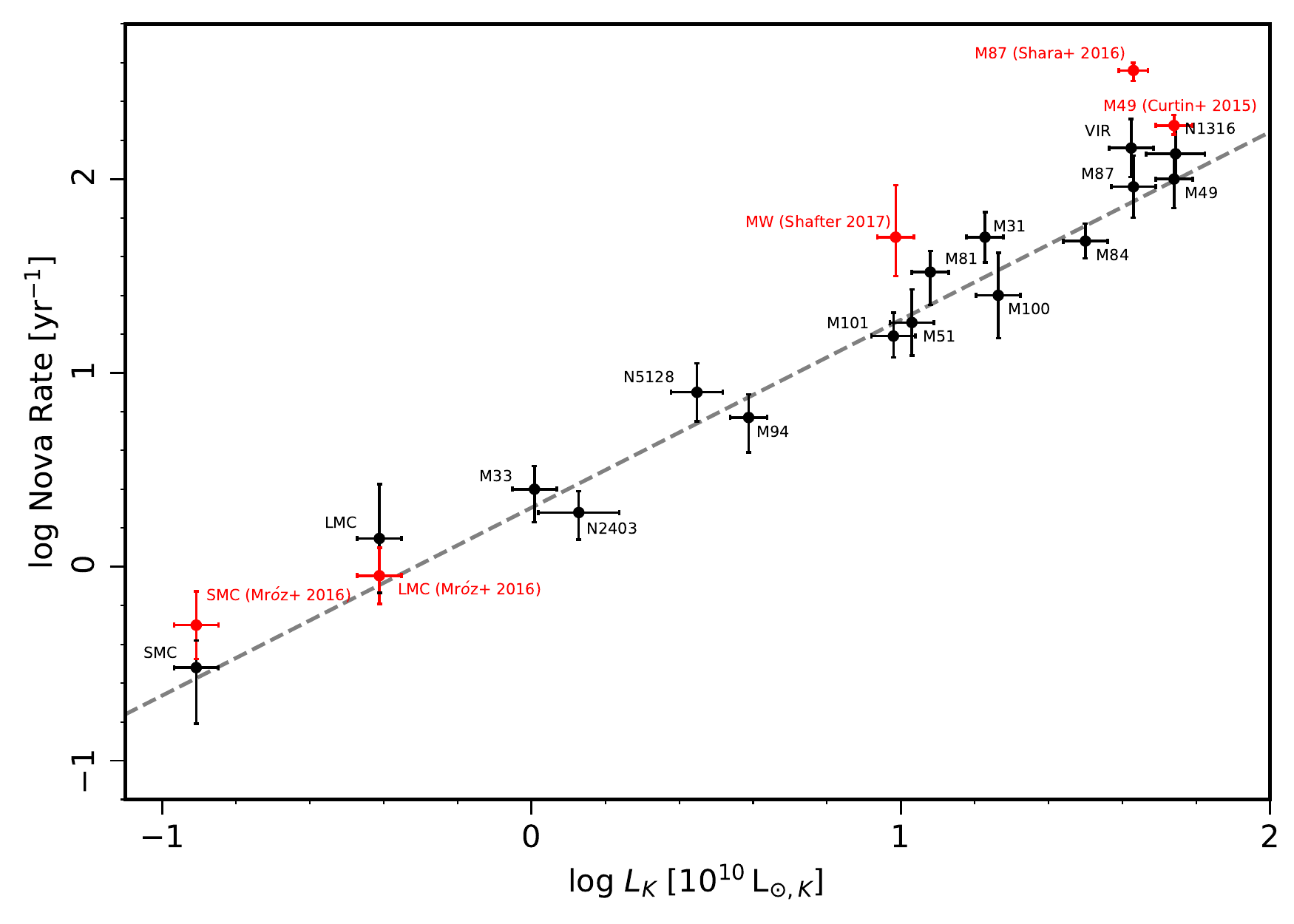}
\end{center}
\caption{`Luminosity Specific Nova Rate' (LSNR) based on the $K$-band luminosity of the host galaxy, using data from \citep[see their Figure~1]{2014ASPC..490...77S}.\label{lsnr-plot}}
\end{figure}

The `Luminosity Specific Nova Rate' (LSNR) was first introduced by \citet{1990ApJ...356..472C,1990AJ.....99.1079C} to compare the nova rates of the M\,31 bulge and the elliptical component of NGC\,5128. The LSNR employs a galaxy's integrated $K$-band luminosity as a proxy for the total stellar mass and permits direct comparison between the nova rates in different galaxies and between galaxies of differing morphological type. \citet{2014ASPC..490...77S} presented a comprehensive review of the evolution and current status of the LSNR. In Figure~\ref{lsnr-plot} we reproduce the LSNR as computed by \citeauthor{2014ASPC..490...77S}\ who concluded that the nova rate is (simply) proportional to the $K$-band luminosity of the host (the grey dashed line). Those authors also found no evidence for the LSNR varying significantly with Hubble type.

Since 2014, new analyses of the Magellanic Clouds \citep{2016ApJS..222....9M}, M\,49 \citep{2015ApJ...811...34C}, M\,87 \citep{2016ApJS..227....1S}, and the Milky Way \citep[see Section~\ref{intro}]{2017ApJ...834..196S} have been published. With the possible exception of the Clouds, each author has reported an elevated nova rate (see the red points in Figure~\ref{lsnr-plot}). As summarised by \citet{2017ApJ...834..196S}, there is now evidence for the LSNR being 3--4 times higher than the adopted value of $\sim2$ novae per year per $10^{10}\,\mathrm{L}_{\odot,K}$ \citep[as computed by][]{2014ASPC..490...77S}. It is, perhaps, the limitations of previous surveys that led to underestimated nova rates. Transient surveys are particularly sensitive to the choice of cadence and the actual temporal sampling achieved, but the depth of extragalactic nova surveys may also be a limiting factor --- one that is perhaps now been bridged by high spatial and temporal resolution {\it HST} surveys of M\,87 \citep{2016ApJS..227....1S}, which can probe any populations of faint yet fast novae.

\section{The `MMRD' and the `faint--fast' novae}\label{sec_mmrd}

No review of nova populations would be complete without a nod to the maximum magnitude---rate of decline relationship (MMRD). \citet{1929ApJ....69..103H} first noted that the brighter an M\,31 nova appeared at peak the more rapidly it diminished. \citet{1945PASP...57...69M} confirmed \citeauthor{1929ApJ....69..103H}'s result Galactically and dubbed the correlation the ``life---luminosity relation''. Over time, the concept that the brightest novae fade the fastest was accepted, the MMRD was refined and, seemingly being invariant to the host population, enabled novae to be touted as primary distance indicators \citep[see, for e.g.,][]{1956AJ.....61...15A,1957ZA.....41..182S,1976A&A....50..113P,1978ApJ...223..351D,1985ApJ...292...90C,1995ApJ...452..704D,2000AJ....120.2007D}. Being brighter than Cepheids at maximum, extragalactic novae seemed like a promising rung on the cosmic distance ladder.  

But the MMRD has always been fraught with problems. Despite the best attempts, a scatter of $\sim0.5$\,mag has persisted. This and the long-held knowledge that the MMRD does not work well for all novae, particularly the RNe \citep[see][]{2010ApJS..187..275S}\footnote{\citet{1945PASP...57...69M} noted that the Galactic recurrent  RS\,Ophiuchi ``may not be typical'' and excluded that system from his analysis.}, hamstrings the relationship for distance determinations to Galactic systems. Even at its best, the MMRD is a population relationship and should not be used to estimate the distance to individual novae, Galactic or otherwise. Thus, despite some advantages over Cepheid variables, in practice employing novae as (extragalactic) distance indicators had never been observationally efficient. 

In the last decade, evidence has slowly started to mount questioning even the concept of an MMRD. The Fast Transients In Nearest Galaxies (P60-FasTING) survey (a forerunner to the Palomar Transient Factory; PTF) was undertaken by the Palomar 60-inch telescope \citep{2011ApJ...735...94K}. This deep and high-cadence survey targeted extragalactic novae, particularly in M\,31. \citeauthor{2011ApJ...735...94K}\ reported the discovery of a `new' population of ``faint--fast'' novae --- novae that populated the lower left quadrant of the MMRD phase-space. As pointed out by \citeauthor{2011ApJ...735...94K}, the M\,31 faint--fast novae occupied a similar locus in MMRD-space as the Galactic RNe. We do note that the \citeauthor{2011ApJ...735...94K}\ sample were corrected for reddening internal to M\,31 by use of the Balmer decrement. As shown specifically in \citet{2017MNRAS.472.1300W}, this decrement should not be used to estimate reddening toward nova eruptions. But it seems unlikely that this should have severely affected the result.

It should be noted that now, almost a decade after the \citeauthor{2011ApJ...735...94K}\ study, a sizeable population of spectroscopically confirmed faint--fast novae has failed to materialise in M\,31. This is despite an increased frequency and depth of coverage by professional and amateur observers alike. A compilation of previous surveys produced by \citet[see their Figure~B.1]{2015A&A...583A.140S} may also indicate a population of faint--fast novae, but also illustrates a very scattered distribution both above and below the traditional MMRD. Faint--fast novae are inherently challenging to discover, let alone observe spectroscopically, it is clear that more work is required locally to understand the true extent of the faint--fast population.

While the situation in M\,31 remains puzzling, there is more serious trouble brewing for the MMRD in a galaxy further away: \citet{2017ApJ...839..109S} published an updated MMRD plot based on a daily-cadence {\it HST} M\,87 survey \citep[see][]{2016ApJS..227....1S}. The M\,87 sample is much less likely (than M\,31 novae) to be affected by reddening internal to M\,87. Here \citeauthor{2017ApJ...839..109S}\ propose (see their Figure~1) that the faint--fast population seen in {\it both} M\,31 and M\,87 severely undermines the validity of the MMRD relationship. To quote \citeauthor{2017ApJ...839..109S}\ directly, ``The fact that these (faint--fast) novae are both common and ubiquitous demonstrates that complete samples of extragalactic novae are not reliable standard candles, and that the MMRD should not be used in the era of precision cosmology either for cosmic distance determinations or the distances of Galactic novae.''

With the availability of Gaia DR2 parallax distances, \citet{2018MNRAS.481.3033S} was the first to re-assess the Galactic MMRD and came to similar conclusions. However, interestingly, \citet{2019A&A...622A.186S} undertook a similar Gaia DR2 analysis using a different (but overlapping) sample to \citeauthor{2018MNRAS.481.3033S}, and concluded that Gaia {\it strengthened} the viability of the MMRD. \citeauthor{2019A&A...622A.186S} also demonstrated that the bolometric luminosity of novae correlates to the optical decline time (a `maximum bolometric magnitude --- rate of decline' relationship?), however, given that the same bolometric correction was used for each of their novae, this is perhaps not surprising --- but we will return to this concept below. \citeauthor{2019A&A...622A.186S}\ went on to explore correlations between other nova system parameters, including $\dot{M}$, finding a correlation between $\dot{M}$ and decline time. The jury is still out on the Galactic MMRD; there is a clear need to understand the sample biases and extinction uncertainties. But the less-biased extragalactic samples indicate that the {\it original} MMRD concept -- a monotonic relation between luminosity and decay rate -- is flawed.

A number of authors have referred to the models of \citet{1995ApJ...445..789P}, which were later built on by \citet{2005ApJ...623..398Y}, for theoretical grounding of the `faint--fast' population. Those models indicate that the original MMRD novae, the ``bright--fast'' and ``faint--slow'' populations are powered by a combination of a high $M_\mathrm{WD}$ and low $\dot{M}$, or a low $M_\mathrm{WD}$ and high $\dot{M}$, respectively. The \citeauthor{2005ApJ...623..398Y}\ models show that faint--fast novae may belong to a population of systems with high $M_\mathrm{WD}$ and high $\dot{M}$, the same fundamental system parameters as the RNe \citep[as noted by][]{2011ApJ...735...94K}. However, we note that \citet[see particularly their Figure~5]{2017ApJ...839..109S} pointed out that the \citeauthor{2005ApJ...623..398Y}\ grids could suggest that the total accreted envelope mass (rather than $\dot{M}$ explicitly) acts along with $M_\mathrm{WD}$ to explain the MMRD position of a given nova. When using grids of models we must consider the relative contribution to the observed population from a particular configuration, e.g. $M_\mathrm{WD}$ and $\dot{M}$. As shorter $P_\mathrm{rec}$ systems inherently produce more eruptions, we would expect faint--fast novae to always have a substantial contribution from RNe.

\citeauthor{2005ApJ...623..398Y}\ also indicated that high $M_\mathrm{WD}$---high $\dot{M}$ novae have low accreted envelope masses, therefore low mass ejecta. As we will see in Section~\ref{12a}, such a low ejected mass may lead to high velocity ejecta and a rapidly evolving eruption. But as shown by \citet{2016ApJ...833..149D}, unlike CNe, the maximum PP radius for faint--fast novae corresponds to a much higher effective temperature \citep[cf.\ $\sim8000$\,K for CNe; see][]{2010AN....331..160B}. Therefore, the peak energy output of the faint--fast novae occurs in the FUV or even EUV, compared to the optical for CNe.

Extending this argument, there is one quadrant of the MMRD that appears unpopulated, the upper right or ``bright--slow'' regime; where one might expect the eruptions of low $M_\mathrm{WD}$ with low $\dot{M}$ to reside. By comparison to faint--fast novae; bright--slow novae should have massive, slowly evolving, ejecta. As such, one might expect their peak to occur somewhere in the IR. But what exactly would such a slowly evolving IR-bright nova actually look like? Would we even identify it as a nova? Galactic examples of such novae {\it could} include systems like the epically-slow evolving V1280\,Scorpii \citep[see, e.g.,][]{2012A&A...545A..63C}; or V723\,Cassiopeiae, which exhibited a SSS so long it was considered a `persistent SSS' \citep{2008AJ....135.1328N,2011ApJS..197...31S} until it abruptly turned off in September 2015 \citep[a SSS phase of almost 10 years;][]{2015ATel.8053....1N}. But more tantalising possibilities present themselves extragalactically. \citet{2017ApJ...839...88K} published the initial results from `SPIRITS', an extragalacitic IR transient survey undertaken with {\it Spitzer}. That paper presented 14 unusual transients those authors dubbed `SPRITES' (eSPecially Red Intermediate-luminosity Transient Events). \citeauthor{2017ApJ...839...88K}\ noted that SPRITES sat in the IR luminosity gap between CNe and SNe, with some SPRITES exhibiting exceptionally slow evolution. With no discovered optical counterparts, perhaps some of the SPRITES fit the criteria of bright--slow novae from low $M_\mathrm{WD}$---low $\dot{M}$ systems? \citet{2010ApJ...725..831S} made similar claims regarding low $M_\mathrm{WD}$---low $\dot{M}$ novae and predicted that {\it some} \citep[particularly `M31-RV'; see][]{1989ApJ...341L..51R} of the `luminous red novae' \citep[LRNe; see, e.g.,][]{2002A&A...389L..51M,2015ApJ...805L..18W} could be extremely slowly evolving CN eruptions\footnote{We note that \citet{2011A&A...528A.114T} presented strong evidence for the LRN V1309\,Scorpii being the merger of a compact binary.}.

Faint--fast novae evaded detection for years because faint--fast transients are just hard to find! But if they arise from high $M_\mathrm{WD}$---high $\dot{M}$ systems they are not inherently faint, they are just optically faint. Likewise, bright--slow novae may be IR bright but optically faint. As such, might there be hope for the MMRD concept yet? Perhaps some time should be taken to further explore the viability of the `maximum {\it bolometric} magnitude---rate of decline' relationship, or the extension of the concept into a multi-parameter space spanned by the luminosity in different energy bands.

\section{Recurring and rapidly recurring novae}\label{sec:rrne}

A combination of a high $M_\mathrm{WD}$ and high $\dot{M}$ is required to drive a RN --- by definition any nova that has been observed in eruption at least twice. The Galactic population of RNe has grown slowly and has remained at ten \citep[see][for a comprehensive review]{2010ApJS..187..275S} since the addition of V2487\,Ophiuchi a decade ago \citep{2009AJ....138.1230P}. The small number is almost certainly a selection effect based mainly on increasing incompleteness as one looks back in time. It is probably not a coincidence that many of Galactic RNe have bright peak apparent magnitudes. The majority of the Galactic RNe are thought to contain a high $M_\mathrm{WD}$ and a high $\dot{M}$ maintained by an evolved donor; a sub-giant or red giant \citep{2012ApJ...746...61D,2014ASPC..490...49D}.

When LMCN\,1968-12a erupted for a second time in 1990 it was widely claimed to be the first extragalactic RN \citep{1991ApJ...370..193S}. It was in fact only the first spectroscopically confirmed extragalactic RN. The honour of the first lies with M31N\,1926-06a \citep[the original eruption discovered by][]{1929ApJ....69..103H}, whose recurrent nature was observed in 1962 independently by \citet{1964AnAp...27..498R} and \citet[see \citealt{2008A&A...477...67H}]{1968AN....291...19B}. Since then, extragalactic RNe have only been discovered in the LMC and M\,31, despite searches within the Local Group and beyond. In Table~\ref{LMC_RNe} we summarise the four currently known LMC RNe and the 18 within M\,31.

\begin{table}[!ht]
\caption{RNe in the LMC (top) and M\,31 (bottom).\label{LMC_RNe}}
\begin{center}
\begin{tabular}{llll}
\hline
\hline
{Nova} & {Known} & {$P_\mathrm{rec}$$^{\dag}$} &  {Refs} \\
{} & {eruptions} & {[yr$^{-1}$]}\\
\hline
LMCN\,1968-12a & 4 & $6.7\pm1.2$ & 1\\
LMCN\,1971-08a & 2 & $\sim38$ & 2 \\
LMCN\,1996 & 2 & $\sim22$ & 3 \\
YY\,Doradus & 2 & $\sim67$ & 4, 5 \\
\hline
M31N\,1919-09a & 2 & $\sim79$ & 6 \\
M31N\,1923-12c & 2 & $\sim88$ & 6 \\
M31N\,1926-06a & 2 & $\sim37$ & 6 \\
M31N\,1926-07c & 3 & $\sim11$ & 6 \\
M31N\,1945-09c & 2 & $\sim27$ & 6 \\
M31N\,1953-09b & 2 & $\sim51$ & 6 \\
M31N\,1960-12a & 3 & $\sim6$ & 6--8 \\
M31N\,1961-11a & 2 & $\sim44$ & 6 \\
M31N\,1963-09c & 4 & $\sim5$ &  6, 9\\
M31N\,1966-09e & 2 & $\sim41$ &  6 \\
M31N\,1982-08b & 2 & $\sim14$ & 6 \\
M31N\,1984-07a & 3 & $\sim8$ & 6 \\
M31N\,1990-10a & 3 & $\sim9$ & 10 \\
M31N\,1997-11k & 3 & $\sim4$ & 6 \\
M31N\,2006-11c & 2 & $\sim8$ & 11 \\
M31N\,2007-10b & 2 & $\sim10$ & 12, 13\\
M31N\,2007-11f & 2 & $\sim9$ & 14 \\
M31N\,2008-12a & 14 & $0.99\pm0.02$ & 15, 16 \\
\hline
\end{tabular}
\end{center}
\begin{minipage}{\columnwidth}
\setstretch{0.75}
{\footnotesize For the equivalent Galactic table, see \citet[their Table~21]{2010ApJS..187..275S}.}\\
{\footnotesize $^{\dag}${To estimate $P_\mathrm{rec}$ we have (excluding LMCN\,1968-12a and M31N\,2008-12a) simply taken the shortest observed inter-eruption period.}}\\
{\footnotesize References --- {(1)~\citet{2019arXiv190903281K}, (2)~\citet{2016ApJ...818..145B}, (3)~\citet{2018ATel11384....1M}, (4)~\citet{2004IAUC.8424....1B}, (5)~\citet{2004IAUC.8424....2M}, (6)~\citet{2015ApJS..216...34S}, (7)~\citet{2019ATel12915....1V}, (8)~\citet{2019ATel12943....1S}, (9)~\citet{1996ApJ...473..240D}, (10)~\citet{2016ATel.9276....1H}, (11)~\citet{2015ATel.7116....1H}, (12)~\citet{Sch2017}, (13)~\citet{2017ATel11088....1W}, (14)~\citet{2017ATel10001....1S}, (15)~\citet{2017ASPC..509..515D}, (16)~this work.}}
\end{minipage}
\end{table}

The first catalogue of M\,31 RN candidates was produced by \citet[see their Table~3]{1996ApJ...473..240D}, who also assessed the RN populations of the LMC and Milky Way. \citeauthor{1996ApJ...473..240D}\ concluded that RNe could only contribute at the few percent level to the SN\,Ia rate in those hosts.

The majority (all but three) of the most recent M\,31 RN catalogue were identified by a monumental search of archival observations by \citet{2015ApJS..216...34S}. Those authors published a catalogue of 16 strong RN candidates, many of which were also spectroscopically confirmed, by virtue of astrometric arguments. Subsequently, three more M\,31 RNe have been identified, M31N\,2006-11c \citep{2015ATel.7116....1H}, 2007-10b \citep{Sch2017,2017ATel11088....1W} and 2007-11f \citep{2017ATel10001....1S}; 1990-10a has erupted again and halved its estimated $P_\mathrm{rec}$ \citep{2016ATel.9276....1H}, as has 1960-12a reducing its $P_\mathrm{rec}$ from $\sim53$ to $\sim6$\,years \citep{2019ATel12915....1V,2019ATel12943....1S}, and 1966-08a has been confirmed to not be a RN.

M31N\,1966-08a and its second eruption 1968-10c both hailed from the \citet{1973A&AS....9..347R} survey. \citeauthor{1973A&AS....9..347R}\ noted ``This star (1966-08a) coincides beyond any doubt with (1968-10c)'', a fact on which \citet{2015ApJS..216...34S} agrees. However, probably due to its (then) unprecedentedly short `recurrence' period there were doubts. \citet{1989SvAL...15..382S} suggested that 1966-08a was more likely a foreground dwarf nova (DN) outburst. Both the 1966 and 1968 events were not observed spectroscopically, and nothing was seen from this system for decades. \citet{2017RNAAS...1a..44S} recovered the progenitor system in archival Local Group Galaxies Survey \citep[LGGS;][]{2006AJ....131.2478M} and 2MASS \citep{2006AJ....131.1163S} data, which indicated a very low eruption amplitude for a nova (even given the potentially short $P_\mathrm{rec}$). Follow-up spectroscopy, also reported by \citeauthor{2017RNAAS...1a..44S}\ indicated that the progenitor was a dwarf not a giant and was therefore incompatible with being in M\,31.  \citeauthor{2017RNAAS...1a..44S}\ proposed, based on the low amplitude and spectroscopy, that 1966-08a and its recurrence were the result of a Galactic flare star. Somewhat ironically, just days after that proposal by \citeauthor{2017RNAAS...1a..44S}, another flare (the first in sixty years) from 1966-08a was discovered \citep{2017Con,2017ATel11094....1A}, followed soon after by another \citep{2019ATel12513....1C}.

Long hailed as a RN, PT\,Andromedae (aka M31N\,1957-10b) had been noted to recur five times \citep{2000IBVS.4909....1A,2010CBET.2574....2R,2010CBET.2574....1Z}. Following spectroscopy  of the 2010 event, \citet{2012ApJ...752..133C} suggested that PT\,And may be an M\,31 RN. However, \citet[and references therein]{2015ApJS..216...34S} instead proposed a Galactic DN origin. Following another detection in 2017 \citep{2017TNSTR.867....1C}, spectroscopy confirmed that PT\,And was not an M\,31 nova but was consistent with a Galactic DN outburst \citep{2017ATel10692....1W,2017ATel10647....1W}.

Based on their statistical analysis of the M\,31 RN and CN populations, \citet{2015ApJS..216...34S} reported that 1/25$^\mathrm{th}$ of detected M\,31 eruptions arose from {\it known} RNe. Their completeness exercise indicated that as many as a third of M\,31 eruptions could be from RNe ($P_\mathrm{rec}\leq100$\,yrs), broadly consistent with the independent findings of \citet{2014ApJ...788..164P} and \citet{2016ApJ...817..143W}. Although relying upon a number of assumptions, \citeauthor{2015ApJS..216...34S}\ used their estimated M\,31 RN population to compute the potential contribution to the SN\,Ia rate in that host, concluding it is unlikely that RNe provide a significant channel ($\sim2\%$).

But if RNe play any important role in the production of SNe\,Ia, the key systems to find are those with WDs already close to the Chandrasekhar mass and accreting at a high rate. Those systems must be the ones with the shortest $P_\mathrm{rec}$ \citep{2005ApJ...623..398Y,2014ApJ...793..136K,2016ApJ...819..168H,2017ApJ...844..143K}. Prior to 2013, the shortest confirmed $P_\mathrm{rec}$ belonged to the Galactic RN U\,Scorpii, which erupts every $\sim10$\,yrs \citep{2010ApJS..187..275S}. But, starting with the discovery of M31N\,2008-12a (see Section~\ref{12a}), a population of `rapid recurrent novae' (RRNe) has been uncovered. 

\begin{figure}
\begin{center}
\includegraphics[width=0.9\columnwidth]{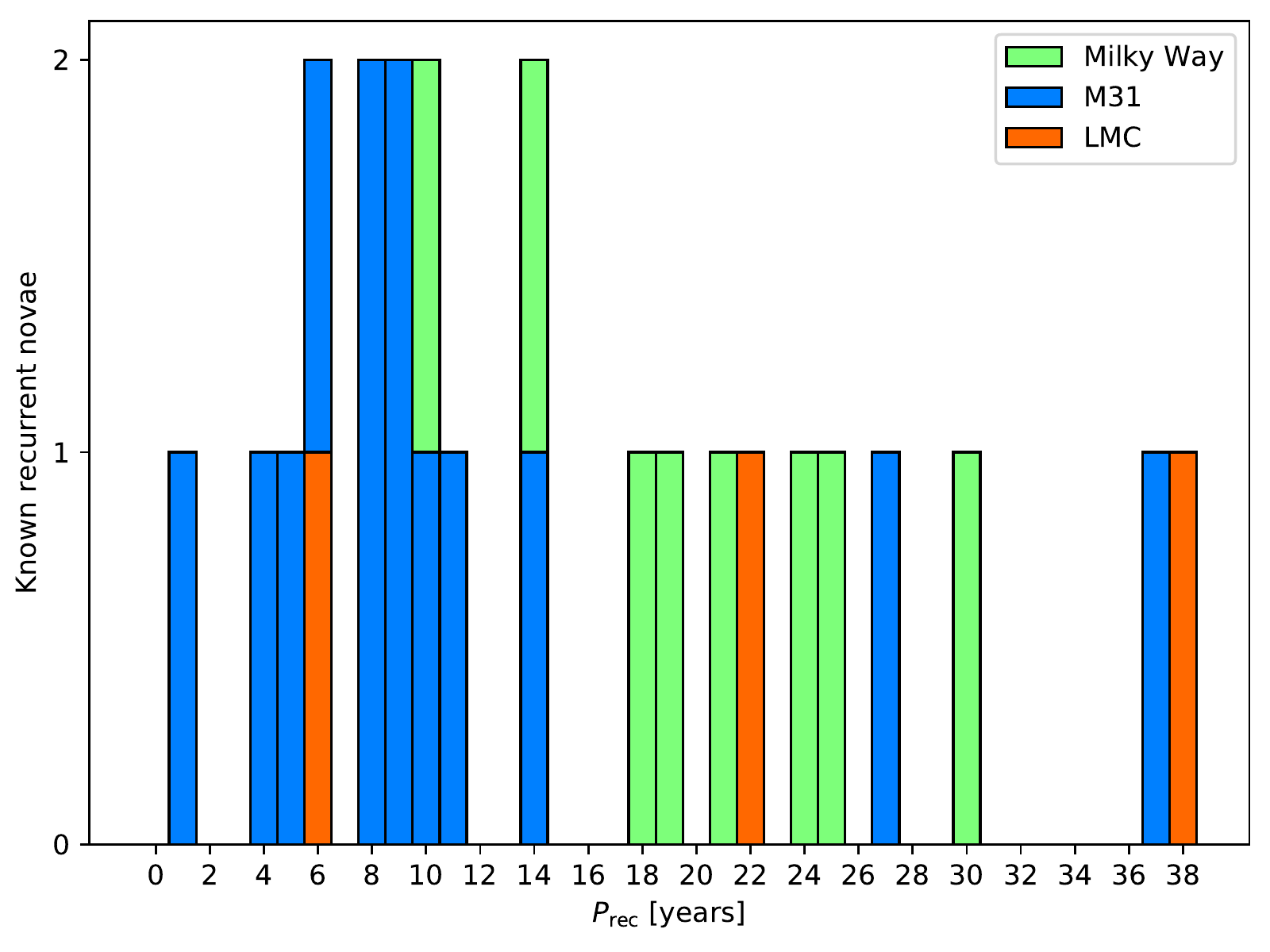}
\end{center}
\caption{Distribution of recurrence periods of all known RNe ($P_\mathrm{rec}\leq40$\,yrs). RRNe are all those with $P_\mathrm{rec}\leq10$\,yrs and they largely exist within M\,31. Data for Galactic RNe are from \citet{2010ApJS..187..275S}. The most rapidly recurring Galactic systems are the prototype sub-giant and red giant donor systems, U\,Sco and RS\,Oph, respectively.\label{RRN_hist}}
\end{figure}

We hereby, and admittedly arbitrarily, define a RRN as a system that has undergone eruptions less than a decade apart. Galactically, the only known example is U\,Sco,  and in the LMC there is LMCN\,1968-12a \citep[$P_\mathrm{rec}=6.7\pm1.2$;][]{2019arXiv190903281K}. But in M\,31 there are eight (see Table~\ref{LMC_RNe}) --- {\it almost half of all known M\,31 RNe}. Indeed, all new M\,31 RNe since M31N\,1984-07a are RRNe. In Figure~\ref{RRN_hist} we illustrate the distribution of $P_\mathrm{rec}$ for the known RNe. These data indicate that the distribution of $P_\mathrm{rec}$ is relatively uniform across the three galaxies. We fervently note that these data likely suffer from multiple selection effects, there is no evidence to support that RRNe should only exist in M31, or that the `peak' at $P_\mathrm{rec}\sim10$\,yr is real.

This ten year threshold creates a phenomenological `watch list' of RNe to study closely through multiple eruptions. Consistent analysis and comparison of multiple eruptions from individual systems will be a key future driver for extragalactic nova science. It would not be the last time that classification based on observed characteristics revealed physical insights.

So where are the Galactic RRNe? To date, the study of these systems has been largely confined to M\,31 \citep[but also see][for a detailed analysis of LMCN\,1968-12a]{2019arXiv190903281K}, which despite the advances in recent years severely limits observation opportunities to just optical to soft X-ray light curves and, in all but the most extreme case (see Section~\ref{12a}), optical spectroscopy.  Given the high $M_\mathrm{WD}$---high $\dot{M}$ requirements for a short $P_\mathrm{rec}$, RRNe should be faint--fast novae. So it is not unlikely that the rapid-fire eruptions from Galactic RRNe might have been mistaken for other transients or even quasi-periodic variables, e.g., flare stars or DNe (the majority of which are not spectroscopically confirmed) --- particularly before the discovery of the prototype system, M31N\,2008-12a. 

An open question -- with direct connection to their ultimate fate -- is just how many RRNe exist? Have we already uncovered the majority, or just scraped the surface? The rapidly approached era of all-sky, wide-field, (multi-messenger,) time-domain astronomy is key to addressing this question. Surveys such as the Zwicky Transient Facility \citep[ZTF;][]{2019PASP..131a8002B} and the Large Synoptic Survey Telescope \citep[LSST;][]{2019ApJ...873..111I} are ideally placed to detect eruptions of RRNe Galactically, in the Local Group, and beyond. To classify and interrogate those eruptions, we are at the mercy of the availability of timely (and in this case, rapid) follow-up observations.

\section{M31N 2008-12a --- a remarkable recurrent nova}\label{12a}

\subsection{Innocuous beginnings}

In-line with predictions \citep{2006MNRAS.369..257D}, there are now regularly over thirty novae discovered in M\,31 each year\footnote{\url{http://www.mpe.mpg.de/~m31novae/opt/m31/index.php}}. So when a new eruption from a previously unknown system, M31N\,2008-12a (hereafter `12a') was announced in 2008, there was nothing remarkable about this event except, perhaps, the date of the eruption, Christmas Day. The discovery note, written by \citet{2008Nis}, simply contains a few sentences about the brightness of the eruption. No known follow-up observations were taken and the event was not spectroscopically confirmed.

In 2011, another eruption was detected by \citet{2011Kor}, while there were a handful of follow-up observations \citep{2011ATel.3725....1B} there was no successful spectroscopy. In the available on-line material, a connection isn't made to the 2008 event, but the statement, ``In SIMBAD object RX\,J0045.4+4154 located at a distance 3.79 asec'' is made. RX\,J0045.4+4154 is, as we will see, intimately associated with 12a.

When the transient reappeared in 2012, again discovered by \citet{2012Nis}, a single spectrum was obtained using the Hobby-Eberly Telescope (HET) by \citet{2012ATel.4503....1S}. That spectrum \citep[reproduced in][]{2014A&A...563L...9D} confirmed that the 2012 event is clearly a nova eruption, within M\,31, and revealed the characteristics of the He/N taxonomic class. In the reporting telegram, \citeauthor{2012ATel.4503....1S}\ make the link between the 2008, 2011, and 2012 events, and the first suggestion that the system may be a RN -- despite the ``unusually short interval between brightenings''. Based on the RN hypothesis, an attempt was made to detect the SSS phase using {\it Swift}, but a series of four XRT \citep{2005SSRv..120..165B} observations beginning 20 days post-eruption failed to detect a source.

\subsection{The realisation}

The intermediate Palomar Transient Factory \citep[iPTF;][]{2009PASP..121.1395L} reported the discovery of the 2013 event, which erupted on Nov 26 \citep{2013ATel.5607....1T}. Upon discovery, iPTF triggered follow-up spectroscopy, which again confirmed the eruptive nova nature \citep{2014ApJ...786...61T}. {\it Swift} observations began only six days post-eruption and found that the SSS was already visible \citep{2014A&A...563L...8H,2014ApJ...786...61T}. At the time, this was the earliest on-set nova SSS to have been observed\footnote{This was surpassed by the 2014 eruption of the RN V745\,Scorpii, whose SSS turned-on 4 days post-discovery \citep{2015MNRAS.454.3108P}.}.

\citet{2014A&A...563L...9D}\ and \citet{2014ApJ...786...61T}\ compiled optical photometry of the 2013 eruption, which confirmed the `under-luminous' nature of the eruptions (as reported in 2008, 2011, and 2012), and indicated an extremely rapid decline, i.e.\ faint--fast. Both \citeauthor{2014A&A...563L...9D}\ and \citeauthor{2014ApJ...786...61T}\ utilised archival {\it HST} data to identify the likely progenitor system -- a very blue system whose SED was consistent with a luminous accretion disk. In those initial analyses, no evidence for the mass donor was recovered.

\citet{2014A&A...563L...8H} and \citet{2014ApJ...786...61T} both found three previous eruptions in the archives of {\it ROSAT} (1992 and 1993) and {\it Chandra} (2001), noting that the system was initially discovered as the ``recurrent supersoft X-ray transient'' RX\,J0045.4+4154 \citep{1995ApJ...445L.125W}. \citeauthor{2014ApJ...786...61T}\ also revealed that PTF detected an eruption in 2009.

Eruptions had been detected in 1992, 1993, 2001, 2008, 2009, 2011, 2012, and 2013. It seemed clear that the 2010 eruption had been missed, probably occurring during a gap in PTF coverage \citep{2012ApJ...752..133C}. The evidence presented by \citeauthor{2014A&A...563L...9D}, \citeauthor{2014A&A...563L...8H}, and \citeauthor{2014ApJ...786...61T}\ was strongly suggestive that 12a was a RN undergoing annual eruptions, and that it had been (at the time) doing so for at least twenty years. Therefore, it was concluded that 12a must contain a particularly massive WD and must be accreting an an elevated rate \citep{2014A&A...563L...9D,2014A&A...563L...8H,2014ApJ...786...61T}, and all three publications ended with a prediction for the 2014 eruption.

\subsection{2010, 2014 and 2015, and a six month recurrence?}

In light of predictions for a 2014 eruption, a programme was put together to monitor the 12a region of M\,31. This was undertaken predominately by the Liverpool Telescope \citep[LT;][]{2004SPIE.5489..679S}, which detected the eruption on October 2.  Upon detection, a pre-planned follow-up campaign was instigated that included multiple ground-based optical telescopes obtaining high-cadence photometry and a number of spectroscopic observations, and space-based UV and X-ray observations by {\it Swift}. \citet[who addressed the optical and UV observations]{2015A&A...580A..45D} reported that the 2014 eruption was similar to that of 2013 and that the nova evolved extremely rapidly ($t_2=1.8\pm0.1$\,days) --- faster than all known Galactic RNe. The first tentative evidence for a light-curve plateau, synonymous with the RN phenomenon \citep{2008ASPC..401..206H,2010ApJS..187..275S,2010AJ....140...34S,2014ApJ...788..164P}, the low peak optical luminosity was consistent with a low ejected mass, and the SEDs indicated a high photospheric temperature at maximum light. \citeauthor{2015A&A...580A..45D}\ also reported that seemingly low ejection velocity, obtained spectroscopically was consistent with models of a high $M_\mathrm{WD}$ and short cycle RNe \cite[see, e.g.,][]{2005ApJ...623..398Y}. The spectra also hinted at possible ejecta deceleration, similar to that seen in RS\,Oph \citep[first noted after the 1958 eruption;][]{1964AnAp...27..555D} when the ejecta interact with pre-existing circumbinary material due to the red giant wind of that system's donor \citep{1967BAN....19..227P,1985MNRAS.217..205B,2006ApJ...652..629B}. \citet{2015A&A...580A..46H} reported on X-ray observations that showed a bright and rapidly evolving SSS with a fast turn-on ($t_\mathrm{on}=5.9\pm0.5$\,days) and short extent ($t_\mathrm{off}=18.4\pm0.5$\,d) --- like the optical and UV, the 2014 X-ray evolution was very similar to that seen in 2013. \citeauthor{2015A&A...580A..46H}\ revealed that a BB parameterisation of the X-ray spectrum indicated a very high effective temperature ($k_\mathrm{B}T=120\pm5$\,eV) and that the X-ray light curve showed substantial variation over the first 10\,days following the unveiling of the SSS. The derived X-ray parameters were also consistent with those predicted based on the M\,31 population \citep[Section~\ref{xrayprop}]{2010A&A...523A..89H,2011A&A...533A..52H,2014A&A...563A...2H}, and were consistent with a near-Chandrasekhar mass WD.

The most interesting finding reported by \citet{2015A&A...580A..45D} was the discovery of extended nebulosity surrounding the system, which is discussed in more detail in Section~\ref{SR}. \citeauthor{2015A&A...580A..45D}\ and \citeauthor{2015A&A...580A..46H}\ predicted that the 2015 eruption would occur between October and December.

The 2015 eruption represented a sea change. A concerted campaign was put together utilising facilities all around the globe including large numbers of observers from the American Association of Variable Star Observers (AAVSO\footnote{\url{https://www.aavso.org}}), the British Astronomical Association (BAA\footnote{\url{https://www.britastro.org}}), and the Variable Star Observers League in Japan (VSOLJ\footnote{\url{http://vsolj.cetus-net.org}}) --- a nova campaign not seen since the 2010 eruption of U\,Sco \citep[see][]{2010AJ....140..925S}. A space-based detection campaign was undertaken by {\it Swift} \citep{2016ApJ...830...40K} in an attempt to capture the long-predicted nova precursor X-ray flash \citep[XRF;][]{1990LNP...369..306S,2002AIPC..637..345K,2015ApJ...808...52K,2016ApJ...824...22H}. Early detection of the 2015 eruption was critical to the follow-up campaigns, which included rapid-response UV spectroscopy and photometry by {\it HST} and late-time ground-based spectroscopy from a number of 8m+ facilities\footnote{Due to unfortunate weather conditions around the globe, none of the late-time spectroscopy was possible. This was all rescheduled for the 2016 eruption and those data remain under analysis.}.

The Las Cumbres Observatory network \citep[LCO;][]{2013PASP..125.1031B} made the discovery on August 28, however, {\it Swift} UVOT had detected the 2015 eruption marginally earlier\footnote{The {\it Swift} observations were hampered by a longer data retrieval time and were received and processed after the LCO data.}. The 2015 eruption occurred sooner than anticipated; the {\it Swift} XRF campaign had only just begun. \citet{2016ApJ...830...40K} reported a failed attempt to capture the XRF, citing the short lead-in time among the possible explanations.  Another possibility presented is that although the XRF could `escape' from the natal nova eruption it was largely absorbed by substantial circumbinary material. However, additional scenarios include insufficient {\it Swift} cadence or the XRF energy being incompatible with the {\it Swift} XRT (particularly given that instrument's low sensitivity to hard X-rays and the distance to M\,31).

The follow-up campaign of the 2015 eruption obtained the most detailed optical (photometric and spectroscopic), UV, and X-ray datasets of any M\,31 nova to date. \citet{2016ApJ...833..149D} presented and analysed a combined dataset from the 2013--2015 eruptions, which showed remarkable similarity at all energies, as suggested for RN eruptions by \citet{2010ApJS..187..275S}. \citeauthor{2016ApJ...833..149D}\ reported that the colour evolution was suggestive of a red giant donor, which was also supported by the strong evidence now seen for ejecta deceleration \citep[see, e.g.,][]{2006ApJ...652..629B}. Tentative evidence for high-excitation coronal lines was also presented, as might be expected in the presence of a shocked donor wind. Detailed SEDs provided no evidence for an optically thick photosphere, even at early times, indicating that the photospheric emission must peak in the FUV or even EUV. \citeauthor{2016ApJ...833..149D}\ went on to describe the extremely high velocity material ($\mathrm{FWHM}\approx13000$\,km\,s$^{-1}$) seen fleetingly in the early-time (pre-maximum) spectra, described as ``indicative of outflows along the polar direction---possibly highly collimated outflows or jets''. There was also evidence for a mid-point (day 11) dip in the X-ray light curve across all three eruptions, which we will address further in Section~\ref{2016}. \citeauthor{2016ApJ...833..149D}\ ended on a prediction for the 2016 eruption occurring in mid-September ($\pm1$\,month).

The {\it HST} observations of the 2015 eruption were successful in tying down a number of the outstanding `unknowns' about the system. The NUV spectra (taken 4 and 5 days post-eruption) finally constrained the extinction toward the system, but otherwise revealed very limited features. The FUV spectrum, taken 3.32 days post-eruption was much more fruitful. \citet{2017ApJ...847...35D} reported that the FUV spectrum was broadly consistent with that expected from a CO WD \citep[see, e.g.,][]{2012BASI...40..185S} and importantly there was no evidence for any neon in the ejecta at that time. The FUV lines also exhibited very high velocities and the resonance lines remained optically thick (and saturated in some cases), the profile of the N\,{\sc v} line (the highest ionisation energy line observed) was shown to be consistent with optically thick outflows or jets. 

Newly obtained {\it HST} optical--NUV photometry was used to explore the late-decline and was coupled with archival observations to explore the quiescent system. \citet{2017ApJ...849...96D} found that 12a takes only $\sim75$ days to reach quiescence following an eruption, showing a possible increase in luminosity toward the onset of the next event. The quiescent photometry were used to model the accretion disk \citep[see, e.g.,][]{2017ApJ...846...52G}, with those models indicating an extremely high $\dot{M}$. By extrapolating the quiescent disk models back to the late-decline, and with comparison to a late-time Keck spectrum of the 2014 eruption, \citet{2017ApJ...849...96D} presented evidence for the 12a accretion disk surviving each eruption and possibly dominating the optical--NUV flux as early as the plateau phase. The quiescent accretion rates presented were in the region of $(0.6-1.4)\times10^{-6}\,\mathrm{M}_\odot\,\mathrm{yr}^{-1}$ -- once the additional effects of a considerable disk wind/outflow had also been considered \citep[see, e.g.,][]{2015MNRAS.450.3331M}. 

With the assistance of the PHAT survey team \citep{2014ApJS..215....9W}, \citet{2017ApJ...849...96D} recovered the mass donor, an M\,31 `red clump' star, most likely a low-luminosity red giant --- with the donor constrained a limit could also be placed on the orbital period ($\gtrsim5$\,days). That paper concluded by assessing all the parameters of the system and made a {\it conservative estimate} of the time remaining for the WD to reach the Chandrasekhar mass of $<20$\,kyr.

The observations of the 2015 eruption \citep{2016ApJ...833..149D,2017ApJ...849...96D,2017ApJ...847...35D} allowed us to complete the basic picture of 12a, placing numbers or strong constraints on most of the key system parameters, which are summarised in Table~\ref{vital_stats}. A few gaps remained, including the 2010 eruption!

\begin{table}
\caption{Key parameters of the M31N\,2008-12a system.\label{vital_stats}}
\begin{center}
\begin{tabular}{lll}
\hline
\hline
{Parameter} & {Value} &  {References} \\
\hline
$P_\mathrm{rec}$ & $347\pm10$\,days & 1 \\
$M_\mathrm{WD}$ & $\simeq1.38$\,M$_\odot$ & 2 \\
$\dot{M}_\mathrm{SSS}$$^{a}$ & $1.6\times10^{-7}$\,M$_\odot$\,yr$^{-1}$ & 2 \\ 
$\dot{M}_\mathrm{disk}$$^{b}$ & $\left(6-14\right)\times10^{-7}$\,M$_\odot$\,yr$^{-1}$ & 3\\
$M_\mathrm{ejected,H}$ & $\left(0.26\pm0.04\right)\times10^{-7}$\,M$_\odot$ & 4 \\
$\eta$$^{c}$ & $+63\%$ & 2\\
$L_\mathrm{donor}$ & $103^{+12}_{-11}\,\mathrm{L}_\odot$ & 3 \\
$R_\mathrm{donor}$ & $14.14^{+0.46}_{-0.47}\,\mathrm{R}_\odot$ & 3 \\
$T_\mathrm{eff,donor}$ & $4890\pm110\,\mathrm{K}$ & 3 \\
$P_\mathrm{orb}$ & $\gtrsim5$\,days & 3 \\
\hline
$d$ & $752\pm17$\,kpc & 5\\
$E\left(B-V\right)$ & $0.10\pm0.03$ & 6 \\
\hline
\end{tabular}
\end{center}
\begin{minipage}{\columnwidth}
\setstretch{0.75}
{\footnotesize $^{a}${Derived by modelling the SSS development of M31N\,2008-12a, it is assumed to be constant throughout a complete eruption cycle.}}\\
{\footnotesize $^{b}${Derived by fitting accretion disk models to the optical and UV quiescence SEDs. Here, the range during quiescence is presented.}}\\
{\footnotesize $^{c}${WD accretion efficiency; as $\eta>0$, $M_\mathrm{WD}$ in increasing.}}\\
{\footnotesize References --- {(1)~\citet{2015A&A...582L...8H,2018ApJ...857...68H}, (2)~\citet{2015ApJ...808...52K}, (3)~\citet{2017ApJ...849...96D}, (4)~\citet{2015A&A...580A..46H}, (5)~\citet{2001ApJ...553...47F}, (6)~\citet{2017ApJ...847...35D}.}}
\end{minipage}
\end{table}

With a hole in the eruption history, an archival search for the `missing' 2010 eruption was undertaken. It did not take long to find the culprit contained within a pair of observations taken on Nov 20, right in the PTF coverage gap \citep{2015A&A...582L...8H}. The timing of the 2008--2014 events suggested that eruptions occurred slightly earlier each year --- i.e.\ a recurrence cycle just under one year (see Figure~\ref{rec_gaps}).  However, when the archival X-ray eruptions from 1992, 1993, and 2001 were included they appeared to break this pattern. The simplest solution to this apparent problem was a shorter recurrence period, half that observed between 2008--2014. \citet{2015A&A...582L...8H} therefore adopted  $P_\mathrm{rec}=175\pm11$\,days. Under this scenario, each of the observed eruptions in 2008--2014 was the second eruption that calendar year, the earlier eruption happened during the M\,31 Sun constraint. But as the proposed  $P_\mathrm{rec}$ was still just under six months, the eruptions would still creep earlier each year --- following the 2015 eruption, \citet{2016ApJ...833..149D} predicted that by 2020/21 there would be a good probability of detecting the earlier eruption and confirming the six month cycle. However, if the eruption pattern remained unchanged, it would be substantially longer until a six month cycle could be confidently excluded.

\subsection{The `peculiar' 2016 eruption}\label{2016}

The lead-in to 2016  focussed on a dedicated attempt to detect the `early' eruption \citep[{\it confidentially} predicted for 2016 March $23\pm1$ month; see][]{HQ}. The results of that work will be published in due course in \citet{HQ}, but, it would not be considered a `spoiler' to report here that the early 2016 eruption was (despite the heroic efforts of some of the observers involved) not recovered. 

While the existence of the `early' eruption remained unproved, attention was focussed to the `normal' later-year event, and again a global detection effort was employed. The mid-September prediction came and went, as did the extended window \citep[ending on October 13;][]{2016ApJ...833..149D}. The 2016 eruption finally occured on December 12 \citep{2016ATel.9848....1I}. The results of the 2016 eruption campaign are presented in full detail in \citet{2018ApJ...857...68H}. In general, despite its lateness, the 2016 eruption proceeded largely as those preceding it, and with the earliest spectrum yet obtained, even stronger evidence of the short-lived high-velocity outflows or jets were seen.  However, the 2016 eruption differed from its forerunners in two aspects.

Firstly, \citeauthor{2018ApJ...857...68H}\ revealed a short-timescale cusp-like peak that preceded and outshone the `normal' eruption peak (at day 1). While the paper discusses possible links between this cusp and the delayed eruption, it was noted that the timing of the 2016 cusp was coincident with holes in the light curves from 2013--2015 --- so no strong connection could be made to the delayed eruption. The 2010 detection provided limited evidence for a similar event that year; an eruption otherwise deemed `typical' \citep{2015A&A...582L...8H}.

Secondly, the SSS phase, which in 2013--2015 had continued until day 18--19 post-eruption, began to turn-off at day 11 and was last detected by {\it Swift} on day 14 \citep{2018ApJ...857...68H}. Prior to turn-off, the unveiling of the 2016 SSS proceeded in a similar manner to that in 2013--2015, that and the similarity in the optical behaviour (sans the `cusp') strongly implied that the eruptions themselves were similar --- a similar ejected mass, with a similar velocity, and a similar peak luminosity, therefore a similar ignition (or accreted) mass must have been involved. How could a late eruption generate essentially a `normal' eruption but a truncated SSS phase? The inter-eruption period of 12a had always varied, but the SSS phase had been consistent.

Given the generally low ejected mass and $\dot{M}$ of novae \citep[see, e.g.,][]{2005ApJ...623..398Y}, $M_\mathrm{WD}$ must be approximately constant between successive eruptions (whether long-term $M_\mathrm{WD}$ increases or decreases) and hence successive eruptions would always have the same ignition mass and similar eruptions. The logical explanation of a late eruption is a decrease in the average inter-eruption $\dot{M}$. This in turn would lead to a less massive accretion disk. A less massive disk would be more readily disrupted during an eruption. \citeauthor{2018ApJ...857...68H}\ also noted that the 2016 eruption began to turn off (at day 11) at around the same time as the X-ray light curve dip seen in 2013--2015. Therefore it was proposed that day 11 was the natural turn off time of the SSS in 12a, given the ignition mass. The higher average $\dot{M}$ in 2013--2015 meant the disk was minimally disrupted, allowing accretion on the WD to resume once the surface nuclear burning first began to wane (at around day 11), the availability of additional H-rich fuel artificially extended the SSS-phase by a week or so. \citeauthor{2018ApJ...857...68H}\ proposed that in 2016 a less massive disk was more substantially disrupted and accretion onto the WD only resumed once the nuclear burning had ceased. In making this proposal, \citeauthor{2018ApJ...857...68H}\ strongly recommended further study of this `re-feeding' concept. Subsequently, \citet{2018MNRAS.480..572A} suggested that SSS re-feeding might explain the unexpected longevity of the SSS phase in the recent nova V407 Lupi.

Following 2016, the eruption pattern proposed in \citet{2015A&A...582L...8H} had been thrown into disarray. Was the 2016 event a statistical anomaly, or was the assumed model incorrect? Moreover, what had caused the decreased $\dot{M}$ between the 2015 and 2016 eruptions, which must have dropped by $\sim25\%$ from the `norm' during this period?

\subsection{2017 and 2018 --- back on track?}

It is fair to write that we really weren't sure when to expect the 2017 eruption, with predictions from within the  `12a collaboration' ranging from the normal pattern to a T\,Pyxidis-style subsidence \citep{2010ApJS..187..275S,2018ApJ...862...89G}. It was not even certain that the `2017 eruption' (aka the next eruption) would occur in 2017.

Leaving it very late, the 2017 eruption was detected on December 31.77 UT\footnote{The time recorded in Table~\ref{12a-erupt} is the estimate of eruption itself.} at the West Challow Observatory in the UK \citep{2017ATel11116....1B}. The 2018 event, was less suspenseful, and was detected on November 6 by the LT \citep{2018ATel12177....1D}. In Table~\ref{12a-erupt} we provide a summary of the past 14 detected eruptions of 12a \citep[based on similar Tables in][]{2017ApJ...849...96D,2017ApJ...847...35D}. 

\begin{table}[!ht]
\caption{Summary of the 14 observed eruptions of M31N\,2008-12a.\label{12a-erupt}}
\begin{center}
\begin{tabular}{lll}
\hline
\hline
{Eruption date$^{\dag}$} & {Inter-eruption} &  {References} \\
{[UT]} & {timescale [days]$^{\ddag}$} & \\
\hline
(1992 Jan.\ 28) & \ldots & 1, 2 \\
(1993 Jan.\ 03) & 341 & 1,2  \\
(2001 Aug.\ 27) & \ldots & 2, 3 \\
2008 Dec.\ 25 & \ldots & 4 \\
2009 Dec.\ 02 & 342 & 5 \\
2010 Nov.\ 19 & 352 & 2 \\
2011 Oct.\ 22.5 & 337.5 & 6 \\
2012 Oct.\ 18.7 & 362.2 & 7 \\
2013 Nov.\ $26.95\pm0.25$ & 403.5 & 4, 8, 9  \\
2014 Oct.\ $02.69\pm0.21$ & $309.8\pm0.7$ & 10, 11 \\
2015 Aug.\ $28.28\pm0.12$ & $329.6\pm0.3$ & 12\\
2016 Dec.\ $12.32\pm0.17$ & $471.7\pm0.2$ & 13\\
2017 Dec.\ $31.3\pm0.1$ & $384.0\pm0.2$ & 14, 15 \\
2018 Nov.\ 06 & $\sim310$ & 15, 16 \\
\hline
\end{tabular}
\end{center}
\begin{minipage}{\columnwidth}
\setstretch{0.75}
{\footnotesize {Updated version of Table~1 from \protect \citet{2017ApJ...849...96D,2017ApJ...847...35D}.}}\\
{\footnotesize $^{\dag}${Those in parentheses are extrapolated from X-ray data.}}\\
{\footnotesize $^{\ddag}${Only quoted for consecutive detections in consecutive years.}}\\
{\footnotesize References --- {(1)~\citet{1995ApJ...445L.125W}, (2)~\citet{2015A&A...582L...8H},  (3)~\citet{2004ApJ...609..735W}, (4)~\citet{2008Nis}, (5)~\citet{2014ApJ...786...61T}, (6)~\citet{2011Kor}, (7)~\citet{2012Nis}, (8)~\citet{2014A&A...563L...9D}, (9)~\citet{2014A&A...563L...8H}, (10)~\citet{2015A&A...580A..45D}, (11)~\citet{2015A&A...580A..46H}, (12)~\citet{2016ApJ...833..149D}, (13)~\citet{2018ApJ...857...68H}, (14)~\citet{2017ATel11116....1B}, (15)~\citet{12a1718}, (16)~\citet{2018ATel12177....1D}.}}
\end{minipage}
\end{table}

\begin{figure*}[!ht]
\begin{center}
\includegraphics[width=0.9\textwidth]{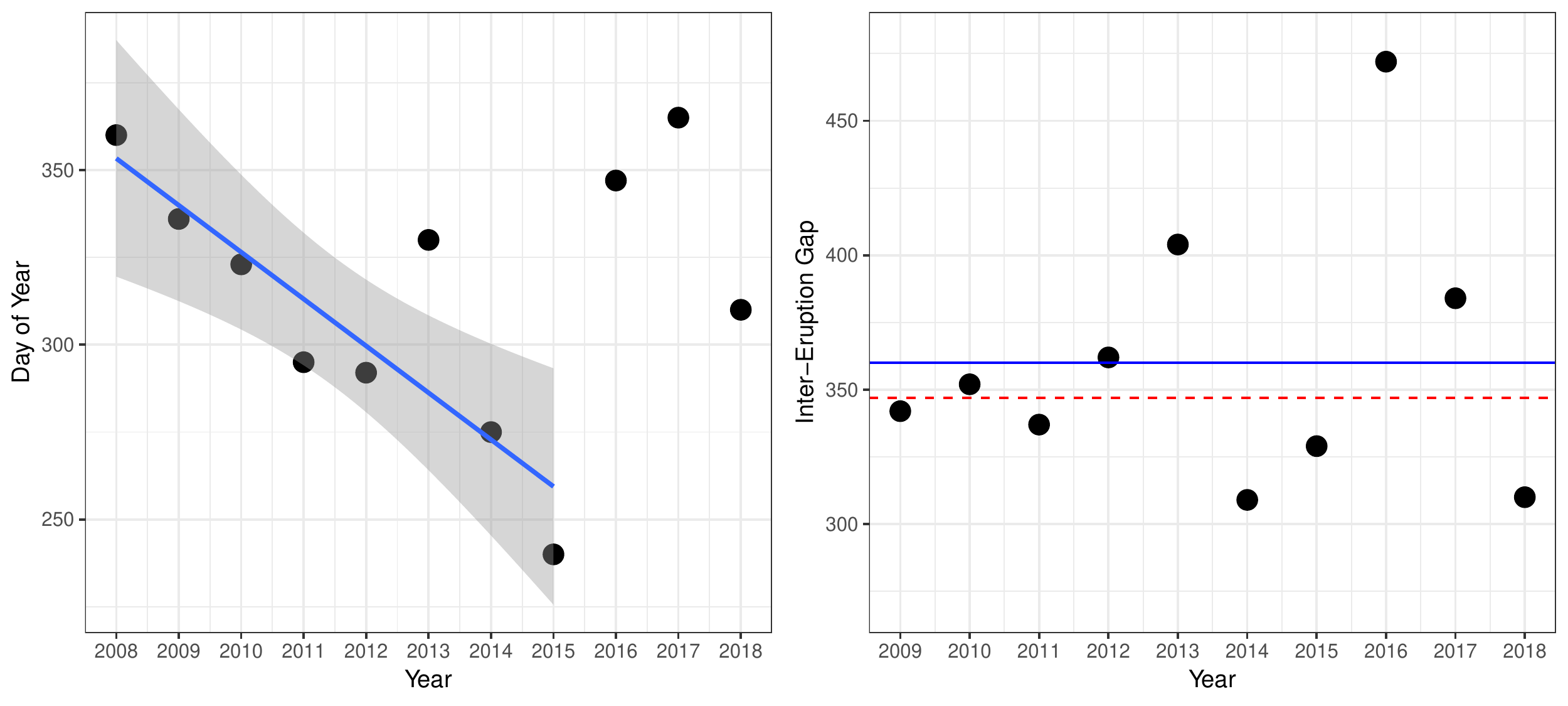}
\end{center}
\caption{(Left) Distribution of M31N\,2008-12a eruption dates since 2008, the blue line and the grey shaded region indicate the original timing model \citep{2015A&A...582L...8H}.  (Right) Distribution of 12a inter-eruption gaps, the blue line shows the mean, the red-dashed line the median.\label{rec_gaps}}
\end{figure*}

In Figure~\ref{rec_gaps} we show (left-hand panel) the original 12a eruption `model' \citep{2015A&A...582L...8H} that was consistent with either a $\sim6$\,month or $\sim12$\,month $P_\mathrm{rec}$ and described the 2008--2015 eruption timings reasonably well. But with the inclusion of the 2016--2018 eruptions, which possibly also emphasise the 2013 event, it seems clear that this original model does not well describe the eruptions. In the right-hand panel, we show the distribution of inter-eruption gaps. The solid blue line shows the mean, 360\,days, the red-dashed line the median, 347\,days, the standard deviation is 50\,days. If we are concerned with how well the mean inter-eruption time is known, then $\overline{P_\mathrm{rec}}=0.99\pm0.02$\,years\footnote{The authors agonised about whether to comment on this number, but decided to leave any speculation to the reader.}.

The 2017 and 2018 eruptions will be presented in a joint paper \citep{12a1718} --- both eruptions appear, at least superficially, to be very similar to the 2013--2015 events. And, so as not to break with tradition, in terms of predictions for the 2019 eruption, then at the time of writing, we would expect it to occur $360\pm50$\,days after the 2018 event:\ between September 12 and December 21.

\subsection{The super-remnant}\label{SR}

We continue to investigate archival observations to attempt the recovery of past -- missed -- eruptions. Our key collaborator, Allen Shafter, alerted us to a H$\alpha$ image of the 12a field that he and Karl Misselt had taken using the Steward 2.3m Bok Telescope \citep[see][]{2008ApJ...686.1261C,2012ApJ...760...13F} as part of an earlier M\,31 nova survey. These data did not reveal a previous eruption, but they did show evidence for {\it vastly} extended nebulosity around 12a \citep{2015A&A...580A..45D}. This discovery was soon confirmed via narrowband data from LGGS \citep{2007AJ....134.2474M} and the LT. Fortuitously, the LT SPRAT \citep[see][]{2014SPIE.9147E..8HP} long-slit spectra of the 2014 eruption contained H$\alpha$+[N\,{\sc ii}] and [S\,{\sc ii}] emission (but little else) from a bright knot in this nebula \citep{2015A&A...580A..45D}.

\begin{figure}[!ht]
\begin{center}
\includegraphics[width=0.9\columnwidth]{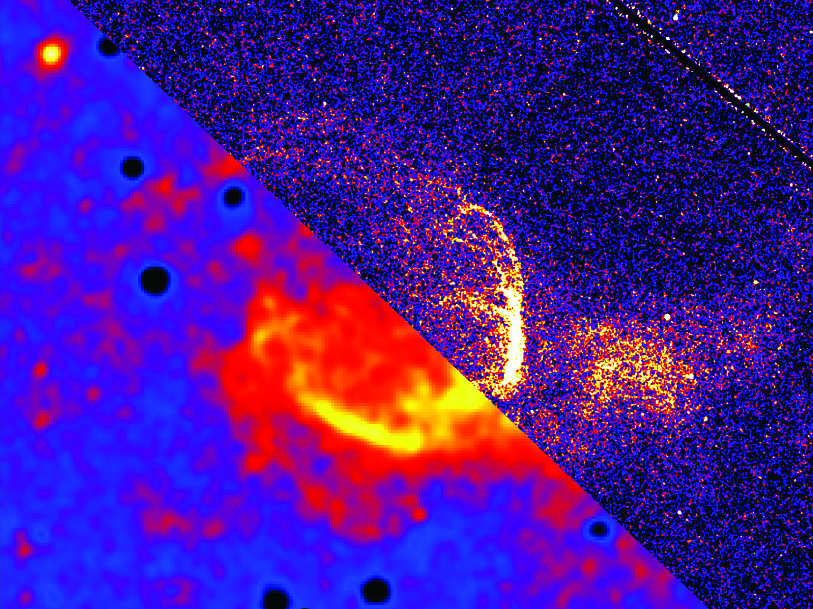}
\end{center}
\caption{The nova super-remnant surrounding M31N\,2008-12a. The lower left of the image shows the LT ground-based narrow-band H$\alpha$ data, the upper right the high spatial resolution {\it HST} H$\alpha$+[N\,{\sc ii}] imaging. Here the colour-scale is based on brightness, this image has been recreated based on the data published in \citet{2019Natur.565..460D}.\label{NSR_img}}
\end{figure}

To follow-up, high-spatial resolution H$\alpha$+[N\,{\sc ii}] imaging was obtained with {\it HST}, and deep low-resolution spectroscopy from the Gran Telescopio Canarias and HET. {\it HST} imaging clearly revealed the shell-like nature of the nebula with the spectra confirming that the phenomenon was not a SN remnant, yet was consistent with being predominately swept-up ISM \citep{2019Natur.565..460D}, see Figure~\ref{NSR_img}.

With semi-major and -minor axes of 67 and 45\,pc, respectively, and a swept-up mass of $\sim10^{5-6}\,\mathrm{M}_\odot$ \citep{2015A&A...580A..45D,2019Natur.565..460D}, a serious question remained, could sustained RN eruptions produce such a vast structure? 

The expanding nebulosity around the Galactic nova GK\,Persei was first noted by \citet{1901ApJ....14..293R,1901ApJ....14..167R}, with that nova's ejecta first photographed by \citeauthor{1916BHarO.621....1B}\footnote{Who identifies a ``Miss (Vera Marie) Gushee'' as the photographer.} in 1916\ \citep[see][]{2008clno.book.....B}. Nebular ejecta have been discovered and investigated around $\sim10\%$ of Galactic novae \citep[see, for e.g.,][]{1990LNP...369..179W,1995MNRAS.276..353S,2007ApJ...665L..63B}, but the largest of these are less than a parsec across \citep{2004ApJ...600L..63B,2007Natur.446..159S,2012ApJ...758..121S}. Evidence for interacting ejecta from successive RN eruptions has been presented for T\,Pyxidis \citep{1997AJ....114..258S,2013ApJ...768...48T}.

As a proof of concept, \citet{2019Natur.565..460D} presented a hydrodynamical simulation \citep[based on the {\tt Morpheus} code;][]{2007ApJ...665..654V} of 100,000 annual eruptions of 12a. This simulation followed each set of ejecta separately, including their self-interaction and interaction with the surrounding ISM. \citeauthor{2019Natur.565..460D}\ showed that such recurrent eruptions create a vast evacuated region around the central system while `piling' up the ISM in a thick expanding shell. Unlike remnants of single explosions/eruptions, this `super-remnant' contained a continually shock-heated region, inside the outer shell, where ejecta from successive eruptions collide. The properties of the simulated super-remnant were consistent with the observational constraints. Given the observed size of the super-remnant, \citeauthor{2019Natur.565..460D}\ suggested an age of $6\times10^{6}$\,yrs:\ assuming $\dot{M}=1.6\times10^{-7}\,\mathrm{M}_\odot\,\mathrm{yr}^{-1}$ \citep{2015ApJ...808...52K}, an accretion efficiency of only $\approx40\%$ would be required to grow a WD from 1\,M$_\odot$ to very close to the Chandrasekhar mass in that time.

At the time of writing, the authors see no reason why similar or `natal' super-remnants should not surround all RNe, particularly those with the shortest inter-eruption periods. However, given their vast scale and expected low surface brightness ($m_{\mathrm{H}\alpha}\gtrsim24$\,arcsec$^{-2}$) searches for additional examples are perhaps best conducted extragalactically. The existence of super-remnants around near-Chandrasekhar mass, rapidly accreting, WDs has potentially interesting consequences for any  catastrophic event that may subsequently befall the central system. For example, the interaction of a SN\,Ia explosion with a super-remnant environment should be explored to identify potential observational signatures for the nova pathway to a SN\,Ia.

\section{Open questions for the next few decades}

Numerous questions about novae still require answers. Those related to ISM enrichment, $\gamma$-rays, dust formation, and the ejecta geometries are probably best broached Galactically. But when it comes to populations and the link to SNe\,Ia, extragalactic work is vital. Are the faint--fast novae related to the RNe, are they all RRNe? How large is the RRN population, are systems like M31N\,2008-12a rare, or is 12a the tip of the iceberg? Can (or should) the MMRD concept be salvaged? Do RRNe provide a substantial SN\,Ia channel? What is the ratio of CO to ONe WDs in novae? Does the nova population vary between and within host galaxies? Do local stellar population,  star formation history, and metallicity affect novae? How do novae affect their environment, is the 12a nova super-remnant unique? 

The hugely anticipated high-cadence all-sky surveys, such as LSST, could be a game changer, particularly for the faint--fast and RRNe (although the anticipated LSST observing cadence might be an issue). When launched, the {\it James Webb Space Telescope} could revolutionise the IR studies of novae and their ejecta, but we may also lose {\it HST} and its unparalleled UV capability. Novae will have to compete for their share of new facilities, and while discovery of new novae is one thing, follow-up capability is another. But despite the onslaught of automated all-sky surveys, it was the amateur community that discovered 12a and continues to provide invaluable support to its study. A huge proportion of extragalactic nova discoveries still come from amateur astronomers, these individuals and groups must not be under-valued. The future is bright, we have a lot of work still to do, observational nova work is likely to become more rewarding, but much more challenging. 

\section*{Acknowledgements}

The authors would like to express there sincere gratitude to the following for their input to, proofing reading of, and discussion of the content of this Review:\ {\c S}{\"o}len Balman, Mike Bode, \'Eamonn Harvey, Mike Healy, Phil James, Fiona Murphy-Glaysher, Conor Ransome, Allen Shafter, and Steve Williams. Darnley \& Henze are grateful to Allen Shafter and Mike Healy for sharing the data and code, respectively, behind Figure~\ref{lsnr-plot}.  Matt Darnley acknowledges funding from the UK Science \& Technology Facilities Council (STFC). Finally, we would like to thank the two referees for their careful reading and thoughtful reports.

\section*{Bibliography}

\bibliographystyle{cospar_new}
\bibliography{refs}

\begin{thebibliography}{259}
\providecommand{\natexlab}[1]{#1}

\bibitem[{{Abdo} et~al.(2010){Abdo}, {Ackermann}, {Ajello}
  et~al.}]{2010Sci...329..817A}
{Abdo}, A.~A., {Ackermann}, M., {Ajello}, M., et~al., 2010.
\newblock {Gamma-Ray Emission Concurrent with the Nova in the Symbiotic Binary
  V407 Cygni}.
\newblock Science, 329, 817-821.

\bibitem[{{Ackermann} et~al.(2014){Ackermann}, {Ajello}, {Albert}
  et~al.}]{2014Sci...345..554A}
{Ackermann}, M., {Ajello}, M., {Albert}, A., et~al., 2014.
\newblock {Fermi establishes classical novae as a distinct class of gamma-ray
  sources}.
\newblock Science, 345, 554-558.

\bibitem[{{Alksnis} \& {Zharova}(2000)}]{2000IBVS.4909....1A}
{Alksnis}, A., {Zharova}, A.~V., 2000.
\newblock {PT Andromedae:\ the Recent Outburst and Earlier Ones}.
\newblock Information Bulletin on Variable Stars, 4909, 1.

\bibitem[{{An} et~al.(2004){An}, {Evans}, {Kerins}
  et~al.}]{2004ApJ...601..845A}
{An}, J.~H., {Evans}, N.~W., {Kerins}, E., et~al., 2004.
\newblock {The Anomaly in the Candidate Microlensing Event PA-99-N2}.
\newblock \apj, 601, 845-857.

\bibitem[{{Arce-Tord} et~al.(2017){Arce-Tord}, {Esteban-Gutierrez},
  {Garcia-Broock} et~al.}]{2017ATel11094....1A}
{Arce-Tord}, C., {Esteban-Gutierrez}, A., {Garcia-Broock}, E., et~al., 2017.
\newblock {INT WFC photometry of a Galactic flare star spatially coincident
  with the recurrent nova candidate M31N 1966-08a = 1968-10c}.
\newblock The Astronomer's Telegram, 11094.

\bibitem[{{Arp}(1956)}]{1956AJ.....61...15A}
{Arp}, H.~C., 1956.
\newblock {Novae in the Andromeda nebula.}
\newblock \aj, 61, 15-34.

\bibitem[{{Auri{\`e}re} et~al.(2001){Auri{\`e}re}, {Baillon}, {Bouquet}
  et~al.}]{2001ApJ...553L.137A}
{Auri{\`e}re}, M., {Baillon}, P., {Bouquet}, A., et~al., 2001.
\newblock {A Short-Timescale Candidate Microlensing Event in the POINT-AGAPE
  Pixel Lensing Survey of M31}.
\newblock \apjl, 553, L137-L140.

\bibitem[{{Aydi} et~al.(2018{\natexlab{a}}){Aydi}, {Orio}, {Beardmore}
  et~al.}]{2018MNRAS.480..572A}
{Aydi}, E., {Orio}, M., {Beardmore}, A.~P., et~al., 2018{\natexlab{a}}.
\newblock {Multiwavelength observations of V407 Lupi (ASASSN-16kt) --- a very
  fast nova erupting in an intermediate polar}.
\newblock \mnras, 480, 572-609.

\bibitem[{{Aydi} et~al.(2018{\natexlab{b}}){Aydi}, {Page}, {Kuin}
  et~al.}]{2018MNRAS.474.2679A}
{Aydi}, E., {Page}, K.~L., {Kuin}, N.~P.~M., et~al., 2018{\natexlab{b}}.
\newblock {Multiwavelength observations of nova SMCN 2016-10a --- one of the
  brightest novae ever observed}.
\newblock \mnras, 474, 2679-2705.

\bibitem[{{Barnard}(1916)}]{1916BHarO.621....1B}
{Barnard}, E.~E., 1916.
\newblock {Nebulosity around Nova Persei}.
\newblock Harvard College Observatory Bulletin, 621, 1-1.

\bibitem[{{Barsukova} et~al.(2011){Barsukova}, {Fabrika}, {Hornoch}
  et~al.}]{2011ATel.3725....1B}
{Barsukova}, E., {Fabrika}, S., {Hornoch}, K., et~al., 2011.
\newblock {Spectroscopy of nova 2011-10d and photometry of nova candidate
  2011-10e in M31}.
\newblock The Astronomer's Telegram, 3725, 1.

\bibitem[{{Bellm} et~al.(2019){Bellm}, {Kulkarni}, {Graham}
  et~al.}]{2019PASP..131a8002B}
{Bellm}, E.~C., {Kulkarni}, S.~R., {Graham}, M.~J., et~al., 2019.
\newblock {The Zwicky Transient Facility: System Overview, Performance, and
  First Results}.
\newblock \pasp, 131, 1, 018002.

\bibitem[{{Bode}(2010)}]{2010AN....331..160B}
{Bode}, M.~F., 2010.
\newblock {The outbursts of classical and recurrent novae}.
\newblock Astronomische Nachrichten, 331, 160-167.

\bibitem[{{Bode} et~al.(2016){Bode}, {Darnley}, {Beardmore}
  et~al.}]{2016ApJ...818..145B}
{Bode}, M.~F., {Darnley}, M.~J., {Beardmore}, A.~P., et~al., 2016.
\newblock {Pan-chromatic Observations of the Recurrent Nova LMC 2009a (LMC
  1971b)}.
\newblock \apj, 818, 145.

\bibitem[{{Bode} et~al.(2009){Bode}, {Darnley}, {Shafter}
  et~al.}]{2009ApJ...705.1056B}
{Bode}, M.~F., {Darnley}, M.~J., {Shafter}, A.~W., et~al., 2009.
\newblock {Optical and X-ray Observations of M31N 2007-12b: An Extragalactic
  Recurrent Nova with a Detected Progenitor?}
\newblock \apj, 705, 1056-1062.

\bibitem[{{Bode} \& {Evans}(2008)}]{2008clno.book.....B}
{Bode}, M.~F., {Evans}, A., editors, 2008.
\newblock {Classical Novae, 2nd Edition}, volume~43 of Cambridge Astrophysics
  Series.
\newblock Cambridge University Press, Cambridge.

\bibitem[{{Bode} et~al.(2007){Bode}, {Harman}, {O'Brien}
  et~al.}]{2007ApJ...665L..63B}
{Bode}, M.~F., {Harman}, D.~J., {O'Brien}, T.~J., et~al., 2007.
\newblock {Hubble Space Telescope Imaging of the Expanding Nebular Remnant of
  the 2006 Outburst of the Recurrent Nova RS Ophiuchi}.
\newblock \apjl, 665, L63-L66.

\bibitem[{{Bode} \& {Kahn}(1985)}]{1985MNRAS.217..205B}
{Bode}, M.~F., {Kahn}, F.~D., 1985.
\newblock {A model for the outburst of nova RS Ophiuchi in 1985}.
\newblock \mnras, 217, 205-215.

\bibitem[{{Bode} et~al.(2006){Bode}, {O'Brien}, {Osborne}
  et~al.}]{2006ApJ...652..629B}
{Bode}, M.~F., {O'Brien}, T.~J., {Osborne}, J.~P., et~al., 2006.
\newblock {Swift Observations of the 2006 Outburst of the Recurrent Nova RS
  Ophiuchi. I. Early X-Ray Emission from the Shocked Ejecta and Red Giant
  Wind}.
\newblock \apj, 652, 629-635.

\bibitem[{{Bode} et~al.(2004){Bode}, {O'Brien}, \&
  {Simpson}}]{2004ApJ...600L..63B}
{Bode}, M.~F., {O'Brien}, T.~J., {Simpson}, M., 2004.
\newblock {Echoes of an Explosive Past: Solving the Mystery of the First
  Superluminal Source}.
\newblock \apjl, 600, L63-L66.

\bibitem[{{Bond} et~al.(2004){Bond}, {Walter}, {Espinoza}
  et~al.}]{2004IAUC.8424....1B}
{Bond}, H.~E., {Walter}, F., {Espinoza}, J., et~al., 2004.
\newblock {Nova in the Large Magellanic Cloud 2004}.
\newblock \iaucirc, 8424.

\bibitem[{{B{\"o}rngen}(1968)}]{1968AN....291...19B}
{B{\"o}rngen}, F., 1968.
\newblock {Novae in M 31 auf Tautenburger Schmidt-Aufnahmeni (Novae in M 31 on
  Tautenburg Schmidt Plates)}.
\newblock Astronomische Nachrichten, 291, 19-24.

\bibitem[{{Boyd} et~al.(2017){Boyd}, {Hornoch}, {Henze}
  et~al.}]{2017ATel11116....1B}
{Boyd}, D., {Hornoch}, K., {Henze}, M., et~al., 2017.
\newblock {Recurrent Nova M31N 2008-12a: discovery of the 2017 eruption}.
\newblock The Astronomer's Telegram, 11116.

\bibitem[{{Brown} et~al.(2013){Brown}, {Baliber}, {Bianco}
  et~al.}]{2013PASP..125.1031B}
{Brown}, T.~M., {Baliber}, N., {Bianco}, F.~B., et~al., 2013.
\newblock {Las Cumbres Observatory Global Telescope Network}.
\newblock \pasp, 125, 1031-1055.

\bibitem[{{Burrows} et~al.(2005){Burrows}, {Hill}, {Nousek}
  et~al.}]{2005SSRv..120..165B}
{Burrows}, D.~N., {Hill}, J.~E., {Nousek}, J.~A., et~al., 2005.
\newblock {The Swift X-Ray Telescope}.
\newblock \ssr, 120, 165-195.

\bibitem[{{Cameron}(1959)}]{1959ApJ...130..916C}
{Cameron}, A.~G.~W., 1959.
\newblock {Pycnonuclear Reations and Nova Explosions.}
\newblock \apj, 130, 916-940.

\bibitem[{{Cao} et~al.(2012){Cao}, {Kasliwal}, {Neill}
  et~al.}]{2012ApJ...752..133C}
{Cao}, Y., {Kasliwal}, M.~M., {Neill}, J.~D., et~al., 2012.
\newblock {Classical Novae in Andromeda: Light Curves from the Palomar
  Transient Factory and GALEX}.
\newblock \apj, 752, 133.

\bibitem[{{Capaccioli} et~al.(1989){Capaccioli}, {Della Valle}, {Rosino}
  et~al.}]{1989AJ.....97.1622C}
{Capaccioli}, M., {Della Valle}, M., {Rosino}, L., et~al., 1989.
\newblock {Properties of the nova population in M31}.
\newblock \aj, 97, 1622-1633.

\bibitem[{{Carey} et~al.(2019){Carey}, {Hornoch}, \&
  {Shafter}}]{2019ATel12513....1C}
{Carey}, G., {Hornoch}, K., {Shafter}, A.~W., 2019.
\newblock {A fourth observed outburst of a Galactic flare star in the field of
  M31}.
\newblock The Astronomer's Telegram, 12513.

\bibitem[{{Chandrasekhar}(1931)}]{1931ApJ....74...81C}
{Chandrasekhar}, S., 1931.
\newblock {The Maximum Mass of Ideal White Dwarfs}.
\newblock \apj, 74, 81-82.

\bibitem[{{Chen} et~al.(2016){Chen}, {Woods}, {Yungelson}
  et~al.}]{2016MNRAS.458.2916C}
{Chen}, H.-L., {Woods}, T.~E., {Yungelson}, L.~R., et~al., 2016.
\newblock {Modelling nova populations in galaxies}.
\newblock \mnras, 458, 2916-2927.

\bibitem[{{Chesneau} et~al.(2012){Chesneau}, {Lagadec}, {Otulakowska-Hypka}
  et~al.}]{2012A&A...545A..63C}
{Chesneau}, O., {Lagadec}, E., {Otulakowska-Hypka}, M., et~al., 2012.
\newblock {The expanding dusty bipolar nebula around the nova V1280 Scorpi}.
\newblock \aap, 545, A63.

\bibitem[{{Chomiuk} et~al.(2014){Chomiuk}, {Linford}, {Yang}
  et~al.}]{2014Natur.514..339C}
{Chomiuk}, L., {Linford}, J.~D., {Yang}, J., et~al., 2014.
\newblock {Binary orbits as the driver of {$\gamma$}-ray emission and mass
  ejection in classical novae}.
\newblock \nat, 514, 339-342.

\bibitem[{{Ciardullo} et~al.(1987){Ciardullo}, {Ford}, {Neill}
  et~al.}]{1987ApJ...318..520C}
{Ciardullo}, R., {Ford}, H.~C., {Neill}, J.~D., et~al., 1987.
\newblock {The spatial distribution and population of novae in M31}.
\newblock \apj, 318, 520-530.

\bibitem[{{Ciardullo} et~al.(1990{\natexlab{a}}){Ciardullo}, {Shafter}, {Ford}
  et~al.}]{1990ApJ...356..472C}
{Ciardullo}, R., {Shafter}, A.~W., {Ford}, H.~C., et~al., 1990{\natexlab{a}}.
\newblock {The H-alpha light curves of novae in M31}.
\newblock \apj, 356, 472-482.

\bibitem[{{Ciardullo} et~al.(1990{\natexlab{b}}){Ciardullo}, {Tamblyn},
  {Jacoby} et~al.}]{1990AJ.....99.1079C}
{Ciardullo}, R., {Tamblyn}, P., {Jacoby}, G.~H., et~al., 1990{\natexlab{b}}.
\newblock {The nova rate in the elliptical component of NGC 5128}.
\newblock \aj, 99, 1079-1087.

\bibitem[{{Coelho} et~al.(2008){Coelho}, {Shafter}, \&
  {Misselt}}]{2008ApJ...686.1261C}
{Coelho}, E.~A., {Shafter}, A.~W., {Misselt}, K.~A., 2008.
\newblock {The Rate and Spatial Distribution of Novae in M101 (NGC 5457)}.
\newblock \apj, 686, 1261-1268.

\bibitem[{{Cohen}(1985)}]{1985ApJ...292...90C}
{Cohen}, J.~G., 1985.
\newblock {Nova shells. II - Calibration of the distance scale using novae}.
\newblock \apj, 292, 90-103.

\bibitem[{{Conseil}(2017{\natexlab{a}})}]{2017Con}
{Conseil}, E., 2017{\natexlab{a}}.
\newblock {CBAT}.
\newblock
  \url{http://www.cbat.eps.harvard.edu/unconf/followups/J00412371+4114594.html}.

\bibitem[{{Conseil}(2017{\natexlab{b}})}]{2017TNSTR.867....1C}
{Conseil}, E., 2017{\natexlab{b}}.
\newblock {Transient Discovery Report for 2017-08-15}.
\newblock Transient Name Server Discovery Report, 2017-867, 1.

\bibitem[{{Curtin} et~al.(2015){Curtin}, {Shafter}, {Pritchet}
  et~al.}]{2015ApJ...811...34C}
{Curtin}, C., {Shafter}, A.~W., {Pritchet}, C.~J., et~al., 2015.
\newblock {Exploring the Role of Globular Cluster Specific Frequency on the
  Nova Rates in Three Virgo Elliptical Galaxies}.
\newblock \apj, 811, 34.

\bibitem[{{Dalcanton} et~al.(2012){Dalcanton}, {Williams}, {Lang}
  et~al.}]{2012ApJS..200...18D}
{Dalcanton}, J.~J., {Williams}, B.~F., {Lang}, D., et~al., 2012.
\newblock {The Panchromatic Hubble Andromeda Treasury}.
\newblock \apjs, 200, 18.

\bibitem[{{Darnley}(2017)}]{2017ASPC..509..515D}
{Darnley}, M.~J., 2017.
\newblock {M31N 2008-12a --- The Remarkable Recurrent Nova in M31}.
\newblock In P.-E. {Tremblay}, B.~{Gaensicke}, T.~{Marsh}, editors, 20th
  European White Dwarf Workshop, volume 509 of Astronomical Society of the
  Pacific Conference Series, pages 515--520. San Francisco.

\bibitem[{{Darnley} et~al.(2014{\natexlab{a}}){Darnley}, {Bode}, {Harman}
  et~al.}]{2014ASPC..490...49D}
{Darnley}, M.~J., {Bode}, M.~F., {Harman}, D.~J., et~al., 2014{\natexlab{a}}.
\newblock {On the Galactic Nova Progenitor Population}.
\newblock In P.~A. {Woudt}, V.~A.~R.~M. {Ribeiro}, editors, Stellar Novae: Past
  and Future Decades, volume 490 of Astronomical Society of the Pacific
  Conference Series, pages 49--55. San Francisco.

\bibitem[{{Darnley} et~al.(2004){Darnley}, {Bode}, {Kerins}
  et~al.}]{2004MNRAS.353..571D}
{Darnley}, M.~J., {Bode}, M.~F., {Kerins}, E., et~al., 2004.
\newblock {Classical novae from the POINT-AGAPE microlensing survey of M31 - I.
  The nova catalogue}.
\newblock \mnras, 353, 571-588.

\bibitem[{{Darnley} et~al.(2006){Darnley}, {Bode}, {Kerins}
  et~al.}]{2006MNRAS.369..257D}
{Darnley}, M.~J., {Bode}, M.~F., {Kerins}, E., et~al., 2006.
\newblock {Classical novae from the POINT-AGAPE microlensing survey of M31 -
  II. Rate and statistical characteristics of the nova population}.
\newblock \mnras, 369, 257-271.

\bibitem[{{Darnley} et~al.(2016{\natexlab{a}}){Darnley}, {Henze}, {Bode}
  et~al.}]{2016ApJ...833..149D}
{Darnley}, M.~J., {Henze}, M., {Bode}, M.~F., et~al., 2016{\natexlab{a}}.
\newblock {M31N 2008-12a - The Remarkable Recurrent Nova in M31: Panchromatic
  Observations of the 2015 Eruption.}
\newblock \apj, 833, 149.

\bibitem[{{Darnley} et~al.(2018){Darnley}, {Henze}, {Shafter}
  et~al.}]{2018ATel12177....1D}
{Darnley}, M.~J., {Henze}, M., {Shafter}, A.~W., et~al., 2018.
\newblock {Recurrent Nova M31N 2008-12a: discovery of the 2018 eruption}.
\newblock The Astronomer's Telegram, 12177.

\bibitem[{{Darnley} et~al.(2019{\natexlab{a}}){Darnley}, {Henze}, {Shafter}
  et~al.}]{12a1718}
{Darnley}, M.~J., {Henze}, M., {Shafter}, A.~W., et~al., 2019{\natexlab{a}}.
\newblock {Back on Track? --- The 2017 and 2018 eruptions of M31N 2008-12a}.
\newblock In preparation, for publication in A\&A.

\bibitem[{{Darnley} et~al.(2015){Darnley}, {Henze}, {Steele}
  et~al.}]{2015A&A...580A..45D}
{Darnley}, M.~J., {Henze}, M., {Steele}, I.~A., et~al., 2015.
\newblock {A remarkable recurrent nova in M31: Discovery and optical/UV
  observations of the predicted 2014 eruption}.
\newblock \aap, 580, A45.

\bibitem[{{Darnley} et~al.(2017{\natexlab{a}}){Darnley}, {Hounsell}, {Godon}
  et~al.}]{2017ApJ...849...96D}
{Darnley}, M.~J., {Hounsell}, R., {Godon}, P., et~al., 2017{\natexlab{a}}.
\newblock {Inflows, Outflows, and a Giant Donor in the Remarkable Recurrent
  Nova M31N 2008-12a?---Hubble Space Telescope Photometry of the 2015
  Eruption}.
\newblock \apj, 849, 96.

\bibitem[{{Darnley} et~al.(2017{\natexlab{b}}){Darnley}, {Hounsell}, {Godon}
  et~al.}]{2017ApJ...847...35D}
{Darnley}, M.~J., {Hounsell}, R., {Godon}, P., et~al., 2017{\natexlab{b}}.
\newblock {No Neon, but Jets in the Remarkable Recurrent Nova M31N
  2008-12a?---Hubble Space Telescope Spectroscopy of the 2015 Eruption}.
\newblock \apj, 847, 35.

\bibitem[{{Darnley} et~al.(2019{\natexlab{b}}){Darnley}, {Hounsell}, {O'Brien}
  et~al.}]{2019Natur.565..460D}
{Darnley}, M.~J., {Hounsell}, R., {O'Brien}, T.~J., et~al., 2019{\natexlab{b}}.
\newblock {A recurrent nova super-remnant in the Andromeda galaxy}.
\newblock \nat, 565, 460-463.

\bibitem[{{Darnley} et~al.(2016{\natexlab{b}}){Darnley}, {Kuin}, {Page}
  et~al.}]{2016ATel.8587....1D}
{Darnley}, M.~J., {Kuin}, N.~P.~M., {Page}, K.~L., et~al., 2016{\natexlab{b}}.
\newblock {Swift observations of the early development of the 2016 eruption of
  the recurrent nova LMCN 1968-12a (OGLE-2016-NOVA-01)}.
\newblock The Astronomer's Telegram, 8587.

\bibitem[{{Darnley} et~al.(2012){Darnley}, {Ribeiro}, {Bode}
  et~al.}]{2012ApJ...746...61D}
{Darnley}, M.~J., {Ribeiro}, V.~A.~R.~M., {Bode}, M.~F., et~al., 2012.
\newblock {On the Progenitors of Galactic Novae}.
\newblock \apj, 746, 61.

\bibitem[{{Darnley} et~al.(2014{\natexlab{b}}){Darnley}, {Williams}, {Bode}
  et~al.}]{2014A&A...563L...9D}
{Darnley}, M.~J., {Williams}, S.~C., {Bode}, M.~F., et~al., 2014{\natexlab{b}}.
\newblock {A remarkable recurrent nova in M 31: The optical observations}.
\newblock \aap, 563, L9.

\bibitem[{{de Vaucouleurs}(1958)}]{1958ApJ...128..465D}
{de Vaucouleurs}, G., 1958.
\newblock {Photoelectric photometry of the Andromeda nebula in the UBV system}.
\newblock \apj, 128, 465-488.

\bibitem[{{de Vaucouleurs}(1978)}]{1978ApJ...223..351D}
{de Vaucouleurs}, G., 1978.
\newblock {The extragalactic distance scale. I - A review of distance
  indicators - Zero points and errors of primary indicators}.
\newblock \apj, 223, 351-363.

\bibitem[{{de Vaucouleurs} \& {Corwin}(1985)}]{1985ApJ...295..287D}
{de Vaucouleurs}, G., {Corwin}, H.~G., Jr., 1985.
\newblock {S Andromedae 1885 - A centennial review}.
\newblock \apj, 295, 287-304.

\bibitem[{{Della Valle} et~al.(1992){Della Valle}, {Bianchini}, {Livio}
  et~al.}]{1992A&A...266..232D}
{Della Valle}, M., {Bianchini}, A., {Livio}, M., et~al., 1992.
\newblock {On the possible existence of two classes of progenitors for
  classical novae}.
\newblock \aap, 266, 232-236.

\bibitem[{{Della Valle} \& {Gilmozzi}(2002)}]{2002Sci...296.1275D}
{Della Valle}, M., {Gilmozzi}, R., 2002.
\newblock {Rebirth of Novae as Distance Indicators due to Efficient Large
  Telescopes}.
\newblock Science, 296, 1275-1276.

\bibitem[{{Della Valle} \& {Livio}(1995)}]{1995ApJ...452..704D}
{Della Valle}, M., {Livio}, M., 1995.
\newblock {The Calibration of Novae as Distance Indicators}.
\newblock \apj, 452, 704-709.

\bibitem[{{Della Valle} \& {Livio}(1996)}]{1996ApJ...473..240D}
{Della Valle}, M., {Livio}, M., 1996.
\newblock {On the Frequency of Occurrence of Recurrent Novae and Their Role as
  Type IA Supernova Progenitors}.
\newblock \apj, 473, 240-243.

\bibitem[{{Della Valle} \& {Livio}(1998)}]{1998ApJ...506..818D}
{Della Valle}, M., {Livio}, M., 1998.
\newblock {The Spectroscopic Differences between Disk and Thick-Disk/Bulge
  Novae}.
\newblock \apj, 506, 818-823.

\bibitem[{{Downes} \& {Duerbeck}(2000)}]{2000AJ....120.2007D}
{Downes}, R.~A., {Duerbeck}, H.~W., 2000.
\newblock {Optical Imaging of Nova Shells and the Maximum Magnitude-Rate of
  Decline Relationship}.
\newblock \aj, 120, 2007-2037.

\bibitem[{{Duerbeck}(1990)}]{1990LNP...369...34D}
{Duerbeck}, H.~W., 1990.
\newblock {Galactic Distribution and Outburst Frequency of Classical Novae}.
\newblock In A.~{Cassatella}, R.~{Viotti}, editors, IAU Colloq. 122: Physics of
  Classical Novae, volume 369 of Lecture Notes in Physics, pages 34--41.
  Springer Verlag, Berlin.

\bibitem[{{Dufay} et~al.(1964){Dufay}, {Bloch}, {Bertaud}
  et~al.}]{1964AnAp...27..555D}
{Dufay}, J., {Bloch}, M., {Bertaud}, C., et~al., 1964.
\newblock {{\'E}lvolution du spectre de Nova RS Ophiuchi apr{\`e}s l'explosion
  de 1958 (Evolution of the spectrum of Nova RS Ophiuchi after the 1958
  explosion)}.
\newblock Annales d'Astrophysique, 27, 555-586.

\bibitem[{{Feeney} et~al.(2005){Feeney}, {Belokurov}, {Evans}
  et~al.}]{2005AJ....130...84F}
{Feeney}, S.~M., {Belokurov}, V., {Evans}, N.~W., et~al., 2005.
\newblock {Automated Detection of Classical Novae with Neural Networks}.
\newblock \aj, 130, 84-94.

\bibitem[{{Ferrarese} et~al.(1996){Ferrarese}, {Livio}, {Freedman}
  et~al.}]{1996ApJ...468L..95F}
{Ferrarese}, L., {Livio}, M., {Freedman}, W., et~al., 1996.
\newblock {Discovery of a Nova in the Virgo Galaxy M100}.
\newblock \apjl, 468, L95-L97.

\bibitem[{{Franck} et~al.(2012){Franck}, {Shafter}, {Hornoch}
  et~al.}]{2012ApJ...760...13F}
{Franck}, J.~R., {Shafter}, A.~W., {Hornoch}, K., et~al., 2012.
\newblock {The Nova Rate in NGC 2403}.
\newblock \apj, 760, 13.

\bibitem[{{Freedman} et~al.(2001){Freedman}, {Madore}, {Gibson}
  et~al.}]{2001ApJ...553...47F}
{Freedman}, W.~L., {Madore}, B.~F., {Gibson}, B.~K., et~al., 2001.
\newblock {Final Results from the Hubble Space Telescope Key Project to Measure
  the Hubble Constant}.
\newblock \apj, 553, 47-72.

\bibitem[{{Gaia Collaboration} et~al.(2018){Gaia Collaboration}, {Brown},
  {Vallenari} et~al.}]{2018A&A...616A...1G}
{Gaia Collaboration}, {Brown}, A.~G.~A., {Vallenari}, A., et~al., 2018.
\newblock {Gaia Data Release 2. Summary of the contents and survey properties}.
\newblock \aap, 616, A1.

\bibitem[{{Gaia Collaboration} et~al.(2016){Gaia Collaboration}, {Prusti}, {de
  Bruijne} et~al.}]{2016A&A...595A...1G}
{Gaia Collaboration}, {Prusti}, T., {de Bruijne}, J.~H.~J., et~al., 2016.
\newblock {The Gaia mission}.
\newblock \aap, 595, A1.

\bibitem[{{Godon} et~al.(2017){Godon}, {Sion}, {Balman}
  et~al.}]{2017ApJ...846...52G}
{Godon}, P., {Sion}, E.~M., {Balman}, {\c S}., et~al., 2017.
\newblock {Modifying the Standard Disk Model for the Ultraviolet Spectral
  Analysis of Disk-dominated Cataclysmic Variables. I. The Novalikes MV Lyrae,
  BZ Camelopardalis, and V592 Cassiopeiae}.
\newblock \apj, 846, 52.

\bibitem[{{Godon} et~al.(2018){Godon}, {Sion}, {Williams}
  et~al.}]{2018ApJ...862...89G}
{Godon}, P., {Sion}, E.~M., {Williams}, R.~E., et~al., 2018.
\newblock {The Long-term Secular Mass Accretion Rate of the Recurrent Nova T
  Pyxidis}.
\newblock \apj, 862, 89.

\bibitem[{{Greiner} et~al.(2004){Greiner}, {Di Stefano}, {Kong}
  et~al.}]{2004ApJ...610..261G}
{Greiner}, J., {Di Stefano}, R., {Kong}, A., et~al., 2004.
\newblock {Supersoft X-Ray Sources in M31. II. ROSAT-detected Supersoft Sources
  in the ROSAT, Chandra, and XMM-Newton Eras}.
\newblock \apj, 610, 261-268.

\bibitem[{{Greiner} et~al.(1996){Greiner}, {Supper}, \&
  {Magnier}}]{1996LNP...472...75G}
{Greiner}, J., {Supper}, R., {Magnier}, E.~A., 1996.
\newblock {Supersoft X-Ray Sources in M 31}.
\newblock In J.~{Greiner}, editor, Supersoft X-Ray Sources, volume 472 of
  Lecture Notes in Physics, Berlin Springer Verlag, pages 75--82.

\bibitem[{{Gurevitch} \& {Lebedinsky}(1957)}]{1957IAUS....3...77G}
{Gurevitch}, L.~E., {Lebedinsky}, A.~I., 1957.
\newblock {On the causes of stellar bursts}.
\newblock In G.~H. {Herbig}, editor, Non-stable stars, volume~3 of IAU
  Symposium, pages 77--82. Cambridge Univ. Press, Cambridge, UK.

\bibitem[{{G{\"u}th} et~al.(2010){G{\"u}th}, {Shafter}, \&
  {Misselt}}]{2010ApJ...720.1155G}
{G{\"u}th}, T., {Shafter}, A.~W., {Misselt}, K.~A., 2010.
\newblock {The Nova Rate in M94 (NGC 4736)}.
\newblock \apj, 720, 1155-1160.

\bibitem[{{Gutierrez} et~al.(1996){Gutierrez}, {Garcia-Berro}, {Iben}
  et~al.}]{1996ApJ...459..701G}
{Gutierrez}, J., {Garcia-Berro}, E., {Iben}, I., Jr., et~al., 1996.
\newblock {The Final Evolution of ONeMg Electron-Degenerate Cores}.
\newblock \apj, 459, 701-705.

\bibitem[{{Hachisu} et~al.(2006){Hachisu}, {Kato}, {Kiyota}
  et~al.}]{2006ApJ...651L.141H}
{Hachisu}, I., {Kato}, M., {Kiyota}, S., et~al., 2006.
\newblock {The Hydrogen-Burning Turnoff of RS Ophiuchi (2006)}.
\newblock \apjl, 651, L141-L144.

\bibitem[{{Hachisu} et~al.(2008){Hachisu}, {Kato}, {Kiyota}
  et~al.}]{2008ASPC..401..206H}
{Hachisu}, I., {Kato}, M., {Kiyota}, S., et~al., 2008.
\newblock {Optical Light Curves of RS Oph (2006) and Hydrogen Burning Turnoff}.
\newblock In A.~{Evans}, M.~F. {Bode}, T.~J. {O'Brien}, M.~J. {Darnley},
  editors, RS Ophiuchi (2006) and the Recurrent Nova Phenomenon, volume 401 of
  Astronomical Society of the Pacific Conference Series, pages 206--209. San
  Francisco.

\bibitem[{{Hachisu} et~al.(1999{\natexlab{a}}){Hachisu}, {Kato}, \&
  {Nomoto}}]{1999ApJ...522..487H}
{Hachisu}, I., {Kato}, M., {Nomoto}, K., 1999{\natexlab{a}}.
\newblock {A Wide Symbiotic Channel to Type IA Supernovae}.
\newblock \apj, 522, 487-503.

\bibitem[{{Hachisu} et~al.(1999{\natexlab{b}}){Hachisu}, {Kato}, {Nomoto}
  et~al.}]{1999ApJ...519..314H}
{Hachisu}, I., {Kato}, M., {Nomoto}, K., et~al., 1999{\natexlab{b}}.
\newblock {A New Evolutionary Path to Type IA Supernovae: A Helium-rich
  Supersoft X-Ray Source Channel}.
\newblock \apj, 519, 314-323.

\bibitem[{{Hachisu} et~al.(2016){Hachisu}, {Saio}, \&
  {Kato}}]{2016ApJ...824...22H}
{Hachisu}, I., {Saio}, H., {Kato}, M., 2016.
\newblock {Shortest Recurrence Periods of Forced Novae}.
\newblock \apj, 824, 22.

\bibitem[{{Hartwig}(1885)}]{1885AN....112..355H}
{Hartwig}, E., 1885.
\newblock {Ueber den neuen Stern in grossen Andromeda-Nebel (About the new star
  in the large Andromeda nebula)}.
\newblock Astronomische Nachrichten, 112, 355-358.

\bibitem[{{Healy} et~al.(2019){Healy}, {Darnley}, {Copperwheat}
  et~al.}]{2019MNRAS.486.4334H}
{Healy}, M.~W., {Darnley}, M.~J., {Copperwheat}, C.~M., et~al., 2019.
\newblock {AT 2017fvz: a nova in the dwarf irregular galaxy NGC 6822}.
\newblock \mnras, 486, 4334-4347.

\bibitem[{{Henze} et~al.(2015{\natexlab{a}}){Henze}, {Darnley}, {Kabashima}
  et~al.}]{2015A&A...582L...8H}
{Henze}, M., {Darnley}, M.~J., {Kabashima}, F., et~al., 2015{\natexlab{a}}.
\newblock {A remarkable recurrent nova in M 31: The 2010 eruption recovered and
  evidence of a six-month period}.
\newblock \aap, 582, L8.

\bibitem[{{Henze} et~al.(2019){Henze}, {Darnley}, {Kafka} et~al.}]{HQ}
{Henze}, M., {Darnley}, M.~J., {Kafka}, S., et~al., 2019.
\newblock {An M31 recurrent nova seen in quiescence with Swift and the
  Liverpool Telescope --- the remarkable case of M31N 2008-12a}.
\newblock In preparation, for submission to A\&A.

\bibitem[{{Henze} et~al.(2018){Henze}, {Darnley}, {Williams}
  et~al.}]{2018ApJ...857...68H}
{Henze}, M., {Darnley}, M.~J., {Williams}, S.~C., et~al., 2018.
\newblock {Breaking the Habit: The Peculiar 2016 Eruption of the Unique
  Recurrent Nova M31N 2008-12a}.
\newblock \apj, 857, 68.

\bibitem[{{Henze} et~al.(2008){Henze}, {Meusinger}, \&
  {Pietsch}}]{2008A&A...477...67H}
{Henze}, M., {Meusinger}, H., {Pietsch}, W., 2008.
\newblock {A systematic search for novae in M 31 on a large set of digitized
  archival Schmidt plates}.
\newblock \aap, 477, 67-78.

\bibitem[{{Henze} et~al.(2014{\natexlab{a}}){Henze}, {Ness}, {Darnley}
  et~al.}]{2014A&A...563L...8H}
{Henze}, M., {Ness}, J.-U., {Darnley}, M.~J., et~al., 2014{\natexlab{a}}.
\newblock {A remarkable recurrent nova in M 31: The X-ray observations}.
\newblock \aap, 563, L8.

\bibitem[{{Henze} et~al.(2015{\natexlab{b}}){Henze}, {Ness}, {Darnley}
  et~al.}]{2015A&A...580A..46H}
{Henze}, M., {Ness}, J.-U., {Darnley}, M.~J., et~al., 2015{\natexlab{b}}.
\newblock {A remarkable recurrent nova in M 31: The predicted 2014 outburst in
  X-rays with Swift}.
\newblock \aap, 580, A46.

\bibitem[{{Henze} et~al.(2009{\natexlab{a}}){Henze}, {Pietsch}, {Haberl}
  et~al.}]{2009A&A...500..769H}
{Henze}, M., {Pietsch}, W., {Haberl}, F., et~al., 2009{\natexlab{a}}.
\newblock {The first two transient supersoft X-ray sources in M 31 globular
  clusters and the connection to classical novae}.
\newblock \aap, 500, 769-779.

\bibitem[{{Henze} et~al.(2010){Henze}, {Pietsch}, {Haberl}
  et~al.}]{2010A&A...523A..89H}
{Henze}, M., {Pietsch}, W., {Haberl}, F., et~al., 2010.
\newblock {X-ray monitoring of classical novae in the central region of M 31.
  I. June 2006-March 2007}.
\newblock \aap, 523, A89.

\bibitem[{{Henze} et~al.(2011){Henze}, {Pietsch}, {Haberl}
  et~al.}]{2011A&A...533A..52H}
{Henze}, M., {Pietsch}, W., {Haberl}, F., et~al., 2011.
\newblock {X-ray monitoring of classical novae in the central region of M 31.
  II. Autumn and winter 2007/2008 and 2008/2009}.
\newblock \aap, 533, A52.

\bibitem[{{Henze} et~al.(2013){Henze}, {Pietsch}, {Haberl}
  et~al.}]{2013A&A...549A.120H}
{Henze}, M., {Pietsch}, W., {Haberl}, F., et~al., 2013.
\newblock {Supersoft X-rays reveal a classical nova in the M 31 globular
  cluster Bol 126}.
\newblock \aap, 549, A120.

\bibitem[{{Henze} et~al.(2014{\natexlab{b}}){Henze}, {Pietsch}, {Haberl}
  et~al.}]{2014A&A...563A...2H}
{Henze}, M., {Pietsch}, W., {Haberl}, F., et~al., 2014{\natexlab{b}}.
\newblock {X-ray monitoring of classical novae in the central region of M 31
  III. Autumn and winter 2009/10, 2010/11, and 2011/12}.
\newblock \aap, 563, A2.

\bibitem[{{Henze} et~al.(2009{\natexlab{b}}){Henze}, {Pietsch}, {Sala}
  et~al.}]{2009A&A...498L..13H}
{Henze}, M., {Pietsch}, W., {Sala}, G., et~al., 2009{\natexlab{b}}.
\newblock {The very short supersoft X-ray state of the classical nova M31N
  2007-11a}.
\newblock \aap, 498, L13-L16.

\bibitem[{{Henze} et~al.(2016){Henze}, {Williams}, {Darnley}
  et~al.}]{2016ATel.9276....1H}
{Henze}, M., {Williams}, S.~C., {Darnley}, M.~J., et~al., 2016.
\newblock {Confirmation of PNV J00430400+4117079 as another eruption of the
  recurrent nova M31N 1990-10a and additional constraints on the eruption
  date}.
\newblock The Astronomer's Telegram, 9276.

\bibitem[{{Hernanz} \& {Jos{\'e}}(2008)}]{2008NewAR..52..386H}
{Hernanz}, M., {Jos{\'e}}, J., 2008.
\newblock {The recurrent nova RS Oph: A possible scenario for type Ia
  supernovae}.
\newblock \nar, 52, 386-389.

\bibitem[{{Hillebrandt} \& {Niemeyer}(2000)}]{2000ARA&A..38..191H}
{Hillebrandt}, W., {Niemeyer}, J.~C., 2000.
\newblock {Type IA Supernova Explosion Models}.
\newblock \araa, 38, 191-230.

\bibitem[{{Hillman} et~al.(2014){Hillman}, {Prialnik}, {Kovetz}
  et~al.}]{2014MNRAS.437.1962H}
{Hillman}, Y., {Prialnik}, D., {Kovetz}, A., et~al., 2014.
\newblock {Nova multiwavelength light curves: predicting UV precursor flashes
  and pre-maximum halts}.
\newblock \mnras, 437, 1962-1975.

\bibitem[{{Hillman} et~al.(2015){Hillman}, {Prialnik}, {Kovetz}
  et~al.}]{2015MNRAS.446.1924H}
{Hillman}, Y., {Prialnik}, D., {Kovetz}, A., et~al., 2015.
\newblock {Observational signatures of SNIa progenitors, as predicted by
  models}.
\newblock \mnras, 446, 1924-1930.

\bibitem[{{Hillman} et~al.(2016){Hillman}, {Prialnik}, {Kovetz}
  et~al.}]{2016ApJ...819..168H}
{Hillman}, Y., {Prialnik}, D., {Kovetz}, A., et~al., 2016.
\newblock {Growing White Dwarfs to the Chandrasekhar Limit: The Parameter Space
  of the Single Degenerate SNIa Channel}.
\newblock \apj, 819, 168.

\bibitem[{{Hornoch} \& {Shafter}(2015)}]{2015ATel.7116....1H}
{Hornoch}, K., {Shafter}, A.~W., 2015.
\newblock {M31N 2006-11c appears to be spatially coincident with PNV
  J00413317+4110124 and hence a recurrent nova in M31}.
\newblock The Astronomer's Telegram, 7116.

\bibitem[{{Hounsell} et~al.(2010){Hounsell}, {Bode}, {Hick}
  et~al.}]{2010ApJ...724..480H}
{Hounsell}, R., {Bode}, M.~F., {Hick}, P.~P., et~al., 2010.
\newblock {Exquisite Nova Light Curves from the Solar Mass Ejection Imager
  (SMEI)}.
\newblock \apj, 724, 480-486.

\bibitem[{{Hounsell} et~al.(2016){Hounsell}, {Darnley}, {Bode}
  et~al.}]{2016ApJ...820..104H}
{Hounsell}, R., {Darnley}, M.~J., {Bode}, M.~F., et~al., 2016.
\newblock {Nova Light Curves From The Solar Mass Ejection Imager (SMEI) - II.
  The extended catalog}.
\newblock \apj, 820, 104.

\bibitem[{{Hubble}(1929)}]{1929ApJ....69..103H}
{Hubble}, E.~P., 1929.
\newblock {A spiral nebula as a stellar system, Messier 31.}
\newblock \apj, 69, 103-158.

\bibitem[{{Humason}(1932)}]{1932PASP...44..381H}
{Humason}, M.~L., 1932.
\newblock {The Spectra of Two Novae in the Andromeda Nebula}.
\newblock \pasp, 44, 381-385.

\bibitem[{{Itagaki} et~al.(2016){Itagaki}, {Gao}, {Darnley}
  et~al.}]{2016ATel.9848....1I}
{Itagaki}, K., {Gao}, X., {Darnley}, M.~J., et~al., 2016.
\newblock {Recurrent Nova M31N 2008-12a: discovery and constraints of the 2016
  eruption}.
\newblock The Astronomer's Telegram, 9848.

\bibitem[{{Ivezi{\'c}} et~al.(2019){Ivezi{\'c}}, {Kahn}, {Tyson}
  et~al.}]{2019ApJ...873..111I}
{Ivezi{\'c}}, {\v{Z}}., {Kahn}, S.~M., {Tyson}, J.~A., et~al., 2019.
\newblock {LSST: From Science Drivers to Reference Design and Anticipated Data
  Products}.
\newblock \apj, 873, 2, 111.

\bibitem[{{Jos\'e}(2016)}]{Jos16}
{Jos\'e}, J., 2016.
\newblock {Stellar Explosions: Hydrodynamics and Nucleosynthesis}.
\newblock CRC/Taylor and Francis, Boca Raton, FL, USA.

\bibitem[{{Joy}(1954)}]{1954ApJ...120..377J}
{Joy}, A.~H., 1954.
\newblock {Spectroscopic Observations of AE Aquarii.}
\newblock \apj, 120, 377-383.

\bibitem[{{Kasliwal} et~al.(2017){Kasliwal}, {Bally}, {Masci}
  et~al.}]{2017ApJ...839...88K}
{Kasliwal}, M.~M., {Bally}, J., {Masci}, F., et~al., 2017.
\newblock {SPIRITS: Uncovering Unusual Infrared Transients with Spitzer}.
\newblock \apj, 839, 88.

\bibitem[{{Kasliwal} et~al.(2011){Kasliwal}, {Cenko}, {Kulkarni}
  et~al.}]{2011ApJ...735...94K}
{Kasliwal}, M.~M., {Cenko}, S.~B., {Kulkarni}, S.~R., et~al., 2011.
\newblock {Discovery of a New Photometric Sub-class of Faint and Fast Classical
  Novae}.
\newblock \apj, 735, 94.

\bibitem[{{Kato} et~al.(2013){Kato}, {Hachisu}, \&
  {Henze}}]{2013ApJ...779...19K}
{Kato}, M., {Hachisu}, I., {Henze}, M., 2013.
\newblock {Novae in Globular Clusters}.
\newblock \apj, 779, 19.

\bibitem[{{Kato} et~al.(2015){Kato}, {Saio}, \&
  {Hachisu}}]{2015ApJ...808...52K}
{Kato}, M., {Saio}, H., {Hachisu}, I., 2015.
\newblock {Multi-wavelength Light Curve Model of the One-year Recurrence Period
  Nova M31N 2008-12A}.
\newblock \apj, 808, 52.

\bibitem[{{Kato} et~al.(2017){Kato}, {Saio}, \&
  {Hachisu}}]{2017ApJ...844..143K}
{Kato}, M., {Saio}, H., {Hachisu}, I., 2017.
\newblock {A Millennium-long Evolution of the 1 yr Recurrence Period
  Nova---Search for Any Indication of the Forthcoming He Flash}.
\newblock \apj, 844, 143.

\bibitem[{{Kato} et~al.(2014){Kato}, {Saio}, {Hachisu}
  et~al.}]{2014ApJ...793..136K}
{Kato}, M., {Saio}, H., {Hachisu}, I., et~al., 2014.
\newblock {Shortest Recurrence Periods of Novae}.
\newblock \apj, 793, 136.

\bibitem[{{Kato} et~al.(2016){Kato}, {Saio}, {Henze}
  et~al.}]{2016ApJ...830...40K}
{Kato}, M., {Saio}, H., {Henze}, M., et~al., 2016.
\newblock {X-ray Flashes in Recurrent Novae: M31N 2008-12a and the Implications
  of the Swift Nondetection}.
\newblock \apj, 830, 40.

\bibitem[{{Kerins} et~al.(2010){Kerins}, {Darnley}, {Duke}
  et~al.}]{2010MNRAS.409..247K}
{Kerins}, E., {Darnley}, M.~J., {Duke}, J.~P., et~al., 2010.
\newblock {Difference image photometry with bright variable backgrounds}.
\newblock \mnras, 409, 247-258.

\bibitem[{{Korotkiy} \& {Elenin}(2011)}]{2011Kor}
{Korotkiy}, S., {Elenin}, L., 2011.
\newblock {CBAT}.
\newblock
  \url{http://www.cbat.eps.harvard.edu/unconf/followups/J00452885+4154094.html}.

\bibitem[{{Kraft}(1964)}]{1964ApJ...139..457K}
{Kraft}, R.~P., 1964.
\newblock {Binary Stars among Cataclysmic Variables. III. Ten Old Novae.}
\newblock \apj, 139, 457-475.

\bibitem[{{Krautter}(2002)}]{2002AIPC..637..345K}
{Krautter}, J., 2002.
\newblock {X-ray Observations of Novae}.
\newblock In M.~{Hernanz}, J.~{Jos{\'e}}, editors, Classical Nova Explosions,
  volume 637 of American Institute of Physics Conference Series, pages
  345--354.

\bibitem[{{Krautter}(2008)}]{2008ASPC..401..139K}
{Krautter}, J., 2008.
\newblock {The Super-soft Phase in Novae}.
\newblock In A.~{Evans}, M.~F. {Bode}, T.~J. {O'Brien}, M.~J. {Darnley},
  editors, RS Ophiuchi (2006) and the Recurrent Nova Phenomenon, volume 401 of
  Astronomical Society of the Pacific Conference Series, pages 139--149. San
  Francisco.

\bibitem[{{Krzeminski}(1977)}]{1977ivsw.conf..238K}
{Krzeminski}, W., 1977.
\newblock {AM Herculis Type Stars - Magnetic Binaries}.
\newblock In R.~{Kippenhahn}, J.~{Rahe}, W.~{Strohmeier}, editors, IAU Colloq.
  42: The Interaction of Variable Stars with their Environment, pages 238--241.

\bibitem[{{Kuin} et~al.(2019){Kuin}, {Page}, {Mr{\'o}z}
  et~al.}]{2019arXiv190903281K}
{Kuin}, N.~P.~M., {Page}, K.~L., {Mr{\'o}z}, P., et~al., 2019.
\newblock {The January 2016 eruption of recurrent nova LMC 1968}.
\newblock arXiv e-prints, arXiv:1909.03281.

\bibitem[{{Kurtz} \& {Lucas}(1977)}]{StarWars}
{Kurtz}, G., {Lucas}, G., 1977.
\newblock Star Wars:\ Episode IV -- A New Hope.
\newblock Twentieth Century Fox.

\bibitem[{{Law} et~al.(2009){Law}, {Kulkarni}, {Dekany}
  et~al.}]{2009PASP..121.1395L}
{Law}, N.~M., {Kulkarni}, S.~R., {Dekany}, R.~G., et~al., 2009.
\newblock {The Palomar Transient Factory: System Overview, Performance, and
  First Results}.
\newblock \pasp, 121, 1395-1408.

\bibitem[{{Lee} et~al.(2012){Lee}, {Riffeser}, {Seitz}
  et~al.}]{2012AA...537A..43L}
{Lee}, C.-H., {Riffeser}, A., {Seitz}, S., et~al., 2012.
\newblock {The Wendelstein Calar Alto Pixellensing Project (WeCAPP): the M 31
  nova catalogue}.
\newblock \aap, 537, A43.

\bibitem[{{Lindegren} et~al.(2018){Lindegren}, {Hern{\'a}ndez}, {Bombrun}
  et~al.}]{2018A&A...616A...2L}
{Lindegren}, L., {Hern{\'a}ndez}, J., {Bombrun}, A., et~al., 2018.
\newblock {Gaia Data Release 2. The astrometric solution}.
\newblock \aap, 616, A2.

\bibitem[{{Mason}(2013)}]{2013A&A...556C...2M}
{Mason}, E., 2013.
\newblock {U Scorpii 2010 outburst: observational evidence of an underlying
  ONeMg white dwarf (Corrigendum)}.
\newblock \aap, 556, C2.

\bibitem[{{Mason} et~al.(2004){Mason}, {Ederoclite}, {Stefanon}
  et~al.}]{2004IAUC.8424....2M}
{Mason}, E., {Ederoclite}, A., {Stefanon}, M., et~al., 2004.
\newblock {Nova in the Large Magellanic Cloud 2004}.
\newblock \iaucirc, 8424.

\bibitem[{{Massey} et~al.(2007){Massey}, {McNeill}, {Olsen}
  et~al.}]{2007AJ....134.2474M}
{Massey}, P., {McNeill}, R.~T., {Olsen}, K.~A.~G., et~al., 2007.
\newblock {A Survey of Local Group Galaxies Currently Forming Stars. III. A
  Search for Luminous Blue Variables and Other H{$\alpha$} Emission-Line
  Stars}.
\newblock \aj, 134, 2474-2503.

\bibitem[{{Massey} et~al.(2006){Massey}, {Olsen}, {Hodge}
  et~al.}]{2006AJ....131.2478M}
{Massey}, P., {Olsen}, K.~A.~G., {Hodge}, P.~W., et~al., 2006.
\newblock {A Survey of Local Group Galaxies Currently Forming Stars. I. UBVRI
  Photometry of Stars in M31 and M33}.
\newblock \aj, 131, 2478-2496.

\bibitem[{{Matthews} et~al.(2015){Matthews}, {Knigge}, {Long}
  et~al.}]{2015MNRAS.450.3331M}
{Matthews}, J.~H., {Knigge}, C., {Long}, K.~S., et~al., 2015.
\newblock {The impact of accretion disc winds on the optical spectra of
  cataclysmic variables}.
\newblock \mnras, 450, 3331-3344.

\bibitem[{{Mclaughlin}(1945)}]{1945PASP...57...69M}
{Mclaughlin}, D.~B., 1945.
\newblock {The Relation between Light-Curves and Luminosities of Novae}.
\newblock \pasp, 57, 69-80.

\bibitem[{{Mr{\'o}z} \& {Udalski}(2018)}]{2018ATel11384....1M}
{Mr{\'o}z}, P., {Udalski}, A., 2018.
\newblock {OGLE-2018-NOVA-01: a Classical (Recurrent) Nova Candidate in the
  Large Magellanic Cloud}.
\newblock The Astronomer's Telegram, 11384.

\bibitem[{{Mr{\'o}z} et~al.(2016){Mr{\'o}z}, {Udalski}, {Poleski}
  et~al.}]{2016ApJS..222....9M}
{Mr{\'o}z}, P., {Udalski}, A., {Poleski}, R., et~al., 2016.
\newblock {OGLE Atlas of Classical Novae. II. Magellanic Clouds}.
\newblock \apjs, 222, 9.

\bibitem[{{Munari} et~al.(2002){Munari}, {Henden}, {Kiyota}
  et~al.}]{2002A&A...389L..51M}
{Munari}, U., {Henden}, A., {Kiyota}, S., et~al., 2002.
\newblock {The mysterious eruption of V838 Mon}.
\newblock \aap, 389, L51-L56.

\bibitem[{{Neill} \& {Shara}(2004)}]{2004AJ....127..816N}
{Neill}, J.~D., {Shara}, M.~M., 2004.
\newblock {The H{$\alpha$} Light Curves and Spatial Distribution of Novae in
  M81}.
\newblock \aj, 127, 816-831.

\bibitem[{{Neill} \& {Shara}(2005)}]{2005AJ....129.1873N}
{Neill}, J.~D., {Shara}, M.~M., 2005.
\newblock {A Possible High Nova Rate for Two Local Group Dwarf Galaxies: M32
  and NGC 205}.
\newblock \aj, 129, 1873-1885.

\bibitem[{{Ness} et~al.(2015){Ness}, {Goranskij}, {Page}
  et~al.}]{2015ATel.8053....1N}
{Ness}, J.-U., {Goranskij}, V.~P., {Page}, K.~L., et~al., 2015.
\newblock {Nova V723 Cas off in X-rays}.
\newblock The Astronomer's Telegram, 8053.

\bibitem[{{Ness} et~al.(2013){Ness}, {Osborne}, {Henze}
  et~al.}]{2013A&A...559A..50N}
{Ness}, J.-U., {Osborne}, J.~P., {Henze}, M., et~al., 2013.
\newblock {Obscuration effects in super-soft-source X-ray spectra}.
\newblock \aap, 559, A50.

\bibitem[{{Ness} et~al.(2008){Ness}, {Schwarz}, {Starrfield}
  et~al.}]{2008AJ....135.1328N}
{Ness}, J.-U., {Schwarz}, G., {Starrfield}, S., et~al., 2008.
\newblock {V723 CASSIOPEIA Still on in X-Rays a Bright Super Soft Source 12
  Years after Outburst}.
\newblock \aj, 135, 1328-1333.

\bibitem[{{Nishiyama} \& {Kabashima}(2008)}]{2008Nis}
{Nishiyama}, K., {Kabashima}, F., 2008.
\newblock {CBAT}.
\newblock \url{http://www.cbat.eps.harvard.edu/iau/CBAT\_M31.html\#2008-12a}.

\bibitem[{{Nishiyama} \& {Kabashima}(2012)}]{2012Nis}
{Nishiyama}, K., {Kabashima}, F., 2012.
\newblock {CBAT}.
\newblock
  \url{http://www.cbat.eps.harvard.edu/unconf/followups/J00452884+4154095.html}.

\bibitem[{{Orio}(2013)}]{2013AstRv...8a..71O}
{Orio}, M., 2013.
\newblock {Hydrogen burning on massive binary white dwarfs: what we know from
  the observations}.
\newblock The Astronomical Review, 8, 1, 71-89.

\bibitem[{{Osborne}(2015)}]{2015JHEAp...7..117O}
{Osborne}, J.~P., 2015.
\newblock {Getting to know classical novae with Swift}.
\newblock Journal of High Energy Astrophysics, 7, 117-125.

\bibitem[{{{\"O}zd{\"o}nmez} et~al.(2018){{\"O}zd{\"o}nmez}, {Ege}, {G{\"u}ver}
  et~al.}]{2018MNRAS.476.4162O}
{{\"O}zd{\"o}nmez}, A., {Ege}, E., {G{\"u}ver}, T., et~al., 2018.
\newblock {A new catalogue of Galactic novae: investigation of the MMRD
  relation and spatial distribution}.
\newblock \mnras, 476, 4162-4186.

\bibitem[{{Page} et~al.(2015){Page}, {Osborne}, {Kuin}
  et~al.}]{2015MNRAS.454.3108P}
{Page}, K.~L., {Osborne}, J.~P., {Kuin}, N.~P.~M., et~al., 2015.
\newblock {Swift detection of the super-swift switch-on of the super-soft phase
  in nova V745 Sco (2014)}.
\newblock \mnras, 454, 3108-3120.

\bibitem[{{Pagnotta} \& {Schaefer}(2014)}]{2014ApJ...788..164P}
{Pagnotta}, A., {Schaefer}, B.~E., 2014.
\newblock {Identifying and Quantifying Recurrent Novae Masquerading as
  Classical Novae}.
\newblock \apj, 788, 164.

\bibitem[{{Pagnotta} et~al.(2009){Pagnotta}, {Schaefer}, {Xiao}
  et~al.}]{2009AJ....138.1230P}
{Pagnotta}, A., {Schaefer}, B.~E., {Xiao}, L., et~al., 2009.
\newblock {Discovery of a Second Nova Eruption of V2487 Ophiuchi}.
\newblock \aj, 138, 1230-1234.

\bibitem[{{Payne-Gaposchkin}(1964)}]{1964gano.book.....P}
{Payne-Gaposchkin}, C., 1964.
\newblock {The galactic novae}.
\newblock Dover Publication, New York.

\bibitem[{{Pfau}(1976)}]{1976A&A....50..113P}
{Pfau}, W., 1976.
\newblock {Recalibration of the absolute magnitudes of novae and application to
  nova Cygni 1975.}
\newblock \aap, 50, 113-115.

\bibitem[{{Piascik} et~al.(2014){Piascik}, {Steele}, {Bates}
  et~al.}]{2014SPIE.9147E..8HP}
{Piascik}, A.~S., {Steele}, I.~A., {Bates}, S.~D., et~al., 2014.
\newblock {SPRAT: Spectrograph for the Rapid Acquisition of Transients}.
\newblock In Ground-based and Airborne Instrumentation for Astronomy V, volume
  9147 of \procspie, page 91478H.

\bibitem[{{Pietsch}(2010)}]{2010AN....331..187P}
{Pietsch}, W., 2010.
\newblock {X-ray emission from optical novae in M 31}.
\newblock Astronomische Nachrichten, 331, 187-192.

\bibitem[{{Pietsch} et~al.(2005){Pietsch}, {Fliri}, {Freyberg}
  et~al.}]{2005A&A...442..879P}
{Pietsch}, W., {Fliri}, J., {Freyberg}, M.~J., et~al., 2005.
\newblock {Optical novae: the major class of supersoft X-ray sources in M 31}.
\newblock \aap, 442, 879-894.

\bibitem[{{Pietsch} et~al.(2007){Pietsch}, {Haberl}, {Sala}
  et~al.}]{2007A&A...465..375P}
{Pietsch}, W., {Haberl}, F., {Sala}, G., et~al., 2007.
\newblock {X-ray monitoring of optical novae in M 31 from July 2004 to February
  2005}.
\newblock \aap, 465, 375-392.

\bibitem[{{Pietsch} et~al.(2011){Pietsch}, {Henze}, {Haberl}
  et~al.}]{2011A&A...531A..22P}
{Pietsch}, W., {Henze}, M., {Haberl}, F., et~al., 2011.
\newblock {Nova M31N 2007-12b: supersoft X-rays reveal an intermediate polar?}
\newblock \aap, 531, A22.

\bibitem[{{Pottasch}(1967)}]{1967BAN....19..227P}
{Pottasch}, S.~R., 1967.
\newblock {An interpretation of the spectrum of the recurrent nova RS
  Ophiuchi}.
\newblock \bain, 19, 227-238.

\bibitem[{{Prialnik} \& {Kovetz}(1995)}]{1995ApJ...445..789P}
{Prialnik}, D., {Kovetz}, A., 1995.
\newblock {An extended grid of multicycle nova evolution models}.
\newblock \apj, 445, 789-810.

\bibitem[{{Prialnik} et~al.(1978){Prialnik}, {Shara}, \&
  {Shaviv}}]{1978A&A....62..339P}
{Prialnik}, D., {Shara}, M.~M., {Shaviv}, G., 1978.
\newblock {The evolution of a slow nova model with a Z = 0.03 envelope from
  pre-explosion to extinction}.
\newblock \aap, 62, 339-348.

\bibitem[{{Ransome} et~al.(2019){Ransome}, {Darnley}, {Habergham-Mawson}
  et~al.}]{Ransome2019}
{Ransome}, C., {Darnley}, M.~J., {Habergham-Mawson}, S.~M., et~al., 2019.
\newblock {A Spectroscopic and Photometric Survey of Novae in M31 II}.
\newblock In preparation, for submission to MNRAS.

\bibitem[{{Rector} et~al.(1999){Rector}, {Jacoby}, {Corbett}
  et~al.}]{1999AAS...195.3608R}
{Rector}, T.~A., {Jacoby}, G.~H., {Corbett}, D.~L., et~al., 1999.
\newblock {A Search for Novae in the Bulge of M31}.
\newblock In American Astronomical Society Meeting Abstracts, volume~31 of
  Bulletin of the American Astronomical Society, page 1420.

\bibitem[{{Rich} et~al.(1989){Rich}, {Mould}, {Picard}
  et~al.}]{1989ApJ...341L..51R}
{Rich}, R.~M., {Mould}, J., {Picard}, A., et~al., 1989.
\newblock {Luminous M giants in the bulge of M31}.
\newblock \apjl, 341, L51-L54.

\bibitem[{{Ritchey}(1901{\natexlab{a}})}]{1901ApJ....14..293R}
{Ritchey}, G.~W., 1901{\natexlab{a}}.
\newblock {Changes in the nebulosity about Nova Persei.}
\newblock \apj, 14, 293-294.

\bibitem[{{Ritchey}(1901{\natexlab{b}})}]{1901ApJ....14..167R}
{Ritchey}, G.~W., 1901{\natexlab{b}}.
\newblock {Nebulosity about Nova Persei.}
\newblock \apj, 14, 167-168.

\bibitem[{{Ritchey}(1917)}]{1917PASP...29..210R}
{Ritchey}, G.~W., 1917.
\newblock {Novae in Spiral Nebulae}.
\newblock \pasp, 29, 210-212.

\bibitem[{{Rosino}(1964)}]{1964AnAp...27..498R}
{Rosino}, L., 1964.
\newblock {Novae in Messier 31 discovered and observed at Asiago from 1955 to
  1963.}
\newblock Annales d'Astrophysique, 27, 498-505.

\bibitem[{{Rosino}(1973)}]{1973A&AS....9..347R}
{Rosino}, L., 1973.
\newblock {Novae in M 31 discovered and observed at Asiago from 1963 to 1970}.
\newblock \aaps, 9, 347-389.

\bibitem[{{Rosino} et~al.(1989){Rosino}, {Capaccioli}, {D'Onofrio}
  et~al.}]{1989AJ.....97...83R}
{Rosino}, L., {Capaccioli}, M., {D'Onofrio}, M., et~al., 1989.
\newblock {Fifty-two novae in M31 discovered and observed at Asiago from 1971
  to 1986}.
\newblock \aj, 97, 83-96.

\bibitem[{{Ruan} \& {Gao}(2010)}]{2010CBET.2574....2R}
{Ruan}, J., {Gao}, X., 2010.
\newblock {Apparent nova in M31: M31N 2010-12a.}
\newblock Central Bureau Electronic Telegrams, 2574, 2.

\bibitem[{{Sanford}(1949)}]{1949ApJ...109...81S}
{Sanford}, R.~F., 1949.
\newblock {High-Dispersion Spectrograms of T Coronae Borealis.}
\newblock \apj, 109, 81-91.

\bibitem[{{Schaefer}(2010)}]{2010ApJS..187..275S}
{Schaefer}, B.~E., 2010.
\newblock {Comprehensive Photometric Histories of All Known Galactic Recurrent
  Novae}.
\newblock \apjs, 187, 275-373.

\bibitem[{{Schaefer}(2014)}]{2014AAS...22430604S}
{Schaefer}, B.~E., 2014.
\newblock {Nova Discovery Efficiency 1890-2014; Only $43\%\pm6\%$ of the
  Brightest Nova Are Discovered}.
\newblock In American Astronomical Society Meeting Abstracts \#224, volume 224
  of American Astronomical Society Meeting Abstracts, page 306.04.

\bibitem[{{Schaefer}(2018)}]{2018MNRAS.481.3033S}
{Schaefer}, B.~E., 2018.
\newblock {The distances to Novae as seen by Gaia}.
\newblock \mnras, 481, 3033-3051.

\bibitem[{{Schaefer} et~al.(2010){Schaefer}, {Pagnotta}, {Xiao}
  et~al.}]{2010AJ....140..925S}
{Schaefer}, B.~E., {Pagnotta}, A., {Xiao}, L., et~al., 2010.
\newblock {Discovery of the Predicted 2010 Eruption and the Pre-eruption Light
  Curve for Recurrent Nova U Scorpii}.
\newblock \aj, 140, 925-932.

\bibitem[{{Schatzman}(1949)}]{1949AnAp...12..281S}
{Schatzman}, E., 1949.
\newblock {Remarques sur le ph{\'e}nom{\`e}ne de novae (Notes on the phenomenon
  of novae)}.
\newblock Annales d'Astrophysique, 12, 281-286.

\bibitem[{{Schatzman}(1951)}]{1951AnAp...14..294S}
{Schatzman}, E., 1951.
\newblock {Remarques sur le ph{\'e}nom{\`e}ne de Nova: IV. L'onde de
  d{\'e}tonation due {\`a} l'isotope $^{3}$He (Notes on the phenomenon of Nova:
  IV. The detonation wave due to the $^3$He isotope)}.
\newblock Annales d'Astrophysique, 14, 294-304.

\bibitem[{{Schmeer}(2017)}]{Sch2017}
{Schmeer}, P., 2017.
\newblock {CBAT}.
\newblock
  \url{http://www.cbat.eps.harvard.edu/unconf/followups/J00432946+4117137.html}.

\bibitem[{{Schmidt}(1957)}]{1957ZA.....41..182S}
{Schmidt}, T., 1957.
\newblock {Die Lichtkurven-Leuchtkraft-Beziehung Neuer Sterne. Mit 8
  Textabbildungen (The nova light curve-luminosity relationship. With 8
  figures)}.
\newblock \zap, 41, 182-201.

\bibitem[{{Schwarz} et~al.(2011){Schwarz}, {Ness}, {Osborne}
  et~al.}]{2011ApJS..197...31S}
{Schwarz}, G.~J., {Ness}, J.-U., {Osborne}, J.~P., et~al., 2011.
\newblock {Swift X-Ray Observations of Classical Novae. II. The Super Soft
  Source Sample}.
\newblock \apjs, 197, 31.

\bibitem[{{Selvelli} \& {Gilmozzi}(2019)}]{2019A&A...622A.186S}
{Selvelli}, P., {Gilmozzi}, R., 2019.
\newblock {A UV and optical study of 18 old novae with Gaia DR2 distances: mass
  accretion rates, physical parameters, and MMRD}.
\newblock \aap, 622, A186.

\bibitem[{{Shafter}(2007)}]{2007AAS...211.5115S}
{Shafter}, A.~W., 2007.
\newblock {Spectroscopic Classes of Galactic Novae}.
\newblock In American Astronomical Society Meeting Abstracts, volume~39 of
  Bulletin of the American Astronomical Society, page 817.

\bibitem[{{Shafter}(2013)}]{2013AJ....145..117S}
{Shafter}, A.~W., 2013.
\newblock {Photometric and Spectroscopic Properties of Novae in the Large
  Magellanic Cloud}.
\newblock \aj, 145, 117.

\bibitem[{{Shafter}(2017)}]{2017ApJ...834..196S}
{Shafter}, A.~W., 2017.
\newblock {The Galactic Nova Rate Revisited}.
\newblock \apj, 834, 196.

\bibitem[{{Shafter} et~al.(2011{\natexlab{a}}){Shafter}, {Bode}, {Darnley}
  et~al.}]{2011ApJ...727...50S}
{Shafter}, A.~W., {Bode}, M.~F., {Darnley}, M.~J., et~al., 2011{\natexlab{a}}.
\newblock {A Spitzer Survey of Novae in M31}.
\newblock \apj, 727, 50.

\bibitem[{{Shafter} et~al.(2000){Shafter}, {Ciardullo}, \&
  {Pritchet}}]{2000ApJ...530..193S}
{Shafter}, A.~W., {Ciardullo}, R., {Pritchet}, C.~J., 2000.
\newblock {Novae in External Galaxies: M51, M87, and M101}.
\newblock \apj, 530, 193-206.

\bibitem[{{Shafter} et~al.(2014){Shafter}, {Curtin}, {Pritchet}
  et~al.}]{2014ASPC..490...77S}
{Shafter}, A.~W., {Curtin}, C., {Pritchet}, C.~J., et~al., 2014.
\newblock {Extragalactic Nova Populations}.
\newblock In P.~A. {Woudt}, V.~A.~R.~M. {Ribeiro}, editors, Stellar Novae: Past
  and Future Decades, volume 490 of Astronomical Society of the Pacific
  Conference Series, pages 77--84. San Francisco.

\bibitem[{{Shafter} et~al.(2012{\natexlab{a}}){Shafter}, {Darnley}, {Bode}
  et~al.}]{2012ApJ...752..156S}
{Shafter}, A.~W., {Darnley}, M.~J., {Bode}, M.~F., et~al., 2012{\natexlab{a}}.
\newblock {On the Spectroscopic Classes of Novae in M33}.
\newblock \apj, 752, 156.

\bibitem[{{Shafter} et~al.(2011{\natexlab{b}}){Shafter}, {Darnley}, {Hornoch}
  et~al.}]{2011ApJ...734...12S}
{Shafter}, A.~W., {Darnley}, M.~J., {Hornoch}, K., et~al., 2011{\natexlab{b}}.
\newblock {A Spectroscopic and Photometric Survey of Novae in M31}.
\newblock \apj, 734, 12.

\bibitem[{{Shafter} et~al.(2017{\natexlab{a}}){Shafter}, {Henze}, {Darnley}
  et~al.}]{2017RNAAS...1a..44S}
{Shafter}, A.~W., {Henze}, M., {Darnley}, M.~J., et~al., 2017{\natexlab{a}}.
\newblock {The Recurrent Nova Candidate M31N 1966-08a = 1968-10c is a Galactic
  Flare Star}.
\newblock Research Notes of the American Astronomical Society, 1, 1, 44.

\bibitem[{{Shafter} et~al.(2015){Shafter}, {Henze}, {Rector}
  et~al.}]{2015ApJS..216...34S}
{Shafter}, A.~W., {Henze}, M., {Rector}, T.~A., et~al., 2015.
\newblock {Recurrent Novae in M31}.
\newblock \apjs, 216, 34.

\bibitem[{{Shafter} et~al.(2012{\natexlab{b}}){Shafter}, {Hornoch}, {Ciardullo}
  et~al.}]{2012ATel.4503....1S}
{Shafter}, A.~W., {Hornoch}, K., {Ciardullo}, J.~V.~R., et~al.,
  2012{\natexlab{b}}.
\newblock {Spectroscopic Observations of the Unusual Transient TCP
  J00452884+4154095 in M31}.
\newblock The Astronomer's Telegram, 4503, 1.

\bibitem[{{Shafter} \& {Irby}(2001)}]{2001ApJ...563..749S}
{Shafter}, A.~W., {Irby}, B.~K., 2001.
\newblock {On the Spatial Distribution, Stellar Population, and Rate of Novae
  in M31}.
\newblock \apj, 563, 749-767.

\bibitem[{{Shafter} et~al.(2017{\natexlab{b}}){Shafter}, {Kundu}, \&
  {Henze}}]{2017RNAAS...1a..11S}
{Shafter}, A.~W., {Kundu}, A., {Henze}, M., 2017{\natexlab{b}}.
\newblock {On the Nova Rate in M87}.
\newblock Research Notes of the American Astronomical Society, 1, 1, 11.

\bibitem[{{Shafter} \& {Quimby}(2007)}]{2007ApJ...671L.121S}
{Shafter}, A.~W., {Quimby}, R.~M., 2007.
\newblock {M31N-2007-06b: A Nova in the M31 Globular Cluster Bol 111}.
\newblock \apjl, 671, L121-L124.

\bibitem[{{Shafter} et~al.(2009){Shafter}, {Rau}, {Quimby}
  et~al.}]{2009ApJ...690.1148S}
{Shafter}, A.~W., {Rau}, A., {Quimby}, R.~M., et~al., 2009.
\newblock {M31N 2007-11d: A Slowly Rising, Luminous Nova in M31}.
\newblock \apj, 690, 1148-1157.

\bibitem[{{Shapley} \& {Curtis}(1921)}]{1921BuNRC...2..171S}
{Shapley}, H., {Curtis}, H.~D., 1921.
\newblock {The Scale of the Universe}.
\newblock Bulletin of the National Research Council, Vol.~2, Part 3, No.~11,
  p.~171-217, 2, 171-217.

\bibitem[{{Shara} et~al.(2017){Shara}, {Doyle}, {Lauer}
  et~al.}]{2017ApJ...839..109S}
{Shara}, M.~M., {Doyle}, T., {Lauer}, T.~R., et~al., 2017.
\newblock {A Hubble Space Telescope Survey for Novae in M87. II. Snuffing out
  the Maximum Magnitude-Rate of Decline Relation for Novae as a Non-standard
  Candle, and a Prediction of the Existence of Ultrafast Novae}.
\newblock \apj, 839, 109.

\bibitem[{{Shara} et~al.(2016){Shara}, {Doyle}, {Lauer}
  et~al.}]{2016ApJS..227....1S}
{Shara}, M.~M., {Doyle}, T.~F., {Lauer}, T.~R., et~al., 2016.
\newblock {A Hubble Space Telescope Survey for Novae in M87. I. Light and Color
  Curves, Spatial Distributions, and the Nova Rate}.
\newblock \apjs, 227, 1.

\bibitem[{{Shara} et~al.(2007){Shara}, {Martin}, {Seibert}
  et~al.}]{2007Natur.446..159S}
{Shara}, M.~M., {Martin}, C.~D., {Seibert}, M., et~al., 2007.
\newblock {An ancient nova shell around the dwarf nova Z Camelopardalis}.
\newblock \nat, 446, 159-162.

\bibitem[{{Shara} et~al.(2012){Shara}, {Mizusawa}, {Wehinger}
  et~al.}]{2012ApJ...758..121S}
{Shara}, M.~M., {Mizusawa}, T., {Wehinger}, P., et~al., 2012.
\newblock {AT Cnc: A Second Dwarf Nova with a Classical Nova Shell}.
\newblock \apj, 758, 121.

\bibitem[{{Shara} et~al.(2010){Shara}, {Yaron}, {Prialnik}
  et~al.}]{2010ApJ...725..831S}
{Shara}, M.~M., {Yaron}, O., {Prialnik}, D., et~al., 2010.
\newblock {An Extended Grid of Nova Models. III. Very Luminous, Red Novae}.
\newblock \apj, 725, 831-841.

\bibitem[{{Shara} \& {Zurek}(2002)}]{2002AIPC..637..457S}
{Shara}, M.~M., {Zurek}, D.~R., 2002.
\newblock {400 Novae in M87}.
\newblock In M.~{Hernanz}, J.~{Jos{\'e}}, editors, Classical Nova Explosions,
  volume 637 of American Institute of Physics Conference Series, pages
  457--461.

\bibitem[{{Shara} et~al.(1997){Shara}, {Zurek}, {Williams}
  et~al.}]{1997AJ....114..258S}
{Shara}, M.~M., {Zurek}, D.~R., {Williams}, R.~E., et~al., 1997.
\newblock {HST Imagery of the Non-Expanding, Clumped ``Shell'' of the Recurrent
  Nova T Pyxidis}.
\newblock \aj, 114, 258-264.

\bibitem[{{Sharov}(1972)}]{1972SvA....16...41S}
{Sharov}, A.~S., 1972.
\newblock {Estimate for the Frequency of Novae in the Andromeda Nebula and our
  Galaxy.}
\newblock \sovast, 16, 41-44.

\bibitem[{{Sharov} \& {Alksnis}(1991)}]{1991ApSS.180..273S}
{Sharov}, A.~S., {Alksnis}, A., 1991.
\newblock {Novae in M31 discovered with wide field telescopes in Crimea and
  Latvia}.
\newblock \apss, 180, 273-286.

\bibitem[{{Sharov} \& {Alksnis}(1992)}]{1992ApSS.190..119S}
{Sharov}, A.~S., {Alksnis}, A., 1992.
\newblock {Novae in M31 discovered with wide-field telescopes in Crimea and
  Latvia - The maximum magnitude versus rate of decline relation for Novae in
  M31}.
\newblock \apss, 190, 119-130.

\bibitem[{{Sharov} \& {Alksnis}(1989)}]{1989SvAL...15..382S}
{Sharov}, A.~S., {Alksnis}, A.~K., 1989.
\newblock {Putative Novae in M31}.
\newblock Soviet Astronomy Letters, 15, 382-384.

\bibitem[{{Shore}(2012)}]{2012BASI...40..185S}
{Shore}, S.~N., 2012.
\newblock {Spectroscopy of novae -- a user's manual}.
\newblock Bulletin of the Astronomical Society of India, 40, 185-212.

\bibitem[{{Shore} et~al.(1991){Shore}, {Sonneborn}, {Starrfield}
  et~al.}]{1991ApJ...370..193S}
{Shore}, S.~N., {Sonneborn}, G., {Starrfield}, S.~G., et~al., 1991.
\newblock {Multiwavelength observations of Nova LMC 1990 Number 2 - The first
  extragalactic recurrent nova}.
\newblock \apj, 370, 193-197.

\bibitem[{{Sin} et~al.(2017){Sin}, {Henze}, {Sala}
  et~al.}]{2017ATel10001....1S}
{Sin}, P., {Henze}, M., {Sala}, G., et~al., 2017.
\newblock {Additional Photometry for nova M31N 2016-12e and classification as a
  recurrent nova (= M31N 2007-11f)}.
\newblock ATel, No.~10001, 1.

\bibitem[{{Skrutskie} et~al.(2006){Skrutskie}, {Cutri}, {Stiening}
  et~al.}]{2006AJ....131.1163S}
{Skrutskie}, M.~F., {Cutri}, R.~M., {Stiening}, R., et~al., 2006.
\newblock {The Two Micron All Sky Survey (2MASS)}.
\newblock \aj, 131, 1163-1183.

\bibitem[{{Slavin} et~al.(1995){Slavin}, {O'Brien}, \&
  {Dunlop}}]{1995MNRAS.276..353S}
{Slavin}, A.~J., {O'Brien}, T.~J., {Dunlop}, J.~S., 1995.
\newblock {A deep optical imaging study of the nebular remnants of classical
  novae}.
\newblock \mnras, 276, 353-371.

\bibitem[{{Soraisam} et~al.(2019){Soraisam}, {Lee}, {Narayan}
  et~al.}]{2019ATel12943....1S}
{Soraisam}, M., {Lee}, C.-H., {Narayan}, G., et~al., 2019.
\newblock {Confirmation of the 2019 nova outburst from RN M31N
  1960-12a/2013-05b}.
\newblock The Astronomer's Telegram, 12943.

\bibitem[{{Soraisam} \& {Gilfanov}(2015)}]{2015A&A...583A.140S}
{Soraisam}, M.~D., {Gilfanov}, M., 2015.
\newblock {Constraining the role of novae as progenitors of type Ia
  supernovae}.
\newblock \aap, 583, A140.

\bibitem[{{Soraisam} et~al.(2016){Soraisam}, {Gilfanov}, {Wolf}
  et~al.}]{2016MNRAS.455..668S}
{Soraisam}, M.~D., {Gilfanov}, M., {Wolf}, W.~M., et~al., 2016.
\newblock {Population of post-nova supersoft X-ray sources}.
\newblock \mnras, 455, 668-679.

\bibitem[{{Stark} et~al.(1992){Stark}, {Gammie}, {Wilson}
  et~al.}]{1992ApJS...79...77S}
{Stark}, A.~A., {Gammie}, C.~F., {Wilson}, R.~W., et~al., 1992.
\newblock {The Bell Laboratories H I survey}.
\newblock \apjs, 79, 77-104.

\bibitem[{Starrfield(2014)}]{doi:10.1063/1.4866984}
Starrfield, S., 2014.
\newblock The accretion of solar material onto white dwarfs: No mixing with
  core material implies that the mass of the white dwarf is increasing.
\newblock AIP Advances, 4, 4, 041007.

\bibitem[{{Starrfield} et~al.(2019){Starrfield}, {Bose}, {Iliadis}
  et~al.}]{Sum19}
{Starrfield}, S., {Bose}, M., {Iliadis}, C., et~al., 2019.
\newblock {Carbon-Oxygen Classical Novae are Galactic $^{7}$Li Producers and
  Supernova Ia Progenitors}.
\newblock Submitted, for publication in ApJ.

\bibitem[{{Starrfield} et~al.(2008){Starrfield}, {Iliadis}, \& {Hix}}]{Sta08}
{Starrfield}, S., {Iliadis}, C., {Hix}, W.~R., 2008.
\newblock {Thermonuclear processes}.
\newblock In M.~F. {Bode}, A.~{Evans}, editors, {Classical Novae, 2nd Edition},
  pages 77--101. Cambridge University Press, Cambridge.

\bibitem[{{Starrfield} et~al.(2016){Starrfield}, {Iliadis}, \&
  {Hix}}]{2016PASP..128e1001S}
{Starrfield}, S., {Iliadis}, C., {Hix}, W.~R., 2016.
\newblock {The Thermonuclear Runaway and the Classical Nova Outburst}.
\newblock \pasp, 128, 5, 051001.

\bibitem[{{Starrfield} et~al.(2012){Starrfield}, {Iliadis}, {Timmes}
  et~al.}]{2012BASI...40..419S}
{Starrfield}, S., {Iliadis}, C., {Timmes}, F.~X., et~al., 2012.
\newblock {Theoretical studies of accretion of matter onto white dwarfs and the
  single degenerate scenario for supernovae of Type Ia}.
\newblock Bulletin of the Astronomical Society of India, 40, 419-442.

\bibitem[{{Starrfield} et~al.(1976){Starrfield}, {Sparks}, \&
  {Truran}}]{1976IAUS...73..155S}
{Starrfield}, S., {Sparks}, W.~M., {Truran}, J.~W., 1976.
\newblock {The cause of the nova outburst}.
\newblock In P.~{Eggleton}, S.~{Mitton}, J.~{Whelan}, editors, Structure and
  Evolution of Close Binary Systems, volume~73 of IAU Symposium, pages
  155--172.

\bibitem[{{Starrfield} et~al.(1972){Starrfield}, {Truran}, {Sparks}
  et~al.}]{1972ApJ...176..169S}
{Starrfield}, S., {Truran}, J.~W., {Sparks}, W.~M., et~al., 1972.
\newblock {CNO Abundances and Hydrodynamic Models of the Nova Outburst}.
\newblock \apj, 176, 169-176.

\bibitem[{{Starrfield} et~al.(1990){Starrfield}, {Truran}, {Sparks}
  et~al.}]{1990LNP...369..306S}
{Starrfield}, S., {Truran}, J.~W., {Sparks}, W.~M., et~al., 1990.
\newblock {Soft X-Ray Emission from Classical Novae in Outburst}.
\newblock In A.~{Cassatella}, R.~{Viotti}, editors, IAU Colloq. 122: Physics of
  Classical Novae, volume 369 of Lecture Notes in Physics, Berlin Springer
  Verlag, pages 306--312.

\bibitem[{{Steele} et~al.(2004){Steele}, {Smith}, {Rees}
  et~al.}]{2004SPIE.5489..679S}
{Steele}, I.~A., {Smith}, R.~J., {Rees}, P.~C., et~al., 2004.
\newblock {The Liverpool Telescope: performance and first results}.
\newblock In J.~M. {Oschmann}, Jr., editor, Ground-based Telescopes, volume
  5489 of Society of Photo-Optical Instrumentation Engineers (SPIE) Conference
  Series, pages 679--692.

\bibitem[{{Strope} et~al.(2010){Strope}, {Schaefer}, \&
  {Henden}}]{2010AJ....140...34S}
{Strope}, R.~J., {Schaefer}, B.~E., {Henden}, A.~A., 2010.
\newblock {Catalog of 93 Nova Light Curves: Classification and Properties}.
\newblock \aj, 140, 34-62.

\bibitem[{{Tang} et~al.(2014){Tang}, {Bildsten}, {Wolf}
  et~al.}]{2014ApJ...786...61T}
{Tang}, S., {Bildsten}, L., {Wolf}, W.~M., et~al., 2014.
\newblock {An Accreting White Dwarf near the Chandrasekhar Limit in the
  Andromeda Galaxy}.
\newblock \apj, 786, 61.

\bibitem[{{Tang} et~al.(2013){Tang}, {Cao}, \&
  {Kasliwal}}]{2013ATel.5607....1T}
{Tang}, S., {Cao}, Y., {Kasliwal}, M.~M., 2013.
\newblock {A Candidate Recurrent Nova in M31 from iPTF}.
\newblock The Astronomer's Telegram, 5607, 1.

\bibitem[{{Tomaney} \& {Shafter}(1992)}]{1992ApJS...81..683T}
{Tomaney}, A.~B., {Shafter}, A.~W., 1992.
\newblock {The spectroscopic and photometric evolution of novae in the bulge of
  M31}.
\newblock \apjs, 81, 683-714.

\bibitem[{{Toraskar} et~al.(2013){Toraskar}, {Mac Low}, {Shara}
  et~al.}]{2013ApJ...768...48T}
{Toraskar}, J., {Mac Low}, M.-M., {Shara}, M.~M., et~al., 2013.
\newblock {Dynamical Fragmentation of the T Pyxidis Nova Shell During Recurrent
  Eruptions}.
\newblock \apj, 768, 48.

\bibitem[{{Tylenda} et~al.(2011){Tylenda}, {Hajduk}, {Kami{\'n}ski}
  et~al.}]{2011A&A...528A.114T}
{Tylenda}, R., {Hajduk}, M., {Kami{\'n}ski}, T., et~al., 2011.
\newblock {V1309 Scorpii: merger of a contact binary}.
\newblock \aap, 528, A114.

\bibitem[{{Valcheva} et~al.(2019){Valcheva}, {Kostov}, {Minev}
  et~al.}]{2019ATel12915....1V}
{Valcheva}, A., {Kostov}, A., {Minev}, M., et~al., 2019.
\newblock {Rebrightening of the very fast RN M31N 1960-12a}.
\newblock The Astronomer's Telegram, 12915.

\bibitem[{{Vaytet} et~al.(2007){Vaytet}, {O'Brien}, \&
  {Bode}}]{2007ApJ...665..654V}
{Vaytet}, N.~M.~H., {O'Brien}, T.~J., {Bode}, M.~F., 2007.
\newblock {Swift Observations of the 2006 Outburst of the Recurrent Nova RS
  Ophiuchi. II. One-dimensional Hydrodynamical Models of Wind-driven Shocks}.
\newblock \apj, 665, 654-662.

\bibitem[{{Wade}(1990)}]{1990LNP...369..179W}
{Wade}, R.~A., 1990.
\newblock {Optical Imagery of Nova Remnants}.
\newblock In A.~{Cassatella}, R.~{Viotti}, editors, IAU Colloq. 122: Physics of
  Classical Novae, volume 369 of Lecture Notes in Physics, Berlin Springer
  Verlag, pages 179--187.

\bibitem[{{Ward}(1885)}]{1885AReg...23..242W}
{Ward}, I.~W., 1885.
\newblock {Correspondence - New Star in Andromeda.}
\newblock Astronomical register, 23, 242-242.

\bibitem[{{Warner}(1983)}]{1983ASSL..101..155W}
{Warner}, B., 1983.
\newblock {The intermediate polars}.
\newblock In M.~{Livio}, G.~{Shaviv}, editors, IAU Colloq. 72: Cataclysmic
  Variables and Related Objects, volume 101 of Astrophysics and Space Science
  Library, pages 155--171.

\bibitem[{{Warner}(1995)}]{1995cvs..book.....W}
{Warner}, B., 1995.
\newblock {Cataclysmic variable stars}.
\newblock Cambridge Astrophysics Series, Cambridge, New York: Cambridge
  University Press, 1995.

\bibitem[{{Whelan} \& {Iben}(1973)}]{1973ApJ...186.1007W}
{Whelan}, J., {Iben}, I., Jr., 1973.
\newblock {Binaries and Supernovae of Type I}.
\newblock \apj, 186, 1007-1014.

\bibitem[{{White} et~al.(1995){White}, {Giommi}, {Heise}
  et~al.}]{1995ApJ...445L.125W}
{White}, N.~E., {Giommi}, P., {Heise}, J., et~al., 1995.
\newblock {RX J0045.4+4154: A recurrent supersoft x-ray transient in M31}.
\newblock \apjl, 445, L125-L128.

\bibitem[{{Williams} et~al.(2004){Williams}, {Garcia}, {Kong}
  et~al.}]{2004ApJ...609..735W}
{Williams}, B.~F., {Garcia}, M.~R., {Kong}, A.~K.~H., et~al., 2004.
\newblock {A Synoptic X-Ray Study of M31 with the Chandra High Resolution
  Camera}.
\newblock \apj, 609, 735-754.

\bibitem[{{Williams} et~al.(2014{\natexlab{a}}){Williams}, {Lang}, {Dalcanton}
  et~al.}]{2014ApJS..215....9W}
{Williams}, B.~F., {Lang}, D., {Dalcanton}, J.~J., et~al., 2014{\natexlab{a}}.
\newblock {The Panchromatic Hubble Andromeda Treasury. X. Ultraviolet to
  Infrared Photometry of 117 Million Equidistant Stars}.
\newblock \apjs, 215, 9.

\bibitem[{{Williams}(2012)}]{2012AJ....144...98W}
{Williams}, R., 2012.
\newblock {Origin of the ``He/N'' and ``Fe II'' Spectral Classes of Novae}.
\newblock \aj, 144, 98.

\bibitem[{{Williams}(1992)}]{1992AJ....104..725W}
{Williams}, R.~E., 1992.
\newblock {The formation of novae spectra}.
\newblock \aj, 104, 725-733.

\bibitem[{{Williams} \& {Darnley}(2017{\natexlab{a}})}]{2017ATel10692....1W}
{Williams}, S.~C., {Darnley}, M.~J., 2017{\natexlab{a}}.
\newblock {Further spectroscopy of the 2017 outburst of PT And}.
\newblock The Astronomer's Telegram, 10692, 1.

\bibitem[{{Williams} \& {Darnley}(2017{\natexlab{b}})}]{2017ATel11088....1W}
{Williams}, S.~C., {Darnley}, M.~J., 2017{\natexlab{b}}.
\newblock {Spectroscopic classification AT 2017jdm as a nova, and likely
  recurrent eruption of M31N 2007-10b}.
\newblock The Astronomer's Telegram, 11088.

\bibitem[{{Williams} \& {Darnley}(2017{\natexlab{c}})}]{2017ATel10647....1W}
{Williams}, S.~C., {Darnley}, M.~J., 2017{\natexlab{c}}.
\newblock {Spectroscopy of AT 2017gay, another outburst of PT And/M31N
  1957-10b}.
\newblock The Astronomer's Telegram, 10647, 1.

\bibitem[{{Williams} et~al.(2014{\natexlab{b}}){Williams}, {Darnley}, {Bode}
  et~al.}]{2014ApJS..213...10W}
{Williams}, S.~C., {Darnley}, M.~J., {Bode}, M.~F., et~al., 2014{\natexlab{b}}.
\newblock {On the Progenitors of Local Group Novae. I. The M31 Catalog}.
\newblock \apjs, 213, 10.

\bibitem[{{Williams} et~al.(2015){Williams}, {Darnley}, {Bode}
  et~al.}]{2015ApJ...805L..18W}
{Williams}, S.~C., {Darnley}, M.~J., {Bode}, M.~F., et~al., 2015.
\newblock {A Luminous Red Nova in M31 and Its Progenitor System}.
\newblock \apjl, 805, L18.

\bibitem[{{Williams} et~al.(2016){Williams}, {Darnley}, {Bode}
  et~al.}]{2016ApJ...817..143W}
{Williams}, S.~C., {Darnley}, M.~J., {Bode}, M.~F., et~al., 2016.
\newblock {On the Progenitors of Local Group Novae. II. The Red Giant Nova Rate
  of M31}.
\newblock \apj, 817, 143.

\bibitem[{{Williams} et~al.(2017){Williams}, {Darnley}, \&
  {Henze}}]{2017MNRAS.472.1300W}
{Williams}, S.~C., {Darnley}, M.~J., {Henze}, M., 2017.
\newblock {Multiwavelength observations of the 2015 nova in the Local Group
  irregular dwarf galaxy IC 1613}.
\newblock \mnras, 472, 1300-1314.

\bibitem[{{Williams} \& {Shafter}(2004)}]{2004ApJ...612..867W}
{Williams}, S.~J., {Shafter}, A.~W., 2004.
\newblock {On the Nova Rate in M33}.
\newblock \apj, 612, 867-876.

\bibitem[{{Woudt} \& {Ribeiro}(2014)}]{2014ASPC..490.....W}
{Woudt}, P.~A., {Ribeiro}, V.~A.~R.~M., editors, 2014.
\newblock {Stella Novae: Past and Future Decades}, volume 490 of Astronomical
  Society of the Pacific Conference Series. Astronomical Society of the
  Pacific, San Francisco.

\bibitem[{{Yaron} et~al.(2005){Yaron}, {Prialnik}, {Shara}
  et~al.}]{2005ApJ...623..398Y}
{Yaron}, O., {Prialnik}, D., {Shara}, M.~M., et~al., 2005.
\newblock {An Extended Grid of Nova Models. II. The Parameter Space of Nova
  Outbursts}.
\newblock \apj, 623, 398-410.

\bibitem[{{Zheng} et~al.(2010){Zheng}, {Romadan}, {Whallon}
  et~al.}]{2010CBET.2574....1Z}
{Zheng}, W., {Romadan}, A., {Whallon}, N., et~al., 2010.
\newblock {Apparent Nova in M31: M31N 2010-12a}.
\newblock Central Bureau Electronic Telegrams, 2574.

\end{thebibliography}

\end{document}